\documentclass[prd,preprint,eqsecnum,nofootinbib,amsmath,amssymb,
               tightenlines,dvips,superscriptaddress]{revtex4}
\usepackage{graphicx}
\usepackage{bm}
\usepackage{amsfonts}
\usepackage{amssymb}

\usepackage{dsfont}

\def\b{{\bm b}}

\def\p{{\bm p}}

\def\A{{\bm A}}
\def\B{{\bm B}}
\def\C{{\bm C}}
\def\P{{\bm P}}

\def\bcalP{{\bm{\mathcal P}}}
\def\bcalT{{\bm{\mathcal T}}}
\def\uOmega{\underline{\Omega}}
\def\Ybar{\overline{Y}}

\def\Nc{N_{\rm c}}
\def\CA{C_{\rm A}}

\def\alphas{\alpha_{\rm s}}
\def\alphaEM{\alpha_{\scriptscriptstyle\rm EM}}
\def\Re{\operatorname{Re}}

\def\sgn{\operatorname{sgn}}
\def\sh{\operatorname{sh}}

\def\cth{\operatorname{cth}}
\def\grad{{\bm\nabla}}

\def\ix{{\rm i}}
\def\fx{{\rm f}}
\def\xx{{\rm x}}
\def\xbx{{\bar{\rm x}}}
\def\yx{{\rm y}}
\def\ybx{{\bar{\rm y}}}
\def\zx{{\rm z}}
\def\Bx{\xbx}
\def\Ax{\ybx}
\def\bx{\yx}
\def\ax{\xx}

\def\yfrak{{\mathfrak y}}

\def\seq{{\rm seq}}

\def\calX{{\cal X}}

\def\QED{{\scriptscriptstyle\rm QED}}
\def\Beta{\operatorname{B}}

\begin {document}



\title
    {
      The LPM effect in sequential bremsstrahlung: \\
      dimensional regularization
    }

\author{Peter Arnold}
\affiliation
    {%
    Department of Physics,
    University of Virginia,
    Charlottesville, Virginia 22904-4714, USA
    \medskip
    }%
\author{Han-Chih Chang}
\affiliation
    {%
    Department of Physics,
    University of Virginia,
    Charlottesville, Virginia 22904-4714, USA
    \medskip
    }%
\author{Shahin Iqbal}
\affiliation
    {%
    National Centre for Physics, \\
    Quaid-i-Azam University Campus,
    Islamabad, 45320 Pakistan
    \medskip
    }%

\date {\today}

\begin {abstract}%
{%
The splitting processes of bremsstrahlung and pair production in a medium
are coherent over large distances in the very high energy limit,
which leads to a suppression known as the Landau-Pomeranchuk-Migdal
(LPM) effect.
Of recent interest is
the case when the coherence
lengths of two consecutive splitting processes overlap (which is
important for understanding corrections to standard treatments
of the LPM effect in QCD).
In previous papers, we have developed methods for
computing such corrections without making soft-gluon approximations.
However, our methods require consistent
treatment of canceling ultraviolet (UV)
divergences associated with coincident emission times, even
for processes with tree-level amplitudes.
In this paper, we show how to use dimensional regularization to
properly handle the UV contributions.
We also present a simple diagnostic test that any
consistent UV regularization method
for this problem needs to pass.
}%
\end {abstract}

\maketitle
\thispagestyle {empty}

{\def\boldmath{}\tableofcontents}
\newpage


\section{Introduction}
\label{sec:intro}

When passing through matter, high energy particles lose energy by
showering, via the splitting processes of hard bremsstrahlung and pair
production.  At very high energy, the quantum mechanical duration of
each splitting process, known as the formation time, exceeds the mean
free time for collisions with the medium, leading to a significant
reduction in the splitting rate known as the Landau-Pomeranchuk-Migdal
(LPM) effect \cite{LP,Migdal}.  As we will review shortly,
calculations of the LPM effect must typically deal with ultraviolet
(UV) divergences in intermediate steps, associated with effectively-vacuum
evolution between nearly coincident times.  For the case of computing
single splitting rates, these divergences are trivial to deal with
(either by subtracting out the vacuum rate a priori, or by using an
appropriate $i\epsilon$ prescription).  However, for the case of two
consecutive splittings with overlapping formation times (which we will
loosely characterize as ``double bremsstrahlung''), the treatment of
ultraviolet divergences is much more difficult.
In previous work \cite{2brem}, an $i\epsilon$ prescription was proposed
for dealing with this problem.  Here, we will explain why that
prescription was incomplete and missed certain contributions to
the result.
Then we will show how to correctly regulate the
ultraviolet using dimensional regularization
and will use our results to correct the QCD LPM analysis of ref.\ \cite{2brem}.
In addition,
we provide a simple example---QED double bremsstrahlung in an
independent emission approximation---that can be used as a test of
the self-consistency of UV regularization prescriptions.

For simplicity of discussion, and in order to make contact with the
double bremsstrahlung calculation of refs.\ \cite{2brem,seq}, we will restrict
attention in this paper to the case of medium-induced bremsstrahlung from
an (approximately) on-shell particle traversing a thick, uniform medium
in the multiple scattering ($\hat q$) approximation.
(Thick means large compared to the formation time of the
bremsstrahlung radiation.)
However, the
same methods should be useful for double bremsstrahlung
in other situations.


\subsection{Examples of UV divergences}

\subsubsection{Single splitting}

\begin {figure}[t]
\begin {center}
  \includegraphics[scale=0.5]{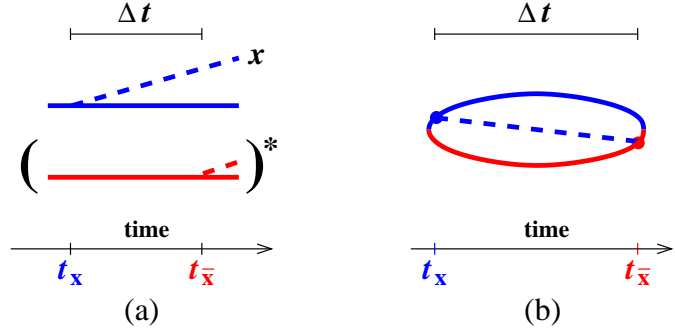}
  \caption{
     \label{fig:single}
     (a) Interference contribution corresponding to the single
     bremsstrahlung rate, and (b) an equivalent diagrammatic
     representation, where the amplitude (blue) and conjugate
     amplitude (red) are drawn sewn together.
     To simplify the drawing, all particles, including
     gluons, are indicated by straight lines.
     Only the high-energy parent and its two daughters are
     shown explicitly; the multiple interactions of these
     particles with the medium are implicit and not shown.
  }
\end {center}
\end {figure}

The standard result for the single splitting rate
in a thick, uniform medium in multiple scattering approximation
is \cite{BDMS}%
\footnote{
  For a discussion specifically in the notation used in this paper,
  see section II of Ref.\ \cite{2brem}.
}
\begin {equation}
   \frac{d\Gamma}{dx} =
   - \frac{\alpha P(x)}{\pi} \Re
   \int_0^\infty d(\Delta t) \> \Omega^2 \csc^2(\Omega\,\Delta t) ,
\label {eq:dGdx}
\end {equation}
where $x$ is the momentum fraction
of one of the daughters, and $P(x)$ is the corresponding (vacuum)
Dokshitzer-Gribov-Lipatov-Altarelli-Parisi (DGLAP)
splitting function.  $\Delta t$ represents the time between emission
in the amplitude and emission in the conjugate amplitude, as depicted
in fig.\ \ref{fig:single}.  $\Omega$ is a complex frequency that
characterizes the evolution and decoherence of this interference
contribution with time $\Delta t$.  In the QCD case of $g \to gg$,
for example, it is given by
\begin {equation}
   \Omega =
   \sqrt{
     - \frac{i \hat q_{\rm A}}{2 E}
     \left( -1 + \frac{1}{1{-}x} + \frac{1}{x} \right)
   } ,
\label {eq:OmegaSingle}
\end {equation}
where $\hat q$ characterizes transverse momentum diffusion
of a high-energy particle due to interactions with the medium
(average $Q_\perp^2 = \hat q t$) and $\hat q_{\rm A}$ indicates
$\hat q$ for an adjoint-color particle, i.e.\ a gluon.
$E$ is the initial particle energy.
Throughout this paper, we will focus on splitting rates that have
been integrated over the final transverse momenta of the
(nearly collinear) daughters.

The negative imaginary part of $\Omega$ accounts for
decoherence of interference (such as fig.\ \ref{fig:single}) at
large time separation, due to random interactions with the medium.
In particular, the integrand in (\ref{eq:dGdx}) falls exponentially
for $|\Omega| \, \Delta t \gg 1$, and so the integral is
infrared ($\Delta t{\to}\infty$) convergent.

The details of the above formulas are not important yet.
What is important is the behavior of the integrand in
(\ref{eq:dGdx}) as $\Delta t\to 0$:
\begin {equation}
   \frac{d\Gamma}{dx} \to
   - \frac{\alpha P(x)}{\pi} \Re
   \int_0 d(\Delta t)
   \left[
     \frac{1}{(\Delta t)^2} + O\bigl( (\Delta t)^0 \bigr)
   \right] ,
\label {eq:dGdxUV}
\end {equation}
which makes the integral UV divergent.  Note that this divergence does
not depend on the medium parameter $\hat q$ and so represents a
purely vacuum contribution to the rate.  One way to deal with it
is to note that an on-shell particle cannot split in vacuum, and
so the purely vacuum contribution must vanish.  One may then sidestep
the technical issue of regulating the divergence by subtracting the
necessarily-vanishing vacuum ($\hat q{\to}0$) contribution from
(\ref{eq:dGdx}) to get the convergent integral
\begin {align}
   \frac{d\Gamma}{dx} &=
   - \frac{\alpha P(x)}{\pi} \Re
   \int_0^\infty d(\Delta t)
   \left[
     \Omega^2 \csc^2(\Omega\,\Delta t)
     - \frac{1}{(\Delta t)^2}
   \right]
\nonumber\\
   &=
   \frac{\alpha P(x)}{\pi} \, \Re(i \Omega) .
\label {eq:dGdx2}
\end {align}

An alternative way to deal with the divergence is to use an
$i\epsilon$ prescription.  Notice that
$\Delta t \equiv t_\xbx - t_\xx$ in fig.\ \ref{fig:single} has the
form of (i) a time in the conjugate amplitude minus (ii) a time
in the amplitude.  The correct $i\epsilon$ prescription here is that
conjugate amplitude times should be thought of as being infinitesimally
displaced in the negative imaginary direction compared to amplitude
times,%
\footnote{
   See, for example, the discussion in section VII.A of ref.\ \cite{2brem}.
}
and so $\Delta t$ in (\ref{eq:dGdx}) should be replaced by
$\Delta t - i\epsilon$.
The $\Delta t\to 0$ behavior (\ref{eq:dGdxUV}) is then
\begin {equation}
   \frac{d\Gamma}{dx} \to
   - \frac{\alpha P(x)}{\pi} \Re
   \int_0 d(\Delta t)
   \left[
     \frac{1}{(\Delta t - i\epsilon)^2} + O\bigl( (\Delta t)^0 \bigr)
   \right] .
\end {equation}
In this case, the UV piece of integration from $\Delta t\to 0$ does
not generate a real part.  That is, we can separate out the
contribution proportional to
\begin {equation}
   \Re \int_0^\infty
     \frac{d(\Delta t)}{(\Delta t - i\epsilon)^2} = 0
\end {equation}
from the calculation (\ref{eq:dGdx}) of $d\Gamma/dx$,
which leaves us again with (\ref{eq:dGdx2}).


\subsubsection{Double splitting}

\begin {figure}[t]
\begin {center}
  \includegraphics[scale=0.5]{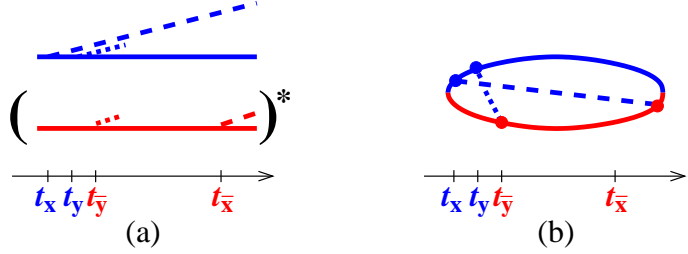}
  \caption{
     \label{fig:pole}
     An example of an interference that contributes to the double
     splitting rate, drawn in the same style as fig.\ \ref{fig:single}.
     A divergence of the form (\ref{eq:pole}) arises from
     either the first three times approaching each
     other (depicted here) or the last three times
     approaching each other.
  }
\end {center}
\end {figure}

Similar purely-vacuum divergences, which may also be easily subtracted,
arise in the calculation of overlapping double splitting
(e.g.\ overlapping double bremsstrahlung).
However, as discussed in ref.\ \cite{2brem}, there are also
sub-leading UV divergences which are not so easily discarded.
These arise from situations such as depicted in
fig.\ \ref{fig:pole}, in the limit where three of the four
emission times become arbitrarily close together.
In that short-time limit, the evolution of the system between the three
close times becomes essentially vacuum evolution.
But the
evolution of the system from there to the further-away
fourth time is not vacuum evolution and depends on $\hat q$,
and so this sub-leading
divergence will not be subtracted away by subtracting the
(vanishing) vacuum result for double splitting of an on-shell
particle.  Ref.\ \cite{2brem} found that the surviving divergence
of each interference diagram could be written in the form of
a sum of terms proportional to
\begin {equation}
   \Re \int_0 d(\Delta t) \> \frac{i\Omega}{\Delta t} ,
\label {eq:pole}
\end {equation}
where
$\Delta t$
characterizes the small separation of the three
emission times that are approaching each other
(e.g.\ $t_\xx$, $t_\yx$, and $t_\ybx$ in fig.\ \ref{fig:pole})
and $\Omega$ is the complex frequency characterizing
the medium evolution associated with the fourth time
(the right-hand part of each diagram in fig.\ \ref{fig:pole}).%
\footnote{
  As an example of a precise formula, see eq.\ (5.46) of
  ref.\ \cite{2brem}, in which $\Delta t \equiv t_\ybx-t_\yx$ is
  the separation of the two intermediate times in fig.\ \ref{fig:pole}
  of this paper.  For an argument that this particular separation also
  characterizes the separation of those two times from $t_\xx$ in
  the calculation of the divergence, see appendix D2 of ref.\ \cite{2brem}.
}

Refs.\ \cite{2brem,seq} found that all the divergences (\ref{eq:pole})
naively cancel each other in the sum over all interference
contributions to the double splitting rate.  But one must be careful,
because finite contributions may still arise from the pole at
$\Delta t=0$.  As a simple mathematical example \cite{2brem},
the unregulated expression
\begin {equation}
  \int_0^\infty \frac{d(\Delta t)}{\Delta t}
  -
  \int_0^\infty \frac{d(\Delta t)}{\Delta t}
\end {equation}
naively looks to be zero, but if it were regularized as
\begin {equation}
  \int_0^\infty \frac{d(\Delta t)}{\Delta t - i\epsilon}
  -
  \int_0^\infty \frac{d(\Delta t)}{\Delta t + i\epsilon}
\end {equation}
then it would instead equal $i\pi$.

Ref.\ \cite{2brem} attempted to find $i\epsilon$ prescriptions for
the poles, replacing each term (\ref{eq:pole}) by
\begin {equation}
   \Re \int_0 d(\Delta t) \> \frac{i\Omega}{\Delta t \pm i\epsilon}
\label {eq:poleeps}
\end {equation}
after arguing what the sign $\pm$ of the $i\epsilon$ prescription
should be for each interference contribution to the double splitting rate.
Unfortunately, this prescription
turns out to miss some additional contributions
from $\Delta t = 0$.
After discussing a relatively simple diagnostic
test, we will explain
(in section \ref{sec:epsbad})
what went wrong with
the substitution (\ref{eq:poleeps}).

For reasons we will discuss later, attempting to fix up the $i\epsilon$ method
for evaluating the pole contributions in double splitting
seems complicated and fraught with subtlety.
Fortunately, there is a cleaner, surer way to deal with the UV regularization
of individual diagrams:
We will show how to use dimensional regularization to compute the
pole contributions from $\Delta t = 0$.
Dimensional regularization will turn the UV-divergent integral
$\int_0 d(\Delta t)/\Delta t$ in (\ref{eq:pole})
into the UV-regularized integral
$\int_0 d(\Delta t)/(\Delta t)^{1-\epsilon/2}$.
Reassuringly, we find that
dimensional regularization passes our diagnostic.

The precise details of exactly how and why, for each diagram, the
earlier work \cite{2brem} chose the sign of $\pm i\epsilon$ in the
denominator of (\ref{eq:poleeps}) will not be very important to the
current discussion.  However, for interested readers, we give a very
brief review in appendix \ref{app:epsnaive}.


\subsection {Outline and Referencing}

\begin {figure}[t]
\begin {center}
  \includegraphics[scale=0.5]{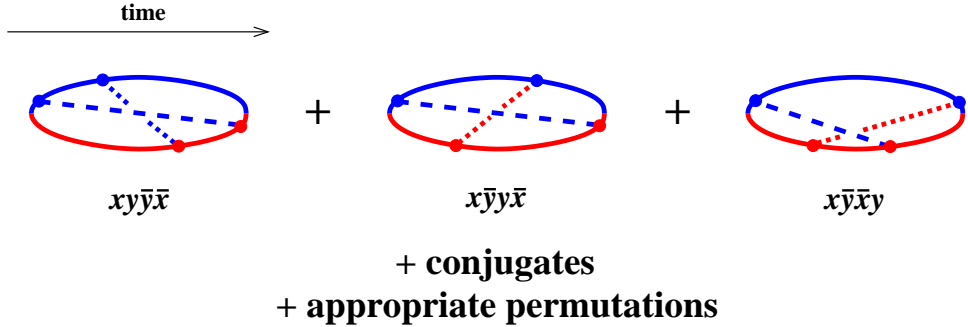}
  \caption{
     \label{fig:subset2}
     Interference contributions contributing to double splitting,
     showing the subset referred to as ``crossed'' diagrams in
     ref.\ \cite{2brem}.
     For QED double bremsstrahlung, the ``appropriate permutations'' are
     simply $x\leftrightarrow y$.  For $g \to ggg$ in QCD, they
     are instead all permutations of $x$, $y$, and $z \equiv 1{-}x{-}y$.
  }
\end {center}
\end {figure}

\begin {figure}[t]
\begin {center}
  \includegraphics[scale=0.5]{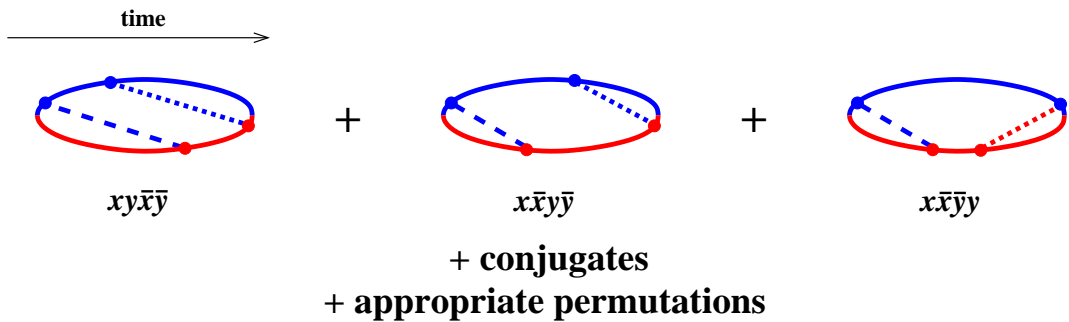}
  \caption{
     \label{fig:seq2}
     As fig.\ \ref{fig:subset2} except for the ``sequential''
     diagrams of ref.\ \cite{seq}.
  }
\end {center}
\end {figure}

In the next section, we present a diagnostic that any consistent
UV regularization scheme should satisfy.  We then discuss what
a successful $i\epsilon$ prescription would have to do to pass
that test and show how the simple prescription (\ref{eq:poleeps})
fails.  We will characterize the type of contributions that
(\ref{eq:poleeps}) misses (which we call ``$1/\pi^2$\kern1pt''
pole pieces).
In section \ref{sec:single}, we turn to dimensional regularization by
warming up with the case of the single splitting rate (\ref{eq:dGdx}).
Sections \ref{sec:cross} and \ref{sec:seq} then apply dimensional
regularization to overlapping double splitting, treating,
respectively, what we call the ``crossed'' interference diagrams
of fig.\ \ref{fig:subset2} and the ``sequential''
diagrams of fig.\ \ref{fig:seq2}.
Section \ref{sec:DRtest} verifies that dimensional regularization
passes the diagnostic test of section \ref{sec:diagnostic}.
Finally, we present a summary of results in
section \ref{sec:summary}.
Various matters along the way are left for appendices.

In this paper, we will occasionally (in footnotes and appendices)
use the author acronyms AI and ACI as shorthand for
Arnold and Iqbal \cite{2brem} and Arnold, Chang and Iqbal \cite{seq}
so that, for example, we may write ``ACI (5.2)'' to refer to
eq.\ (5.2) of ref.\ \cite{seq}.


\section{A diagnostic}
\label{sec:diagnostic}

\subsection{The QED independent emission test}

\begin {figure}[t]
\begin {center}
  \includegraphics[scale=0.5]{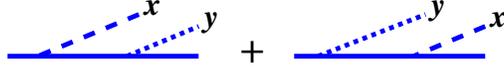}
  \caption{
     \label{fig:2bremQED}
     Amplitude for double bremsstrahlung in QED.
  }
\end {center}
\end {figure}

Consider the case of double bremsstrahlung in QED, such as shown
in fig.\ \ref{fig:2bremQED}.
In particular,
consider the soft limit where the momentum fractions $x$ and $y$
carried by the two photons are small: $x,y \ll 1$.
In that case, the backreaction on the initial high-energy electron is
negligible, and so we might expect that the $x$ and $y$ emission are
independent from each other:
\begin {equation}
   \frac{dI}{dx\,dy} \simeq \frac{dI}{dx} \times \frac{dI}{dy} ,
\label {eq:indep}
\end {equation}
where $dI/dx\,dy$ is the differential probability for double
bremsstrahlung and $dI/dx$ is the differential probability for
single bremsstrahlung.  We will refer to (\ref{eq:indep}), and
similar formulas later for emission rates, as the independent
emission model.

Now consider an idealized Monte Carlo (IMC)
description of shower development, based
on the rates for {\it single} splitting (such as single bremsstrahlung
and single pair production).
In ref.\ \cite{seq}, it is explained that the important quantity
for characterizing corrections to such a Monte Carlo due to overlapping
formation times is the difference of actual double splitting rates
from what such a Monte Carlo would predict for two consecutive splittings:
\begin {equation}
   \Delta\, \frac{d\Gamma}{dx\,dy} \equiv
   \frac{d\Gamma}{dx\,dy}
   -
   \left[ \frac{d\Gamma}{dx\,dy} \right]_{\rm IMC} .
\label{eq:DeltaDef}
\end {equation}
However, the independent emission approximation described above
already assumed that there were no effects from overlapping
formation times, and so
\begin {equation}
   \Delta\, \frac{d\Gamma}{dx\,dy} = 0
\label{eq:Dindep}
\end {equation}
in the independent emission approximation for $d\Gamma/dx\,dy$.
This seems trivial.  Nonetheless, we will obtain below an interesting test
by reorganizing the individual terms that contribute to
(\ref{eq:Dindep}).

\begin {figure}[t]
\begin {center}
  \includegraphics[scale=0.5]{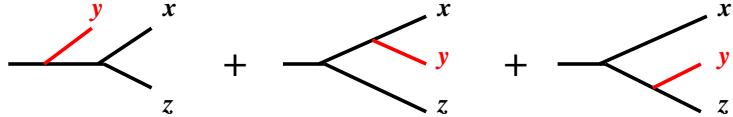}
  \caption{
     \label{fig:MC2}
     Idealized Monte Carlo (IMC) picture of two successive splittings in QCD.
  }
\end {center}
\end {figure}

We have chosen QED rather than QCD bremsstrahlung for our test
because the independent emission approximation is not the same thing
as the Monte Carlo approximation to double bremsstrahlung
in QCD.  QCD Monte Carlo based on single splitting rates
allows for the second bremsstrahlung to be independently emitted from
{\it either} daughter of the first bremsstrahlung process, as shown in
fig \ref{fig:MC2}.  That means
that the Monte Carlo probability of $y$ emission is different
depending on whether the $y$ emission happens before or after the $x$
emission.  For QCD, Monte Carlo is therefore inconsistent
with the independent emission model of (\ref{eq:indep}), since
in (\ref{eq:indep})
the two emissions do not affect each other in any way.%
\footnote{
  For some use of the independent emission approximation
  in QCD, see appendix B3 of ref.\ \cite{seq}.
}

To turn (\ref{eq:Dindep}) into a test, split
$d\Gamma/dx\,dy$ for double bremsstrahlung into
a sum over the different possible time orderings of the
emissions, shown in figs.\ \ref{fig:subset2} and \ref{fig:seq2}.
Then, in the independent emission approximation,
\begin {equation}
   2\Re(
       xy\bar y\bar x + x\bar y y\bar x + x\bar y\bar x y
       + xy\bar x\bar y
       + x\bar x y\bar y + x\bar x \bar y y
   )
   + (x \leftrightarrow y)
   - {\rm IMC}
   = 0 .
\label {eq:test}
\end {equation}
Following refs.\ \cite{2brem,seq}, the notation $xy\bar y\bar x$
indicates that the emissions in the corresponding diagram of
fig.\ \ref{fig:subset2} happen in the time order of (i) $x$ emission
in the amplitude, followed by (ii) $y$ emission in the amplitude,
followed by (iii) $y$ emission in the conjugate amplitude, and
finally (iv) $x$ emission in the conjugate amplitude.

There are two ways to make use of (\ref{eq:test}).
One is to apply it to the {\it full}\/ calculation (as opposed
to the independent emission approximation) of the
different double bremsstrahlung interference diagrams in QED,
expand the results in small $x$ and $y$, and
then check whether the equality (\ref{eq:test}) holds to
leading order in that expansion.%
\footnote{
   A more precise statement: Calculate the (UV finite)
   total crossed diagram
   contribution of fig.\ \ref{fig:subset2} to leading order in
   small $x$ and $y$, which turns out to be order $1/x^{1/2} y$ for
   $y \lesssim x \ll 1$ in QED.
   Similarly calculate the total sequential diagram
   contribution of fig.\ \ref{fig:seq2} minus the idealized Monte Carlo
   prediction.
   Then check that these two
   results cancel each other at order $1/x^{1/2} y$.
}

The other use is to directly investigate how (\ref{eq:test}) works in
the independent emission model itself, using one's favorite
UV regularization of single bremsstrahlung rates.  We'll now do
just that for the case of $i\epsilon$ prescriptions.


\subsection{Application to \boldmath$i\epsilon$ prescriptions}
\label{sec:epsbad}

Consider use of an $i\epsilon$ prescription to UV-regulate the
single splitting rate of (\ref{eq:dGdx}):
\begin {equation}
   \frac{d\Gamma}{dx} =
   - \frac{\alpha P(x)}{\pi} \Re
   \int_0^\infty d(\Delta t_x) \>
   \Omega^2 \csc^2\bigl(\Omega(\Delta t_x-i\epsilon)\bigr) .
\label {eq:dGdxeps}
\end {equation}
Appendix \ref{app:epstest} shows that the test (\ref{eq:test}) is
indeed satisfied (as it must be) if we use (\ref{eq:dGdxeps}) in
the independent emission model.
Here we want to focus on the UV contributions to
that test, which we have summarized in the second column of
Table \ref{tab:epstest}.
That column shows the small $\Delta t$ limit of
the $d(\Delta t)$ integrands for each diagram,
corresponding to the pole pieces in
(\ref{eq:pole}).
In this table, we have further assumed $y \ll x$ just to make the
formulas as simple (and so easy to compare) as possible, and we have
also introduced the short-hand notation
\begin {equation}
   \Delta t_\pm \equiv \Delta t \pm i\epsilon .
\label {eq:Deltatpm}
\end {equation}

\begin {table}[t]
\tabcolsep 10pt
\begin {tabular}{|c|c|c|}
\hline
  & independent emission & $i\epsilon$ method \\
  & approximation & of ref.\ \cite{2brem} \\
\hline
$2\Re(yx\bar x\bar y)$ & 0 & same \\
$2\Re(y\bar xx\bar y)$ & 0 & same \\
$2\Re(xy\bar y\bar x)$
   & $\frac12 \Re\bigl(\frac{i\Omega_x}{\Delta t_-}\bigr)$
   & same \\[4pt]
$2\Re(x\bar y y\bar x)$
   & $\frac12 \Re\bigl(\frac{i\Omega_x\Delta t_-}
                            {(\Delta t_+)^2}\bigr)$
   & $\frac12 \Re\bigl(\frac{i\Omega_x}{\Delta t_+}\bigr)$
   \\[4pt]
$2\Re(x\bar y\bar x y)$
   & $-\frac12 \Re\bigl(\frac{i\Omega_x}{\Delta t_+}\bigr)$
   & same \\[4pt]
$2\Re(\bar y xy\bar x)$
   & $-\frac12 \Re\bigl(\frac{i\Omega_x}{\Delta t_+}\bigr)$
   & same \\[4pt]
\hline
$2\Re(x y\bar x\bar y)$
   & $-\frac12 \Re\bigl(\frac{i\Omega_x}{\Delta t_-}\bigr)$
   & same \\[4pt]
$2\Re(y x\bar y\bar x)$
   & $-\frac12 \Re\bigl(\frac{i\Omega_x}{\Delta t_-}\bigr)$
   & same \\[4pt]
$2\Re(x \bar x y\bar y + x\bar x \bar y y) + (x\leftrightarrow y) - {\rm IMC}$
   & $\Delta t \Re(i\Omega_x)\Re\bigl( \frac{1}{(\Delta t_-)^2} \bigr)$
   & $\Re(i\Omega_x)\Re\bigl( \frac{1}{\Delta t_-} \bigr)$
   \\[4pt]
\hline
\end {tabular}
\caption
    {
    The UV ($\Delta t \to 0$)
    behavior of individual contributions to the test (\ref{eq:test})
    for QED double bremsstrahlung in the limit $y \ll x \ll 1$.
    Purely-vacuum divergences ($\hat q = 0$) have already been subtracted
    and are not shown.  (See appendix \ref{app:epstest} for additional
    clarification on what is included here.)
    Above, $\Delta t_\pm \equiv \Delta t \pm i\epsilon$.
    Each entry is to be integrated over (small) $\Delta t$ as in
    (\ref{eq:pole}) and multiplied by $4\alphaEM^2/\pi^2 x y$, where
    $\alphaEM$ is the fine structure constant.  The small-$x$ complex
    frequency $\Omega_x$ above is
    defined by $\Omega_x^\QED \equiv \sqrt{-i x\hat q/2E}$.
    Following ref.\ \cite{2brem}, $\Delta t$ (no subscript) above is
    defined as the time separation between the middle two emissions for
    all but the last entry: e.g.\ $\Delta t \equiv t_\ybx-t_\yx$ for
    $xy\bar y\bar x$ and $\Delta t \equiv t_\xbx-t_\ybx$ for $x\bar y\bar x y$.
    All other times have been integrated over.
    (For the identification of $\Delta t$ in the last entry, see
    section IIA of ref.\ \cite{seq} or appendix \ref{app:seqINDEPeps} here.)
    \label {tab:epstest}
    }
\end {table}

One may now see the problem with the earlier proposal
(\ref{eq:poleeps}) of ref.\ \cite{2brem} for how to regulate the
$1/\Delta t$ poles in double bremsstrahlung calculations using the
$i\epsilon$ prescription.  It is true that, if you ignore $\epsilon$
prescriptions, then all of the small-$\Delta t$ divergences in
Table \ref{tab:epstest} take the form $1/\Delta t$.  But this does
not mean that putting in $i\epsilon$'s gives $1/(\Delta t \pm i\epsilon)$.
For example, the $2\Re(x\bar yy\bar x)$ entry of the table reads
\begin {equation}
  2\Re \left[ \frac{d\Gamma}{dx\,dy} \right]_{x \bar y y \bar x}
  \simeq
  \frac{4\alphaEM^2}{\pi^2 x y}
  \int_0 d(\Delta t) \>
  \frac12 \Re\Bigl(\frac{i\Omega_x(\Delta t - i\epsilon)}
                            {(\Delta t + i\epsilon)^2}\Bigr) .
\label {eq:xybyxbpole}
\end {equation}
The problem with the
earlier analysis of ref.\ \cite{2brem}
is that it
correctly identified the $i\epsilon$ prescription in denominators,%
\footnote{
  If details desired, see
  the brief review in appendix \ref{app:epsnaive} of this paper,
  which points to appendix D3 of ref.\ \cite{2brem}.
}
but failed to realize that $1/\Delta t = (\Delta t)/(\Delta t)^2$ could
also have different and important $i\epsilon$ dependence in a
{\it numerator}.
The prescription proposed by ref.\ \cite{2brem} is shown in
the last column of table \ref{tab:epstest} and can be obtained
from the second column by replacing the full $\Delta t$ dependence
by just $1/\Delta t$, with the $i\epsilon$ dependence taken from
the denominator in the second column.
So, for instance, the $2\Re(x\bar y y \bar x)$
contribution (\ref{eq:xybyxbpole}) is replaced by
\begin {equation}
  2\Re \left[ \frac{d\Gamma}{dx\,dy} \right]^{\rm naive}_{x \bar y y \bar x}
  \simeq
  \frac{4\alphaEM^2}{\pi^2 x y}
  \int_0 d(\Delta t) \>
  \frac12 \Re\Bigl(\frac{i\Omega_x}
                            {\Delta t + i\epsilon}\Bigr) .
\label {eq:xybyxbpoleAI}
\end {equation}
The difference between the prescriptions
(\ref{eq:xybyxbpole}) and (\ref{eq:xybyxbpoleAI})
gives an integral that is
dominated by $\Delta t \sim \epsilon$ and so is a purely ``pole'' contribution
that can be evaluated without needing to know how the integrand behaves for
large $t$:
\begin {align}
  2\Re \biggl(
  \left[ \frac{d\Gamma}{dx\,dy} \right]_{x \bar y y \bar x}
  - {}&
  \left[ \frac{d\Gamma}{dx\,dy} \right]^{\rm naive}_{x \bar y y \bar x}
  \biggr)
\nonumber\\
  &=
  \frac{2\alphaEM^2}{\pi^2 x y} \Re
  \left\{
    i\Omega_x \int_0^\infty d(\Delta t)
    \left[
      \frac{(\Delta t - i\epsilon)}{(\Delta t + i\epsilon)^2}
      -
      \frac{1}{(\Delta t + i\epsilon)}
    \right]
  \right\}
\nonumber\\
  &=
  \frac{4\alphaEM^2}{\pi^2 x y} \, \Re(-i\Omega_x) .
\end {align}
The one other difference in Table \ref{tab:epstest} can be evaluated
similarly.
For the $y\ll x \ll 1$ limit of
the QED independent emission approximation, the total difference
(summing all contributions) between
the correct $i\epsilon$ prescription and the naive prescription of
(\ref{eq:poleeps}) is
\begin {equation}
  \frac{d\Gamma}{dx\,dy}
  -
  \left[ \frac{d\Gamma}{dx\,dy} \right]^{\rm naive}
  \simeq
  \frac{4\alphaEM^2}{\pi^2 x y} \, 2 \Re(-i\Omega_x) .
\label {eq:discrepancy}
\end {equation}
This is non-zero.
Because the independent emission calculation must (and does) satisfy the test
(\ref{eq:test}),
we see that the naive prescription of ref.\ \cite{2brem} does not.

Rather than only check this failure of the naive prescription in the
context of the independent emission calculation, we have also done
a full QED calculation
(not making any assumptions about the
size of $x$ and $y$) of the LPM effect in double bremsstrahlung,
along lines similar to the QCD calculation in refs.\ \cite{2brem,seq}.
We have verified that the $y \ll x \ll 1$ limit of those full results
reproduce the last column of table \ref{tab:epstest} if we use the
naive $i\epsilon$ prescription of (\ref{eq:poleeps}) following
ref.\ \cite{2brem}.  We have also verified that the diagnostic test
(\ref{eq:test}) fails in this limit by exactly the amount
(\ref{eq:discrepancy}).  The details of the full QED calculation
are not directly relevant to our current task, which is to
find a clear, correct method for determining the UV contributions
which works not just in the context of the QED independent emission
approximation but generalizes to QCD and to any values of $x$ and $y$.
So we will defer presenting the details of full QED results for
double bremsstrahlung to future work.

Why do we not simply fix up the $i\epsilon$ prescription in
the general case, to make it work correctly like in the second column
of Table \ref{tab:epstest}?  Despite various attempts, we were unable
to find a convincing generalization that worked outside of the
limiting case of $y \ll x \ll 1$.  We briefly discuss the issues we
encountered in appendix \ref{app:epsgeneral}.
Here, we will instead turn to dimensional regularization
for the general case.


\subsection{\boldmath$1/\pi$ vs.\ \boldmath$1/\pi^2$ pole terms}

Before moving on, it is interesting to note a qualitative difference
between what the naive $i\epsilon$ prescription does account for
and what it does not.  As an example, consider the naive prescription
small-$\Delta t$ behavior (\ref{eq:xybyxbpoleAI}).  Use the identity
\begin {equation}
   \frac{1}{\Delta t \mp i\epsilon} = 
   \operatorname{P.P.}\frac{1}{\Delta t} \pm i \pi \, \delta(\Delta t) ,
\end {equation}
where ``$\operatorname{P.P.}$'' indicates the principal part
prescription.  The real-valued $1/\Delta t$ integration terms
(the ``principal part'' terms above)
will cancel among all the diagrams, as mentioned earlier, leaving
a finite total result.
The pole contributions (in the naive $i\epsilon$ prescription) are
given by the $i\pi\,\delta(\Delta t)$ term above.  Note that this term is
associated with an extra factor of $\pi$.  So, for instance,
the corresponding pole piece of (\ref{eq:xybyxbpoleAI}) is given by
\begin {equation}
  2\Re \left[ \frac{d\Gamma}{dx\,dy} \right]^{\rm naive~pole}_{x \bar y y \bar x}
  \simeq
  \frac{4\alphaEM^2}{\pi^2 x y}
  \int_0 d(\Delta t) \>
  \frac12 \Re(\Omega_x) \, \pi \, \delta(\Delta t)
  =
  \frac{\alphaEM^2}{\pi x y} \, \Re(\Omega_x)
\label {eq:1pi}
\end {equation}
(in which integrating over only half a $\delta$ function is understood
to give $\frac12$).%
\footnote{
  The factors of 2 are irrelevant to the point here, but
  see section VII.B.1 of ref.\ \cite{2brem}
  if a more convincing discussion
  that does not rely on $\delta$ functions is desired.
}
Looking at the factors of $\pi$ in (\ref{eq:1pi}),
the original, naive $i\epsilon$
prescription (\ref{eq:poleeps})
gives all of what we will call the ``$1/\pi$'' terms for the
pole contribution.  The correct analysis of the problem,
in contrast, introduces additional
$1/\pi^2$ pole contributions such as (\ref{eq:discrepancy}).

We should also mention that if one analyzes all the entries this way,
then the $1/\pi$ terms produced by the third column of table
\ref{tab:epstest} add up to zero.  This turns out to be an artifact
of the $y \ll x \ll 1$ limit, and there is no such cancellation in the
more general case.


\section{Single splitting with dimensional regularization}
\label {sec:single}

We now turn to dimensional regularization and will start with
the single splitting formula (\ref{eq:dGdx}).
As reviewed in the introduction, dealing with the UV divergence for
single splitting by other means is trivial, but it will provide
a simple and useful warm-up example.


\subsection{Straightforward method}
\label {sec:straighforward}

Following Zakharov \cite{Zakharov}, one way to view
the source of the single splitting formula (\ref{eq:dGdx})
is as an effective 2-dimensional non-Hermitian non-relativistic
quantum mechanics problem for the three high-energy particles shown in
fig.\ \ref{fig:single}b.  Using symmetries of the problem,
the three-particle quantum mechanics problem can be reduced to
a one-particle quantum mechanics problem.
In this language, the basic formula
corresponding to fig.\ \ref{fig:single} is
\begin {equation}
   \frac{d\Gamma}{dx} =
   \frac{\alpha P(x)}{[x(1-x)E]^2} \Re
   \int_0^\infty d(\Delta t) \>
   \grad_{\B_\xbx} \cdot \grad_{\B_\xx}
   \langle \B_\xbx,\Delta t | \B_\xx,0 \rangle
   \Bigr|_{\B_\xbx = \B_\xx = 0} ,
\label {eq:dGdxQM}
\end {equation}
where $\B$ is a single, convenient combination of the transverse positions of
the three particles.
For a more complete discussion in the
notation used here, see section II of ref.\ \cite{2brem}.
In the multiple scattering ($\hat q$) approximation appropriate
for high energy particles traversing thick media, the problem
turns out to become a harmonic oscillator problem with complex frequency
$\Omega$ and mass
\begin {equation}
   M = x(1{-}x) E .
\label {eq:M}
\end {equation}
For constant $\Omega$ (appropriate to the case of a thick, homogeneous
medium), the propagator of a 2-dimensional harmonic oscillator is
\begin {equation}
   \langle \B,\Delta t | \B',0 \rangle
   =
   \frac{M\Omega\csc(\Omega \, \Delta t)}{2\pi i} \,
      \exp\Bigl(
        \tfrac{i}2 M\Omega
        \bigl[ (\B^2+\B'^2) \cot(\Omega \Delta t)
              - 2\B\cdot\B' \csc(\Omega \Delta t) \bigr]
      \Bigr) .
\end {equation}
Using this in (\ref{eq:dGdxQM}) reproduces the earlier formula
(\ref{eq:dGdx}).

To implement dimensional regularization, we simply generalize the
$d{=}2$ analysis for two transverse dimensions to an arbitrary number
$d$ of transverse dimensions.  (Note that we are defining
$d$ to be the number of {\it transverse spatial}\/ dimensions, not the
total number of space-time dimensions, and so the real world is $d=2$
in this paper, not $d=4$.)  The propagator for a $d$-dimensional
Harmonic oscillator is
\begin {equation}
   \langle \B,\Delta t | \B',0 \rangle
   =
   \left( \frac{M\Omega\csc(\Omega \, \Delta t)}{2\pi i} \right)^{d/2}
      \exp\Bigl(
        \tfrac{i}2 M\Omega
        \bigl[ (\B^2+\B'^2) \cot(\Omega \Delta t)
              - 2\B\cdot\B' \csc(\Omega \Delta t) \bigr]
      \Bigr) .
\label {eq:1prop}
\end {equation}
The corresponding integral in (\ref{eq:dGdxQM}) is then
\begin {align}
   \int_0^\infty d(\Delta t) \>
   \grad_{\B_\xbx} \cdot \grad_{\B_\xx} &
   \langle \B_\xbx,\Delta t | \B_\xx,0 \rangle
   \Bigr|_{\B_\xbx = \B_\xx = 0}
\nonumber\\ &
   =
   - \int_0^\infty d(\Delta t) \>
   \left( \frac{M\Omega\csc(\Omega \, \Delta t)}{2\pi i} \right)^{d/2}
   i d M\Omega\csc(\Omega \, \Delta t)
\label {eq:integral0}
\end {align}
(using $\grad_\B\cdot\B = d$).
Changing integration variable to $\tau \equiv i\Omega\,\Delta t$, this
becomes
\begin {equation}
   =
   - i d M \left( \frac{M\Omega}{2\pi} \right)^{d/2}
   \int_0^{\infty} \frac{d\tau}{\sh^{1+\frac{d}{2}}\tau} \,.
\label {eq:integral1}
\end {equation}
We've been seemingly cavalier here about the complex phase of the
upper limit of integration---see appendix \ref{app:details} for
a more careful discussion.

The integral (\ref{eq:integral1}) converges for $-2 < d < 0$.
The result for other $d$ is defined by analytic continuation, giving
(see appendix \ref{app:details})
\begin {equation}
   \int_0^\infty d(\Delta t) \>
   \grad_{\B_\xbx} \cdot \grad_{\B_\xx}
   \langle \B_\xbx,\Delta t | \B_\xx,0 \rangle
   \Bigr|_{\B_\xbx = \B_\xx = 0}
   =
   - \frac{i d M}{2} \left( \frac{M\Omega}{2\pi} \right)^{d/2}
   \Beta(\tfrac12{+}\tfrac{d}{4},-\tfrac{d}{4}) ,
\label {eq:integral2}
\end {equation}
where
\begin {equation}
   \Beta(x,y) = \frac{\Gamma(x) \, \Gamma(y)}{\Gamma(x+y)}
\end {equation}
is the Euler beta function.

We can now take the $d\to 2$ limit, giving
\begin {equation}
   \int_0^\infty d(\Delta t) \>
   \grad_{\B_\xbx} \cdot \grad_{\B_\xx}
   \langle \B_\xbx,\Delta t | \B_\xx,0 \rangle
   \Bigr|_{\B_\xbx = \B_\xx = 0}
   \to
   \frac{i M^2\Omega}{\pi}
\label {eq:integral2d2}
\end {equation}
Because this regularized result for the integral is finite for
$d{=}2$, we do not have to worry about the generalization
of the prefactor
$\alpha P(x)/[x(1{-}x)E]^2$ in the $d\Gamma/dx$ formula
(\ref{eq:dGdxQM}) to general
dimension: we can just use the known $d{=}2$ version.
Combining (\ref{eq:dGdxQM}), (\ref{eq:M}) and (\ref{eq:integral2d2})
correctly
reproduces the usual result (\ref{eq:dGdx2}) for single splitting.


\subsection{Alternative derivation}

Before we launch into the complexities of the double bremsstrahlung
calculation, we can introduce another formula that we will need
by repeating the previous calculation in a slightly more roundabout
manner: We will do the $\Delta t$ integral in (\ref{eq:dGdxQM})
{\it before}\/ taking the $\B_\xbx$ derivative and so also before
setting $\B_\xbx$ to zero.
The integral we need has the form
\begin {equation}
   \int_0^\infty d(\Delta t) \>
   \grad_{\B'}
   \langle \B,\Delta t | \B',0 \rangle
   \Bigr|_{\B' = 0} .
\label {eq:tauint0}
\end {equation}
Using the $d$-dimensional propagator (\ref{eq:1prop}), this becomes
\begin {equation}
   =
   - i M \B \left( \frac{M\Omega}{2\pi} \right)^{d/2}
   \int_0^{\infty} \frac{d\tau}{\sh^{1+\frac{d}{2}}\tau} \,
   e^{-\frac12 M \Omega B^2 \cth\tau} ,
\label {eq:tauint}
\end {equation}
with $\tau \equiv i\Omega\,\Delta t$ as before.
The integral converges for $d > -2$ when $\B \not= 0$.
Using
\begin {equation}
   \int_0^\infty d\tau \> \frac{e^{-z \cth\tau}}{\sh^{\frac{d}{2}+1}\tau}
   = \frac{\Gamma(\frac12+\frac{d}{4})}{\sqrt\pi} \,
     \left( \frac{2}{z} \right)^{d/4} \,
     K_{d/4}(z)
\label {eq:dimint}
\end {equation}
(see appendix \ref{app:details}),
we obtain
\begin {equation}
   \int_0^\infty d(\Delta t) \>
     \grad_{\B'}
     \langle \B,\Delta t | \B',0 \rangle
     \Bigr|_{\B' = 0} 
   =
   -
     \frac{iM\B}{\pi^{(d+1)/2}} \left(\frac{M \Omega}{B^2}\right)^{d/4}
     \Gamma(\tfrac12{+}\tfrac{d}{4}) \, K_{d/4}(\tfrac12 M\Omega B^2) .
\label {eq:dimint2}
\end {equation}
This is a formula we will need later for double splitting, where
it will be convenient to rewrite it equivalently as
\begin {equation}
   \int_{-\infty}^t dt' \>
     \grad_{\B'}
     \langle \B,t | \B',t' \rangle
     \Bigr|_{\B' = 0} 
   =
   -
     \frac{iM\B}{\pi^{(d+1)/2}} \left(\frac{M \Omega}{B^2}\right)^{d/4}
     \Gamma(\tfrac12{+}\tfrac{d}{4}) \, K_{d/4}(\tfrac12 M\Omega B^2) .
\label {eq:t1int0}
\end {equation}

Let's finish up the single splitting calculation by showing that
we can use these formulas to get the same answer as before.
We need to take the gradient $\grad_\B$ of (\ref{eq:dimint2}) and
set $\B$ to zero.  The last step raises a subtlety which happily
will not arise in the double splitting calculation later:
the original integral (\ref{eq:tauint}) has convergence problems
for $\B{=}0$ unless $d<0$.  Eventually we want to focus on $d\to 2$,
but we should be cautious about whether we analytically continue to
that limit before or after setting $\B$ to zero.
To see that there is an
issue, use the generic expansion of the Bessel function for
small arguments:
\begin {equation}
   K_\nu(z) =
   \tfrac12 \, \Gamma(\nu) \, (\tfrac{z}{2})^{-\nu}
      \bigl[1 + O(z^2)\bigr]
   + \tfrac12 \, \Gamma(-\nu) \, (\tfrac{z}{2})^\nu
      \bigl[1 + O(z^2)\bigr] .
\label {eq:BesselSmall}
\end {equation}
Which term dominates depends on the sign of $\nu$ and so,
in our case, on the sign of $d$.
Keeping both of the potentially leading terms above,
the small $\B$ behavior of (\ref{eq:dimint2}) is
\begin {multline}
   \int_0^\infty d(\Delta t) \>
     \grad_{\B'}
     \langle \B,\Delta t | \B',0 \rangle
     \Bigr|_{\B' = 0} 
\\
   \simeq
   - \frac{i M\B}{2}
   \left[
     \left(\frac{2}{\pi B^2}\right)^{d/2}
        \frac{\Gamma(\tfrac12{+}\tfrac{d}{4}) \, \Gamma(\tfrac{d}{4})}
             {\pi^{1/2}}
     +
     \left(\frac{M\Omega}{2\pi}\right)^{d/2}
         \Beta(\tfrac12{+}\tfrac{d}{4},-\tfrac{d}{4})
   \right]
   .
\end {multline}
Now dot $\grad_\B$ into the above expression and then set $\B$ to
zero.  The second term gives exactly our earlier
answer (\ref{eq:integral2}), which leads to the correct result
(\ref{eq:integral2d2}) when we analytically continue to $d=2$.
The first term gives a (UV) singularity unless $d \le 0$.
Since the point of dimensional regularization was to regulate
the UV, we learn that in this application to single splitting
we need to keep the
dimension as $d < 0$ until after we take $\B$ to zero.


\section{Crossed diagrams with dimensional regularization}
\label {sec:cross}

We now turn to double bremsstrahlung, focusing on $g \to ggg$
and starting with
the crossed diagrams of
fig.\ \ref{fig:subset2}, which were evaluated for $d{=}2$
(other than the missing $1/\pi^2$ pole terms) in ref.\ \cite{2brem}.
As in ref.\ \cite{2brem}, we will start with the $xy\bar y\bar x$ diagram,
to which the others can be related.


\subsection{First equations}
\label {sec:first}

Our starting point here will be the $d{=}2$ expression%
\footnote{AI (4.40)}
\begin {align}
   \left[\frac{dI}{dx\,dy}\right]_{xy\bar y\bar x}
   = &
   \frac{\CA^2 \alphas^2 }{8 E^4} \,
   \frac{
     ( \alpha \delta^{\bar n n} \delta^{\bar m m}
     {+} \beta \delta^{\bar n \bar m} \delta^{nm}
     {+} \gamma \delta^{\bar n m} \delta^{n \bar m} )
   }{
     |\hat x_1 + \hat x_4| |\hat x_3 + \hat x_4|
   }
   \int_{t_\xx < t_\yx < t_\ybx < t_\xbx}
\nonumber\\ &
   \int_{\B^\Ax,\B^\bx}
   \nabla_{\B^\Bx}^{\bar n}
   \langle\B^\Bx,t_\Bx|\B^\Ax,t_\Ax\rangle
   \Bigr|_{\B^\Bx=0}
\nonumber\\ &\qquad\times
   \nabla_{\C_{12}^\Ax}^{\bar m}
   \nabla_{\C_{23}^\bx}^n
  \langle\C_{34}^\Ax,\C_{12}^\Ax,t_\Ax|\C_{41}^\bx,\C_{23}^\bx,t_\bx\rangle
   \Bigr|_{\C_{12}^\Ax=0=\C_{23}^\bx; ~ \C_{34}^\Ax=\B^\Ax; ~ \C_{41}^\bx=\B^\bx}
\nonumber\\ &\qquad\times
   \nabla_{\B^\ax}^m
   \langle\B^\bx,t_\bx|\B^\ax,t_\ax\rangle
   \Bigr|_{\B^\ax=0}
\label {eq:IIxyyx}
\end {align}
developed for $xy\bar y\bar x$ in section IV of ref.\ \cite{2brem}.
$\langle\B^\bx,t_\bx|\B^\ax,t_\ax\rangle$ and
$\langle\C_{34}^\Ax,\C_{12}^\Ax,t_\Ax|\C_{41}^\bx,\C_{23}^\bx,t_\bx\rangle$
and
$\langle\B^\Bx,t_\Bx|\B^\Ax,t_\Ax\rangle$
represent, respectively, the
(i) 3-particle evolution of the system in the initial
time interval $t_\xx < t < t_\yx$ of the figure, (ii) 4-particle
evolution in the intermediate interval $t_\yx < t < t_\ybx$, and
(iii) 3-particle evolution of the system in the final interval
$t_\ybx < t < t_\xbx$.  Because of the symmetries of the problem,
these have been reduced to effective (i) 1-particle, (ii) 2-particle,
and (iii) 1-particle problems in non-Hermitian $d{=}2$ quantum mechanics.
Each vertex in the diagram is associated with one of the gradients
above.  The dimensionless functions
$\alpha(x,y)$, $\beta(x,y)$, and $\gamma(x,y)$
contain normalization factors
and combinations of spin-dependent DGLAP splitting functions associated
with those vertices.%
\footnote{
  AI (4.30--39)
}
The variables $\hat x_i$ refer to the momentum fractions associated
with the four particles involved in the 4-particle evolution in
this diagram, which are
\begin {equation}
   (\hat x_1,\hat x_2,\hat x_3,\hat x_4) = (-1,y,1{-}x{-}y,x) .
\label {eq:xhat}
\end {equation}
The overall factors of $|\hat x_1 + \hat x_4|^{-1}$
and $|\hat x_3 + \hat x_4|^{-1}$ are additional normalization factors
associated with the vertices at the intermediate times
$t_\yx$ and $t_\ybx$ given our choice of normalization of the transverse
position variables and of corresponding states such as
$|\C_{41},\C_{23}\rangle$ and $|\B\rangle$.%
\footnote{
  AI (4.6) and (4.22-25)
}
The expression (\ref{eq:IIxyyx}) also assumes the large $\Nc$ limit
in order to simplify the color dynamics of the problem associated
with the 4-particle propagation in the medium.
However, we believe that the results for the pole contributions
calculated in this paper do not depend on the assumption of large
$\Nc$ because the 4-particle propagation in pole contributions is
effectively vacuum propagation (as discussed earlier with regards
to fig.\ \ref{fig:pole}).

The $d$-dimensional generalization of (\ref{eq:IIxyyx}) is
\begin {align}
   \left[\frac{dI}{dx\,dy}\right]_{xy\bar y\bar x}
   = &
   \frac{\CA^2 \alphas^2 }{8 E^{2d}} \,
   \frac{
     ( \alpha \delta^{\bar n n} \delta^{\bar m m}
     {+} \beta \delta^{\bar n \bar m} \delta^{nm}
     {+} \gamma \delta^{\bar n m} \delta^{n \bar m} )
   }{
     |\hat x_1 + \hat x_4|^{d/2} |\hat x_3 + \hat x_4|^{d/2}
   }
   \int_{t_\xx < t_\yx < t_\ybx < t_\xbx}
\nonumber\\ &
   \int_{\B^\Ax,\B^\bx}
   \nabla_{\B^\Bx}^{\bar n}
   \langle\B^\Bx,t_\Bx|\B^\Ax,t_\Ax\rangle
   \Bigr|_{\B^\Bx=0}
\nonumber\\ &\qquad\times
   \nabla_{\C_{12}^\Ax}^{\bar m}
   \nabla_{\C_{23}^\bx}^n
  \langle\C_{34}^\Ax,\C_{12}^\Ax,t_\Ax|\C_{41}^\bx,\C_{23}^\bx,t_\bx\rangle
   \Bigr|_{\C_{12}^\Ax=0=\C_{23}^\bx; ~ \C_{34}^\Ax=\B^\Ax; ~ \C_{41}^\bx=\B^\bx}
\nonumber\\ &\qquad\times
   \nabla_{\B^\ax}^m
   \langle\B^\bx,t_\bx|\B^\ax,t_\ax\rangle
   \Bigr|_{\B^\ax=0}
\label {eq:IIxyyxDR}
\end {align}
There are no important changes to this formula other
than the fact that transverse positions $\B$ and $\C$ are now
$d$-dimensional, but we should comment on the other,
mostly unimportant differences between
(\ref{eq:IIxyyx}) and (\ref{eq:IIxyyxDR}):
\begin {itemize}
\item
 The functions $\alpha$, $\beta$, and $\gamma$ will be different in $d$
 dimensions.  For one thing, the number of ``helicities''
 to be summed over depends on $d$.  Except as noted below,
 we absorb all this dependence
 on $d$ into new $d$-dimensional versions of $\alpha$, $\beta$, and
 $\gamma$.  However, similar to the discussion of
 $P(x)$ in section \ref{sec:straighforward},
 we will see when we take $d \to 2$
 at the end of the day that we only need explicit formulas for the
 original $d{=}2$ versions given in ref.\ \cite{2brem}.
\item
 One of the exceptions to absorbing all of the differences into the
 $d$-dimensional definitions of $(\alpha,\beta,\gamma)$:  We
 have changed the overall $E^{-4}$ in (\ref{eq:IIxyyx}) to $E^{-2d}$
 in order to keep $\alpha$, $\beta$, and $\gamma$ dimensionless
 (and so make dimensional analysis of our formulas easier).
 This difference originates with the factors of $E$ associated
 with the vertices (see appendix \ref{app:details}).
\item
 We have also treated separately the generalization of the normalization
 factors $|x_1{+}x_4|^{-1} |x_3{+}x_4|^{-1}$ to
 $|x_1{+}x_4|^{-d/2} |x_3{+}x_4|^{-d/2}$
 (see appendix \ref{app:details}).  The reason that we do not also absorb
 these differences is that, unlike $(\alpha,\beta,\gamma)$, these
 factors will not be the same for the three diagrams shown explicitly
 in fig.\ \ref{fig:subset2}.  Since we will later relate these diagrams
 to each other, we need to keep track of factors that change between
 them.  These factors are nonetheless fairly uninteresting because
 $|x_1{+}x_4|^{-d/2} |x_3{+}x_4|^{-d/2}$ will simply cancel some related
 normalization factors when we later write more explicit formulas for the
 4-particle propagator.
\end {itemize}

We now want to use (\ref{eq:t1int0}) to integrate over the first time
$t_\xx$ in (\ref{eq:IIxyyxDR}).  The derivation of (\ref{eq:t1int0})
relied on the $M$ associated with the 3-particle evolution being
positive, and the associated $\Omega$ having phase $\sqrt{-i}$.
All is well for now, but when we later relate the other crossed diagrams
to $xy\bar y\bar x$, we will encounter situations where instead
both $M$ is negative and $\Omega \propto \sqrt{+i}$.  It's therefore
convenient to use an appropriate generalization of (\ref{eq:t1int0})
to cover both situations:%
\begin {subequations}
\label {eq:t1int}
\begin {equation}
   \int_{-\infty}^t dt' \>
     \grad_{\B'}
     \langle \B,t | \B',t' \rangle
     \Bigr|_{\B' = 0} 
   =
   -
     \frac{iM\B}{\pi^{(d+1)/2}} \left(\frac{|M| \Omega}{B^2}\right)^{d/4}
     \Gamma(\tfrac12{+}\tfrac{d}{4}) \, K_{d/4}(\tfrac12 |M|\Omega B^2)
\label {eq:t1inti}
\end {equation}
(see appendix \ref{app:details}).
The similar result for integration over the final time is
\begin {equation}
   \int_{t'}^\infty dt \>
     \grad_{\B}
     \langle \B,t | \B',t' \rangle
     \Bigr|_{\B = 0} 
   =
   -
     \frac{iM\B'}{\pi^{(d+1)/2}} \left(\frac{|M| \Omega}{B'^2}\right)^{d/4}
     \Gamma(\tfrac12{+}\tfrac{d}{4}) \, K_{d/4}(\tfrac12 |M|\Omega B'^2)
   .
\end {equation}
\end {subequations}
Using these for the initial $t_\xx$ and final $t_\xbx$
time integrations in (\ref{eq:IIxyyxDR}) gives%
\footnote{
  The $d{=}2$ cases of (\ref{eq:t1int}) and (\ref{eq:IIxyyx2DR})
  reproduce AI (5.9) and AI (5.10) respectively.
}
\begin {align}
   \left[\frac{d\Gamma}{dx\,dy}\right]_{xy\bar y\bar x}
   = &
   - \frac{\CA^2 \alphas^2 M_\ix M_\fx}{8 \pi^{d+1} E^{2d}}
   \frac{
     ( \alpha \delta^{\bar n n} \delta^{\bar m m}
     {+} \beta \delta^{\bar n \bar m} \delta^{nm}
     {+} \gamma \delta^{\bar n m} \delta^{n \bar m} )
   }{
     |\hat x_1 + \hat x_4|^{d/2} |\hat x_3 + \hat x_4|^{d/2}
   }
   \> \Gamma^2( \tfrac12{+}\tfrac{d}{4} ) \,
\nonumber\\ & \times
   \int_0^{\infty} d(\Delta t)
   \int_{\B^\Ax,\B^\bx}
   B^{\ybx}_{\bar n} \left(\frac{|M_\fx|\Omega_\fx}{(B^\ybx)^2}\right)^{d/4}
     K_{d/4}\bigl(\tfrac12 |M_\fx| \Omega_\fx (B^\ybx)^2\bigr)
\nonumber\\ &\qquad\times
   B^{\yx}_{m} \left(\frac{|M_\ix|\Omega_\ix}{(B^\yx)^2}\right)^{d/4}
     K_{d/4}\bigl(\tfrac12 |M_\ix| \Omega_\ix (B^\yx)^2\bigr)
\nonumber\\ &\qquad\times
   \nabla_{\C_{12}^\Ax}^{\bar m}
   \nabla_{\C_{23}^\bx}^n
   \langle\C_{34}^\Ax,\C_{12}^\Ax,\Delta t|\C_{41}^\bx,\C_{23}^\bx,0\rangle
   \Bigr|_{\C_{12}^\Ax=0=\C_{23}^\bx; ~ \C_{34}^\Ax=\B^\Ax; ~ \C_{41}^\bx=\B^\bx}
   ,
\label {eq:IIxyyx2DR}
\end {align}
where $(M_\ix,\Omega_\ix)$ are associated with the initial
3-particle evolution ($t_\xx < t < t_\yx$) and $(M_\fx,\Omega_\fx)$
with the final 3-particle evolution ($t_\ybx < t < t_\xbx$).


\subsection{4-particle propagator}

For evaluation of the pole pieces (which are the pieces that require
UV regularization), we only need the small $\Delta t$ limit of
(\ref{eq:IIxyyx2DR}).  In that limit, medium effects on the
propagator
$\langle\C_{34}^\Ax,\C_{12}^\Ax,\Delta t|\C_{41}^\bx,\C_{23}^\bx,0\rangle$
are small.  We will therefore use the vacuum result for
$\langle\C_{34}^\Ax,\C_{12}^\Ax,\Delta t|\C_{41}^\bx,\C_{23}^\bx,0\rangle$.
(Readers who would prefer to see a derivation closer to ref.\ \cite{2brem},
where we first find full expressions for the propagator before taking
the small $\Delta t$ limit to find the poles, may turn instead to
Appendix \ref{app:4particle}.)

As discussed in ref.\ \cite{2brem},%
\footnote{
   Specifically, see the discussion leading up to AI (5.15--18).
}
the effective 4-particle evolution in interference
diagrams such as fig.\ \ref{fig:subset2} is given
(in the high-energy limit) by a Lagrangian of the form
\begin {equation}
   L
   =
   \tfrac12
   \begin{pmatrix} \dot\C_{34} \\ \dot\C_{12} \end{pmatrix}^{\!\top}
   \!{\mathfrak M}
   \begin{pmatrix} \dot\C_{34} \\ \dot\C_{12} \end{pmatrix}
   - V(\C_{34},\C_{12})
\label {eq:L3}
\end {equation}
where
\begin {equation}
   \C_{ij} \equiv \frac{\b_i - \b_j}{x_i+x_j} ,
\label {eq:CijDef}
\end {equation}
$\b_i$ are the transverse positions of the individual particles,
and
\begin {equation}
   {\mathfrak M}
   =
   \begin{pmatrix}
      x_3 x_4 (x_3+x_4) & \\ & x_1 x_2 (x_1+x_2)
   \end {pmatrix} E
   =
   \begin{pmatrix}
      x_3 x_4 & \\ & -x_1 x_2
   \end {pmatrix} (x_3+x_4) E .
\label {eq:frakM}
\end {equation}
The imaginary-valued potential $V$ implements medium effects
which cause decoherence of interference over times $\Delta t$
of order the formation time.  In vacuum, $V=0$.%
\footnote{
  As in refs.\ \cite{2brem,seq}, we have for simplicity assumed that
  the energy is high enough that we may ignore the effects of the
  physical masses of the
  high-energy particles.  If one does not ignore them, their effects
  contribute a real-valued constant to $V$ \cite{Zakharov},
  even in vacuum.
  See the discussion surrounding AI (2.15).
}
Were we to express the propagator associated with (\ref{eq:L3})
solely in terms of the variables
$(\C_{34},\C_{12})$,
it would then (in this limit) simply be
\begin {multline}
  \langle\C_{34},\C_{12},\Delta t|\C'_{34},\C'_{12},0\rangle
  \simeq
\\
  (2\pi i \, \Delta t)^{-d} (\det{\mathfrak M})^{d/2}
  \exp\Biggl[
     \frac{i}{2\,\Delta t}
     \begin{pmatrix} \C_{34}{-}\C'_{34} \\ \C_{12}{-}\C'_{12} \end{pmatrix}^\top
     {\mathfrak M}
     \begin{pmatrix} \C_{34}{-}\C'_{34} \\ \C_{12}{-}\C'_{12} \end{pmatrix}
  \Biggr] .
\label {eq:Cprop00}
\end {multline}
Changing variables from $(\C_{34},\C_{12})$ to $(\C_{41},\C_{23})$
in just the ket $|\C'_{34},\C'_{12},0\rangle$ then gives the
version of this propagator that we need:
\begin {align}
  \langle\C_{34},\C_{12},\Delta t|&\C'_{41},\C'_{23},0\rangle
  \simeq
  (2\pi i \, \Delta t)^{-d}
  (\det{\mathfrak M})^{d/4} (\det{\mathfrak M}')^{d/4}
\nonumber\\ & \times
  \exp\Biggl[
     \frac{i}{2\,\Delta t}
     \begin{pmatrix} \C'_{41} \\ \C'_{23} \end{pmatrix}^\top \!
       {\mathfrak M}'
       \begin{pmatrix} \C'_{41} \\ \C'_{23} \end{pmatrix}
     +
     \frac{i}{2\,\Delta t}
     \begin{pmatrix} \C_{34} \\ \C_{12} \end{pmatrix}^\top \!
       {\mathfrak M}
       \begin{pmatrix} \C_{34} \\ \C_{12} \end{pmatrix}
\nonumber\\ & \qquad\quad
     +
     \frac{i E}{\Delta t}
     \begin{pmatrix} \C'_{41} \\ \C'_{23} \end{pmatrix}^\top \!
       \begin{pmatrix}
         x_1 x_3 x_4 & x_1 x_2 x_4 \\
         x_2 x_3 x_4 & x_1 x_2 x_3
       \end{pmatrix}
       \begin{pmatrix} \C_{34} \\ \C_{12} \end{pmatrix}
  \Biggr]
\label{eq:CpropSmall}
\end {align}
(see appendix \ref{app:details}), where ${\mathfrak M}'$ is the
$1{\leftrightarrow}3$ permutation
of ${\mathfrak M}$ (\ref{eq:frakM}),
\begin {equation}
   {\mathfrak M}'
   =
   \begin{pmatrix}
      x_1 x_4 & \\ & -x_2 x_3
   \end {pmatrix} (x_1+x_4) E .
\label {eq:frakMprime}
\end {equation}
In order to keep notation as close as possible to ref.\ \cite{2brem}, it
will be useful to rewrite the exponential in (\ref{eq:CpropSmall}) as
\begin {multline}
   \exp\Biggl[
     - \frac12
     \begin{pmatrix} \C'_{41} \\ \C'_{23} \end{pmatrix}^\top \!
       \begin{pmatrix} \calX_\bx & Y_\bx \\ Y_\bx & Z_\bx \end{pmatrix}
     \begin{pmatrix} \C'_{41} \\ \C'_{23} \end{pmatrix}
     -
     \frac12
     \begin{pmatrix} \C_{34} \\ \C_{12} \end{pmatrix}^\top \!
       \begin{pmatrix} \calX_\Ax & Y_\Ax \\ Y_\Ax & Z_\Ax \end{pmatrix}
       \begin{pmatrix} \C_{34} \\ \C_{12} \end{pmatrix}
\\
     +
     \begin{pmatrix} \C'_{41} \\ \C'_{23} \end{pmatrix}^\top \!
       \begin{pmatrix} X_{\bx\Ax} & Y_{\bx\Ax} \\
                       \Ybar_{\bx\Ax} & Z_{\bx\Ax} \end{pmatrix}
       \begin{pmatrix} \C_{34} \\ \C_{12} \end{pmatrix}
   \Biggr],
\label {eq:rewrite}
\end {multline}
where%
\footnote{
  \label{foot:calX}
  Eqs.\ (\ref{eq:calXYZsmalldt}) are the same as AI (D2) except
  that ${\cal X}$ here does not contain the $|M_\ix|\Omega_\ix$ and
  $|M_\fx|\Omega_\fx$ terms that $X$ has there.
} 
\begin {subequations}
\label {eq:calXYZsmalldt}
\begin {align}
   \begin{pmatrix} \calX_\bx & Y_\bx \\ Y_\bx & Z_\bx \end{pmatrix}
   &=
     -\frac{i E (x_1{+}x_4)}{\Delta t}
       \begin{pmatrix} x_1 x_4 & 0 \\ 0 & -x_2 x_3 \end{pmatrix}
     + O(\Delta t) ,
\\
   \begin{pmatrix} \calX_\Ax & Y_\Ax \\ Y_\Ax & Z_\Ax \end{pmatrix}
   &=
     -\frac{i E (x_3{+}x_4)}{\Delta t}
       \begin{pmatrix} x_3 x_4 & 0 \\ 0 & -x_1 x_2 \end{pmatrix}
     + O(\Delta t) ,
\\
   \begin{pmatrix} X_{\bx\Ax} & Y_{\bx\Ax} \\ \Ybar_{\bx\Ax} & Z_{\bx\Ax} \end{pmatrix}
   &=
     \frac{i E}{\Delta t}
     \begin{pmatrix} x_1 x_3 x_4 &  x_1 x_2 x_4 \\
                     x_2 x_3 x_4 &  x_1 x_2 x_3 \end{pmatrix}
     + O(\Delta t) .
\end {align}
\end {subequations}
Here, the unspecified $+O(\Delta t)$ contributions represent the
size of effects due to the medium.
Using (\ref{eq:CpropSmall}) and (\ref{eq:rewrite}) in
(\ref{eq:IIxyyx2DR}) then gives%
\footnote{
  It is easy to get confused about cuts associated with the
  fractional exponents.  In the $xy\bar y\bar x$
  case at hand, they are resolved by the facts that (i)
  ${\mathfrak M}$ and ${\mathfrak M}'$ are positive definite, so that
  $(\det{\mathfrak M})^{d/4}(\det{\mathfrak M}')^{d/4} =
   |\det{\mathfrak M}|^{d/4}|\det{\mathfrak M}'|^{d/4}$, and
  (ii) $\hat x_1 \hat x_2 \hat x_3 \hat x_4$ given by (\ref{eq:xhat})
  is negative, so that
  $|\hat x_1 \hat x_2 \hat x_3 \hat x_4|^{d/2} =
   (-\hat x_1 \hat x_2 \hat x_3 \hat x_4)^{d/2}$.
  See appendix \ref{app:sheet} for a more general discussion of cuts. 
  Additionally,
  the $d{=}2$ case of (\ref{eq:DRgeneral}) reproduces AI (5.43)
  with $X_\yx = \calX_\yx + |M_\ix|\Omega_\ix$ and
  $X_\ybx = \calX_\ybx + |M_\fx|\Omega_\fx$, except that we have
  expanded here in the small $\Delta t$ limit.
}
\begin {align}
   \left[\frac{d\Gamma}{dx\,dy}\right]_{xy\bar y\bar x}
   \simeq &
   - \frac{\CA^2 \alphas^2 M_\ix M_\fx}{2^{d+3}\pi^{2d+1}i^dE^d} \,
     \Gamma^2( \tfrac12{+}\tfrac{d}{4} ) \,
     ({-}\hat x_1 \hat x_2 \hat x_3 \hat x_4)^{d/2}
     ( \alpha \delta^{\bar n n} \delta^{\bar m m}
     {+} \beta \delta^{\bar n \bar m} \delta^{nm}
     {+} \gamma \delta^{\bar n m} \delta^{n \bar m} )
\nonumber\\ & \times
   \int_0 \frac{d(\Delta t)}{(\Delta t)^d}
   \int_{\B^\Ax,\B^\bx}
   B^{\ybx}_{\bar n} \left(\frac{|M_\fx|\Omega_\fx}{(B^\ybx)^2}\right)^{d/4}
     K_{d/4}\bigl(\tfrac12 |M_\fx| \Omega_\fx (B^\ybx)^2\bigr)
\nonumber\\ & \times
   B^{\yx}_{m} \left(\frac{|M_\ix|\Omega_\ix}{(B^\yx)^2}\right)^{d/4}
     K_{d/4}\bigl(\tfrac12 |M_\ix| \Omega_\ix (B^\yx)^2\bigr)
\nonumber\\ & \times
   \bigl[
     (Y_\bx \B^\bx - \Ybar_{\bx\Ax} \B^\Ax)_n
     (Y_\Ax \B^\Ax - Y_{\bx\Ax} \B^\bx)_{\bar m}
     + Z_{\bx\Ax} \delta_{n\bar m}
   \bigr]
\nonumber\\ & \times
   \exp\Bigl[
     - \tfrac12
     \calX_\bx (B^\bx)^2
     -
     \tfrac12
     \calX_\Ax (B^\Ax)^2
     +
     X_{\bx\Ax} \B^\bx \cdot \B^\Ax
   \Bigr] .
\label {eq:DRgeneral}
\end {align}


\subsection{Small $\B$ expansion}

We could try to follow the $d{=}2$ analysis of ref.\ \cite{2brem} by
next doing the two $\B$ integrations in (\ref{eq:DRgeneral}).
Unfortunately, the integrals are complicated.  Fortunately, we can
simply the calculation because we need general-$d$
expressions for only the pole pieces, corresponding to $\Delta t \to 0$.
The exponential factors in (\ref{eq:DRgeneral}) become highly
oscillatory for $B \gg \calX^{-1/2} \sim \sqrt{\Delta t / E}$
and so cause the integrals
to be dominated by the scale $B \sim \sqrt{\Delta t/E}$ (where we
are not showing the $x_i$ dependence).  So, for the purpose of
extracting the small $\Delta t$ behavior, we may expand the Bessel
functions in (\ref{eq:DRgeneral}) for small arguments, as in
(\ref{eq:BesselSmall}).  For $-4 < d < 4$ (which includes both the
physical point $d{=}2$ and the region $-2 < d < 0$ where all of
our earlier integrals were convergent), the terms shown
explicitly in (\ref{eq:BesselSmall}) are the leading ones and give
\begin {equation}
   \left(\frac{|M|\Omega}{B^2}\right)^{d/4}
   K_{d/4}(\tfrac12 |M| \Omega B^2)
   \simeq
   \tfrac12 \, \Gamma(\tfrac{d}{4}) \,
     \left( \frac{2}{B^2} \right)^{d/2}
   + \tfrac12 \, \Gamma(-\tfrac{d}{4}) \,
     \left( \frac{|M|\Omega}{2} \right)^{d/2} .
\label {eq:Kfactor}
\end {equation}
We will see later that the integrals we have left to do converge for
$d = 2-\epsilon$ with $\epsilon$ small,
and so we may as well take $d=2-\epsilon$ now.
In that case, the first term in (\ref{eq:Kfactor}) dominates over
the second.  By itself, however, the first term is uninteresting
because it does not depend on $\Omega$ and so does
not depend on $\hat q$.  If we use just the first term for both
of the $(|M|\Omega/B^2)^{d/4} K_{d/4}$ factors in (\ref{eq:DRgeneral}),
we will obtain a contribution that will be canceled when we
subtract away the purely vacuum result.  We should therefore focus
on the next term:
\begin {multline}
   \left(\frac{|M_\fx|\Omega_\fx}{(B^\ybx)^2}\right)^{d/4}
     K_{d/4}\bigl(\tfrac12 |M_\fx| \Omega_\fx (B^\ybx)^2\bigr)
   \times
   \left(\frac{|M_\ix|\Omega_\ix}{(B^\yx)^2}\right)^{d/4}
     K_{d/4}\bigl(\tfrac12 |M_\ix| \Omega_\ix (B^\yx)^2\bigr)
   \simeq
\\
   (\mbox{vacuum}) +
   \tfrac14 \, \Gamma(\tfrac{d}{4}) \, \Gamma(-\tfrac{d}{4}) \,
   \left[
     \left( \frac{|M_\ix|\Omega_\ix}{(B^\ybx)^2} \right)^{d/2}
   +
     \left( \frac{|M_\fx|\Omega_\fx}{(B^\yx)^2} \right)^{d/2}
   \right]
   + (\mbox{sub-leading}) .
\end {multline}
The (non-vacuum) small $\Delta t$ limit of the integrand in
(\ref{eq:DRgeneral}) then gives
\begin {align}
   \left[\frac{d\Gamma}{dx\,dy}\right]_{xy\bar y\bar x}
   \simeq &
   \frac{\CA^2 \alphas^2 M_\ix M_\fx}{2^{d+3}d\pi^{2d}i^dE^d} \,
     \frac{\Gamma^2( \tfrac12{+}\tfrac{d}{4} )}{\sin(\frac{\pi d}{4})} \,
     ({-}\hat x_1 \hat x_2 \hat x_3 \hat x_4)^{d/2}
     ( \alpha \delta^{\bar n n} \delta^{\bar m m}
     {+} \beta \delta^{\bar n \bar m} \delta^{nm}
     {+} \gamma \delta^{\bar n m} \delta^{n \bar m} )
\nonumber\\ & \times
   \int_0 \frac{d(\Delta t)}{(\Delta t)^d} \>
   \int_{\B^\Ax,\B^\bx}
   B^{\ybx}_{\bar n} B^{\yx}_{m}
   \left[
     \left( \frac{|M_\ix|\Omega_\ix}{(B^\ybx)^2} \right)^{d/2}
   +
     \left( \frac{|M_\fx|\Omega_\fx}{(B^\yx)^2} \right)^{d/2}
   \right]
\nonumber\\ &\times
   \bigl[
     (Y_\bx \B^\bx - \Ybar_{\bx\Ax} \B^\Ax)_n
     (Y_\Ax \B^\Ax - Y_{\bx\Ax} \B^\bx)_{\bar m}
     + Z_{\bx\Ax} \delta_{n\bar m}
   \bigr]
\nonumber\\ & \times
   \exp\Bigl[
     - \tfrac12
     \calX_\bx (B^\bx)^2
     -
     \tfrac12
     \calX_\Ax (B^\Ax)^2
     +
     X_{\bx\Ax} \B^\bx \cdot \B^\Ax
   \Bigr] ,
\label {eq:dGamIntDR}
\end {align}
where we've used
$\frac14 \, \Gamma(\tfrac{d}{4}) \, \Gamma(-\tfrac{d}{4})
  = - \frac{\pi}{d} \csc(\frac{\pi d}{4})$
for the sake of compactness.


\subsection {Scaling}

Before diving into the details of explicitly performing the $\B$ integrals
in (\ref{eq:dGamIntDR}), it is worthwhile to see the general structure of
the result using a simple scaling argument.
We can quickly see the $\Delta t$ dependence by
rescaling $\B = \hat\B \sqrt{\Delta t/E}$, where $\hat\B$ is dimensionless,
and noting that $(X,Y,Z) \propto E/\Delta t$ in (\ref{eq:calXYZsmalldt}).
This rescaling pulls out {\it all}\/ of
the $\Delta t$ (and $E$) dependence from the two
$d^dB$ integrals, identifying the small $\Delta t$ behavior as
\begin {align}
   &\left[\frac{d\Gamma}{dx\,dy}\right]_{xy\bar y\bar x}
\nonumber\\ & \qquad
   \propto
   M_\ix M_\fx \left(\frac{|M_\ix| \Omega_\ix}{E^3}\right)^{d/2}
   \int \frac{d(\Delta t)}{(\Delta t)^{d/2}}
   \int_{\hat\B^\Ax,\hat\B^\bx}
   (\mbox{dimensionless function of $\hat\B$s and $\{x_j\}$})
\nonumber\\ & \qquad\qquad\qquad
   + [\mbox{similar for $\ix \leftrightarrow \fx$}]
\nonumber\\ &\qquad
   =
   (\mbox{dimensionless function of $\{x_j\}$})
   \times
   E^2 \left(\frac{\hat q}{E^5}\right)^{d/4}
   \int_0 \frac{d(\Delta t)}{(\Delta t)^{d/2}}
   \,.
\label {eq:structure}
\end {align}
Note that this result has the right dimension to be $d\Gamma/dx\,dy$
and also gives the usual dependence
$d\Gamma/dx\,dy \propto \sqrt{\hat q/E}$ on $\hat q$ and $E$ for $d=2$.

The UV behavior of the $\Delta t$ integral above is convergent and
well-defined for $d = 2-\epsilon$.
On the infrared (IR) end of the integration region (large $\Delta t$), our
small-$\Delta t$ expansion formulas are no longer valid.  But the
formulas above will be good enough to study the contribution we get,
if any, from arbitrarily small $\Delta t$.
Imagine taking the full expression for $d\Gamma/dx\,dy$ as an
integral over $\Delta t$, without having made any small $\Delta t$
approximation.  Then divide the integration region up into
$\Delta t < a$ and $\Delta t > a$ for some $a$ chosen
very small compared to the formation times in the problem:
\begin {equation}
   \int_0^\infty d(\Delta t) \>\cdots
   =
   \int_0^a d(\Delta t) \>\cdots
   +
   \int_a^\infty d(\Delta t) \>\cdots .
\end {equation}
The $\Delta t > a$ integration can be handled with the $d{=}2$
formulas of ref.\ \cite{2brem}: the difference between $d{=}2-\epsilon$
and $d{=}2$ in this region can be ignored as $\epsilon\to 0$.
The UV contributions that required UV regularization appear only in
the $\Delta t < a$ integration.  In (\ref{eq:structure}), that
integration region gives
\begin {equation}
   \int_0^a \frac{d(\Delta t)}{(\Delta t)^{(2-\epsilon)/2}}
   = \frac{2 a^{\epsilon/2}}{\epsilon}
   = \frac{2}{\epsilon} + \ln a + O(\epsilon) .
\label {eq:dtintDR}
\end {equation}
The $1/\epsilon$ and
$\ln a$ terms will cancel between diagrams
for the same reason that $1/\Delta t$ terms cancel between diagrams
in ref.\ \cite{2brem}.

However, there can be finite $O(\epsilon^0)$ contributions to diagrams
that do not cancel.  The UV piece of a diagram $D$ will have the
generic form
\begin {equation}
   f_D(\epsilon,\{x_j\}) \int_0^a \frac{d(\Delta t)}{(\Delta t)^{(2-\epsilon)/2}}
   .
\end {equation}
For the $xy\bar y\bar x$ diagram, for example, $f_D(\epsilon,\{x_j\})$
represents all
the factors in (\ref{eq:structure}) besides the $\Delta t$ integral.
Now consider the sum of some set of diagrams,
\begin {equation}
   \sum_D f_D(\epsilon,\{x_j\})
   \int_0^a \frac{d(\Delta t)}{(\Delta t)^{(2-\epsilon)/2}} \,,
\end {equation}
for which the
the $1/\Delta t$ pieces of the integrand cancel in $d{=}2$:
\begin {equation}
   \sum_D f_D(0,\{x_j\}) = 0 .
\end  {equation}
Using (\ref{eq:dtintDR}), we see that we can still get a non-vanishing
result from (i) the $O(\epsilon)$ pieces of $f_D(\epsilon)$ multiplying
(ii) the $2/\epsilon$ piece of (\ref{eq:dtintDR}):
\begin {equation}
   \sum_D f_D(\epsilon,\{x_j\})
   \int_0^a \frac{d(\Delta t)}{(\Delta t)^{(2-\epsilon)/2}}
   =
   2 \sum_D \frac{\partial f_D}{\partial\epsilon}(0,\{x_j\})
   + O(\epsilon) . 
\end {equation}
This finite piece, which survives in the $a \to 0$ limit, represents
the pole contribution that we are looking for: it is a contribution
associated with $\Delta t=0$ in the limit $d \to 2$.


\subsection {Actually doing the $\B$ integrals}

We now return to the unscaled $\B$'s, just to keep the discussion
as close as possible to the $d{=}2$ analysis of ref.\ \cite{2brem}.%
\footnote{
  \label{foot:expand}
  There is a slight difference with AI
  as far as line-by-line comparisons go.
  The small-argument expansion (\ref{eq:Kfactor})
  for the Bessel functions corresponds (for $d{=}2$)
  to making the expansion
  $e^{-\tfrac12 |M| \Omega B^2} \simeq 1 - \tfrac12 |M| \Omega B^2$
  to the 3-particle factors in AI {\it before} doing the integration
  over the $\B$'s.  In the end, that should get us to the same
  small-$\Delta t$ behavior, but AI did
  the operations in the opposite order: AI did the $\B$ integrals first
  and only then extracted the small $\Delta t$ limit.
}
Contracting the various transverse spatial indices $m$, $n$, $\bar m$,
and $\bar n$ of (\ref{eq:dGamIntDR}) gives
\begin {align}
   \left[\frac{d\Gamma}{dx\,dy}\right]_{xy\bar y\bar x} \simeq &
   \frac{\CA^2 \alphas^2 M_\ix M_\fx}{2^{d+3}d\pi^{2d}i^dE^d} \,
     \frac{\Gamma^2( \tfrac12{+}\tfrac{d}{4} )}{\sin(\frac{\pi d}{4})} \,
     \bigl({-}\hat x_1 \hat x_2 \hat x_3 \hat x_4 |M_\ix| \Omega_\ix\bigr)^{d/2}
   \int \frac{d(\Delta t)}{(\Delta t)^d} \>
\nonumber\\ &\times
   \Bigl\{
     (\beta Y_\bx Y_\Ax + \alpha \Ybar_{\bx\Ax} Y_{\bx\Ax}) J_{\ix 0}
     + (\alpha+\beta+d\gamma) Z_{\bx\Ax} J_{\ix 1}
\nonumber\\ &\quad
     + \bigl[
         (\alpha+\gamma) Y_\bx Y_\Ax
         + (\beta+\gamma) \Ybar_{\bx\Ax} Y_{\bx\Ax}
        \bigr] J_{\ix 2}
     - (\alpha+\beta+\gamma)
       (\Ybar_{\bx\Ax} Y_\Ax J_{\ix 3} + Y_\bx Y_{\bx\Ax} J_{\ix 4})
   \Bigl\}
\nonumber\\ &
   + [ \ix \leftrightarrow \fx] ,
\label {eq:IGammaDR}
\end {align}
where%
\footnote{
  Eq.\ (\ref{eq:IGammaDR}) above
  is the analog of a small-$\Delta t$ expansion of AI (5.45),
  along the lines of the previous footnote.
  The integrals $J$ of (\ref{eq:J}) here correspondingly
  play the roles of the
  integrals $I$ of AI (5.44).
}
\def\vfxx{\vphantom{\bigr)^2}}
\begin {subequations}
\label {eq:J}
\begin {align}
   J_{\ix 0} &\equiv
   \int_{\B^\bx,\B^\Ax}
   \frac{(B^\yx)^2 (B^\ybx)^2}{(B^\ybx)^d}
   \exp\Bigl[
     - \tfrac12 \calX_\bx (B^\bx)^2
     - \tfrac12 \calX_\Ax (B^\Ax)^2
     + X_{\bx\Ax} \B^\bx \cdot \B^\Ax
   \Bigr]
\nonumber\\ &\qquad
   =
   \frac{2^{\frac{d}{2}}\pi^d}{\Gamma(\frac{d}{2})\, \calX_\yx^{d/2}}
     \left[
       \frac{4 \calX_\bx \calX_\Ax}{\bigl(\calX_\bx \calX_\Ax - X_{\bx\Ax}^2\bigr)^2}
       - \frac{4 (1-\frac{d}{2})}{\calX_\bx \calX_\Ax - X_{\bx\Ax}^2\vfxx}
       \right] ,
\displaybreak[0]\\
   J_{\ix 1} &\equiv
   \int_{\B^\bx,\B^\Ax}
   \frac{\B^\bx\cdot\B^\Ax}{(B^\ybx)^d}
   \exp\Bigl[
     - \tfrac12 \calX_\bx (B^\bx)^2
     - \tfrac12 \calX_\Ax (B^\Ax)^2
     + X_{\bx\Ax} \B^\bx \cdot \B^\Ax
   \Bigr]
\nonumber\\ &\qquad
   =
   \frac{2^{\frac{d}{2}}\pi^d}{\Gamma(\frac{d}{2})\, \calX_\yx^{d/2}}
   \,
   \frac{2 X_{\bx\Ax}}{(\calX_\bx \calX_\Ax - X_{\bx\Ax}^2)\vfxx}
   \,,
\displaybreak[0]\\
   J_{\ix 2} &\equiv
   \int_{\B^\bx,\B^\Ax}
   \frac{(\B^\bx\cdot\B^\Ax)^2}{(B^\ybx)^d}
   \exp\Bigl[
     - \tfrac12 \calX_\bx (B^\bx)^2
     - \tfrac12 \calX_\Ax (B^\Ax)^2
     + X_{\bx\Ax} \B^\bx \cdot \B^\Ax
   \Bigr]
\nonumber\\ &\qquad
   =
   \frac{2^{\frac{d}{2}}\pi^d}{\Gamma(\frac{d}{2})\, \calX_\yx^{d/2}}
     \left[
       \frac{4 \calX_\bx \calX_\Ax}{\bigl(\calX_\bx \calX_\Ax - X_{\bx\Ax}^2\bigr)^2}
       - \frac{2}{\calX_\bx \calX_\Ax - X_{\bx\Ax}^2\vfxx}
     \right]
   ,
\displaybreak[0]\\
   J_{\ix 3} &\equiv
   \int_{\B^\bx,\B^\Ax}
   \frac{(B^\ybx)^2 \B^\bx\cdot\B^\Ax}{(B^\ybx)^d}
   \exp\Bigl[
     - \tfrac12 \calX_\bx (B^\bx)^2
     - \tfrac12 \calX_\Ax (B^\Ax)^2
     + X_{\bx\Ax} \B^\bx \cdot \B^\Ax
   \Bigr]
\nonumber\\ &\qquad
   =
   \frac{2^{\frac{d}{2}}\pi^d}{\Gamma(\frac{d}{2})\, \calX_\yx^{d/2}} \,
     \frac{4 \calX_\bx X_{\bx\Ax}}{\bigl(\calX_\bx \calX_\Ax - X_{\bx\Ax}^2\bigr)^2}
   ,
\displaybreak[0]\\
   J_{\ix 4} &\equiv
   \int_{\B^\bx,\B^\Ax}
   \frac{(B^\yx)^2 \B^\bx\cdot\B^\Ax}{(B^\ybx)^d}
   \exp\Bigl[
     - \tfrac12 \calX_\bx (B^\bx)^2
     - \tfrac12 \calX_\Ax (B^\Ax)^2
     + X_{\bx\Ax} \B^\bx \cdot \B^\Ax
   \Bigr]
\nonumber\\ &\qquad
   =
   \frac{2^{\frac{d}{2}}\pi^d}{\Gamma(\frac{d}{2})\, \calX_\yx^{d/2}}
     \left[
       \frac{4\calX_\Ax X_{\bx\Ax}}{\bigl(\calX_\bx \calX_\Ax - X_{\bx\Ax}^2\bigr)^2}
       + \frac{2d X_{\bx\Ax}}{\calX_\bx(\calX_\bx \calX_\Ax - X_{\bx\Ax}^2)\vfxx}
       \right]
   .
\end {align}
\end {subequations}
(See appendix \ref{app:Jints} for how to evaluate the integrals.)
The $J_{\fx n}$ are the same but with the denominator
$(B^\ybx)^d$ replaced by $(B^\yx)^d$ in the integrand.
That's equivalent to
\begin {equation}
   J_{\fx n} = (J_{\ix n} ~ \mbox{with} ~ \calX_\yx \leftrightarrow \calX_\ybx) .
\end {equation}

Because $Y_\yx$ and $Y_\ybx$ are $O(\Delta t)$ in the small $\Delta t$
limit [see (\ref{eq:calXYZsmalldt})], the expression
(\ref{eq:IGammaDR}) simplifies to%
\footnote{
  (\ref{eq:IGammaDR2}) plays a role similar to AI (D4).
}
\begin {align}
   \left[\frac{d\Gamma}{dx\,dy}\right]_{xy\bar y\bar x} \simeq &
   \frac{\CA^2 \alphas^2 M_\ix M_\fx}{2^{d+3}d\pi^{2d}i^dE^d} \,
     \frac{\Gamma^2( \tfrac12{+}\tfrac{d}{4} )}{\sin(\frac{\pi d}{4})} \,
     \bigl({-}\hat x_1 \hat x_2 \hat x_3 \hat x_4 |M_\ix| \Omega_\ix\bigr)^{d/2}
   \int \frac{d(\Delta t)}{(\Delta t)^d} \>
\nonumber\\ &\times
   \Bigl\{
     \Ybar_{\bx\Ax} Y_{\bx\Ax} (\alpha J_{\ix 0} + \beta J_{\ix 2} + \gamma J_{\ix 2})
     + Z_{\bx\Ax} (\alpha+\beta+d\gamma) J_{\ix 1}
   \Bigl\}
\nonumber\\ &
   + \{ \ix \leftrightarrow \fx \} .
\label {eq:IGammaDR2}
\end {align}
Using the small $\Delta t$ limit (\ref{eq:calXYZsmalldt})
for $X$,
and using $M_\ix = x_1 x_4(x_1{+}x_4)E$ and $M_\fx = x_3 x_4(x_3{+}x_4)E$, we have%
\footnote{
  We do not need $J_{\ix 3}$ and $J_{\ix 4}$ in (\ref{eq:IGammaDR}), but
  their small $\Delta$ limits may be found in eqs.\ (\ref{eq:J34small})
  of appendix \ref{app:Jints}.
}
\begin {subequations}
\label {eq:J2}
\begin {align}
   J_{\ix 0} &\simeq
   \frac{2^{\frac{d}{2}}\pi^d}{\Gamma(\frac{d}{2})}
   \left( \frac{\Delta t}{-i M_\ix} \right)^{d/2}
   \left( \frac{\Delta t}{-i E} \right)^2
   \frac{4(x_1x_3 - \frac{d}{2} x_2x_4)}{x_1 x_2^2 x_3 x_4^4}
   \,,
\displaybreak[0]\\
   J_{\ix 1} &\simeq
   \frac{2^{\frac{d}{2}}\pi^d}{\Gamma(\frac{d}{2})}
   \left( \frac{\Delta t}{-i M_\ix} \right)^{d/2}
   \left( \frac{\Delta t}{-i E} \right)
   \frac{2}{x_2 x_4^2}
   \,,
\displaybreak[0]\\
   J_{\ix 2} &\simeq
   \frac{2^{\frac{d}{2}}\pi^d}{\Gamma(\frac{d}{2})}
   \left( \frac{\Delta t}{-i M_\ix} \right)^{d/2}
   \left( \frac{\Delta t}{-i E} \right)^2
   \frac{4(x_1x_3 - \frac{1}{2} x_2x_4)}{x_1 x_2^2 x_3 x_4^4}
   .
\end {align}
\end {subequations}
Eqs.\ (\ref{eq:calXYZsmalldt}) and (\ref{eq:IGammaDR2}) then give%
\footnote{
  Why does the result (\ref{eq:poleDR0}) diverge for $d=4$?
  For that $d$, the Bessel function expansion (\ref{eq:BesselSmall}) used
  in (\ref{eq:Kfactor}) is
  incorrect: for $\nu = d/4 \to 1$, the $z^\nu$ term in (\ref{eq:BesselSmall})
  becomes the same order as the next
  term in the expansion ($z^{-\nu+2}$),
  and those two terms combine to give subleading $z\ln z$ behavior in
  (\ref{eq:BesselSmall}) instead of just $z$.
}
\begin {align}
   \left[\frac{d\Gamma}{dx\,dy}\right]_{xy\bar y\bar x} &\simeq
     \frac{\CA^2 \alphas^2 M_\ix M_\fx}{2^{\frac{d}{2}+2}d\pi^{d}x_4^2E^d} \,
     \frac{\Gamma^2( \tfrac12{+}\tfrac{d}{4} )}
          {\Gamma(\frac{d}{2})\,\sin(\frac{\pi d}{4})} \,
     (i\hat x_1 \hat x_2 \hat x_3 \hat x_4 \Omega_\ix \sgn M_\ix)^{d/2}
   \int \frac{d(\Delta t)}{(\Delta t)^{d/2}}
\nonumber\\ &\qquad\times
       \left[
         \bigl(\alpha + \beta + (2{-}d)\gamma\bigr)
              (\hat x_1 \hat x_3 - \hat x_2 \hat x_4)
         - (d-1)(\alpha + \gamma) \hat x_2 \hat x_4
       \right]
\nonumber\\ &\quad
   + \{\ix\leftrightarrow\fx\} ,
\label {eq:poleDR0}
\end {align}
which (using $x_1+x_2+x_3+x_4=0$) can be rewritten as
\begin {align}
   &
   \left[\frac{d\Gamma}{dx\,dy}\right]_{xy\bar y\bar x} \simeq
\nonumber\\ & \qquad
   \int \frac{d(\Delta t)}{(\Delta t)^{d/2}} \>
   \frac{\CA^2 \alphas^2}{2^{\frac{d}{2}+2}d\pi^{d}E^{d-2}} \,
     \frac{\Gamma^2( \tfrac12{+}\tfrac{d}{4} )}
          {\Gamma(\frac{d}{2})\,\sin(\frac{\pi d}{4})} \,
     (i\hat x_1 \hat x_2 \hat x_3 \hat x_4 \Omega_\ix \sgn M_\ix)^{d/2}
\nonumber\\ &\qquad\times
     \hat x_1 \hat x_3 (\hat x_1+\hat x_4)^2 (\hat x_3+\hat x_4)^2
       \left[
         \bigl(\alpha + \beta + (2{-}d)\gamma\bigr)
         - \frac{(d-1)(\alpha + \gamma) \hat x_2 \hat x_4}
                {(\hat x_1+\hat x_4)(\hat x_3+\hat x_4)}
       \right]
\nonumber\\ &\quad
   + \{\ix\leftrightarrow\fx\} .
\label {eq:poleDR1}
\end {align}
If desired,
the $\Gamma$ functions may be manipulated into the same form as
in the single splitting result (\ref{eq:integral2}) using
\begin {equation}
   \frac{\Gamma^2( \tfrac12{+}\tfrac{d}{4} )}
        {\Gamma(\frac{d}{2})\,\sin(\frac{\pi d}{4})}
   = 
   - \frac{d \Beta(\tfrac12{+}\tfrac{d}{4},-\tfrac{d}{4})}
          {2^{1+\frac{d}{2}}} .
\label {eq:GammaRewrite}
\end {equation}

For $d{=}2$, our result reproduces the corresponding
(unregulated) small-$\Delta t$ behavior
in ref.\ \cite{2brem},%
\footnote{
  AI (5.46)
}
which is, for future reference,
\begin {align}
   &
   \left[\frac{d\Gamma}{dx\,dy}\right]_{xy\bar y\bar x} \simeq
   \int \frac{d(\Delta t)}{\Delta t} \>
   \frac{\CA^2 \alphas^2}{16\pi^2} \,
     (i\Omega_\ix \sgn M_\ix + i\Omega_\fx \sgn M_\fx)
\nonumber\\ &\qquad\times
     \hat x_1^2 \hat x_2 \hat x_3^2 \hat x_4
     (\hat x_1+\hat x_4)^2 (\hat x_3+\hat x_4)^2
       \left[
         (\alpha + \beta)
         - \frac{(\alpha + \gamma) \hat x_2 \hat x_4}
                {(\hat x_1{+}\hat x_4)(\hat x_3{+}\hat x_4)}
       \right]
\label {eq:poled2}
\end {align}
(after subtracting the purely vacuum contribution).


\subsection {Branch cuts}

We will need to be somewhat careful about branch cuts when we expand
in $\epsilon$ for $d=2-\epsilon$.  As an example, $(-i)^d$ could
in principle be
$(e^{-i\pi/2})^{2-\epsilon} = -1 + i \frac{\pi}{2} \epsilon + O(\epsilon^2)$
or
$(e^{3i\pi/2})^{2-\epsilon} = -1 - i \frac{3\pi}{2} \epsilon + O(\epsilon^2)$
or some other variant.
The derivation of (\ref{eq:poleDR1}) took the cavalier
attitude that,
if we keep formulas as simple as possible and never isolate factors
that might be fractional powers of negative real numbers,
then standard choices of branch cuts would give the correct
answer.  That is, we assumed it was okay to write
$(i x_1 x_2 x_3 x_4 \Omega \sgn M)^{d/2}$
but not (without further branch-cut clarification)
$(i \Omega \sgn M)^{d/2} (x_1 x_2 x_3 x_4)^{d/2}$, since
$x_1 x_2 x_3 x_4 < 0$ for the $x y\bar y\bar x$ interference diagram.
Moreover, we want expressions that will also work correctly for
{\it other}\/ diagrams when we later relate them to the
results for $x y\bar y\bar x$, in which case $M$, $\Omega$, and
$x_1 x_2 x_3 x_4$ turn out to have other signs and phases.
In appendix \ref{app:sheet}, we check that the combination
$(i x_1 x_2 x_3 x_4 \Omega \sgn M)^{d/2}$ used in
(\ref{eq:poleDR1}), with the standard choice to run the branch cut
along the negative real axis,
produces the correct overall phase for all cases of
interest.  We will see in section \ref{sec:epsexpand} below that
these complex phases generate the $1/\pi$ pole terms
previously found in ref.\ \cite{2brem} using naive $\epsilon$
prescriptions.

Despite the ambiguity of the expression $(x_1 x_2 x_3 x_4)^{d/2}$
for $x_1 x_2 x_3 x_4 < 0$, it will be convenient to rewrite
\begin {subequations}
\label {eq:sheet}
\begin {equation}
  (i x_1 x_2 x_3 x_4 \Omega \sgn M)^{d/2}
  = (i \Omega \sgn M)^{d/2} (x_1 x_2 x_3 x_4)^{d/2}
\label {eq:sheet1}
\end {equation}
below.  This rewriting works if we adopt the convention that
negative real numbers have phase $e^{-i\pi}$.
See appendix \ref{app:sheet} for verification of (\ref{eq:sheet1})
in all relevant cases.
That is, one should
interpret
\begin {equation}
   x_1 x_2 x_3 x_4 \equiv e^{-i\pi \theta(-x_1 x_2 x_3 x_4)} |x_1 x_2 x_3 x_4|
\label {eq:sheet2}
\end {equation}
\end {subequations}
in what follows,
where $\theta$ is the step function.


\subsection {Expansion in \boldmath$\epsilon$}
\label {sec:epsexpand}

Now take the $\epsilon \to 0$ limit of (\ref{eq:poleDR1}) for
$d = 2-\epsilon$.
To isolate the pole contribution,
introduce a small cut-off $a$ as in (\ref{eq:dtintDR}):
\begin {equation}
   \int_0^a \frac{d(\Delta t)}{(\Delta t)^{(2-\epsilon)/2}}
   = \frac{2 a^{\epsilon/2}}{\epsilon}
   = \frac{2}{\epsilon} + \ln a + O(\epsilon) .
\label {eq:dtintDR2}
\end {equation}
With this cut-off, we can rewrite (\ref{eq:poleDR1}) as
\begin {align}
   \left[\frac{d\Gamma}{dx\,dy}\right]^{(\Delta t < a)}_{xy\bar y\bar x} &=
   (\mbox{almost usual})
\nonumber\\ &
   - \frac{i\CA^2 \alphas^2}{16\pi^2}
     \bigl[ \Omega_\ix \sgn M_\ix + \Omega_\fx \sgn M_\fx \bigr]
     \hat x_1^2 \hat x_2 \hat x_3^2 \hat x_4
     (\hat x_1+\hat x_4)^2 (\hat x_3+\hat x_4)^2
\nonumber\\ &\times
     \left\{
       \left[
         (\alpha + \beta)
         - \frac{(\alpha + \gamma) \hat x_2 \hat x_4}
                {(\hat x_1+\hat x_4)(\hat x_3+\hat x_4)}
       \right]
       \ln(\hat x_1 \hat x_2 \hat x_3 \hat x_4)
       - 2\gamma 
       - \frac{2 (\alpha+\gamma) \hat x_2 \hat x_4}
                {(\hat x_1+\hat x_4)(\hat x_3+\hat x_4)}
    \right\}
\nonumber\\ &
    + O(\epsilon) ,
\label {eq:poleDR2A}
\end {align}
where
\begin {align}
   (\mbox{almost usual}) =&
   \left[\frac{2}{\epsilon} + \ln(E^2 a) + c_1\right]
   \frac{\CA^2 \alphas^2}{16\pi^2}
     \bigl[ (i\Omega_\ix \sgn M_\ix)^{d/2} + (i\Omega_\fx \sgn M_\fx)^{d/2} \bigr]
\nonumber\\ &\times
     \hat x_1^2 \hat x_2 \hat x_3^2 \hat x_4
     (\hat x_1+\hat x_4)^2 (\hat x_3+\hat x_4)^2
       \left[
         (\alpha + \beta)
         - \frac{(\alpha + \gamma) \hat x_2 \hat x_4}
                {(\hat x_1+\hat x_4)(\hat x_3+\hat x_4)}
       \right]
\label {eq:usualish}
\end {align}
and $c_1 = 1+\ln(2\pi^2)$ is an uninteresting numerical constant that will
not appear in our final results.
It's convenient not to explicitly expand the factors of
$(i\Omega \sgn M)^{d/2}$ above.

\begin {figure}[t]
\begin {center}
  \includegraphics[scale=0.5]{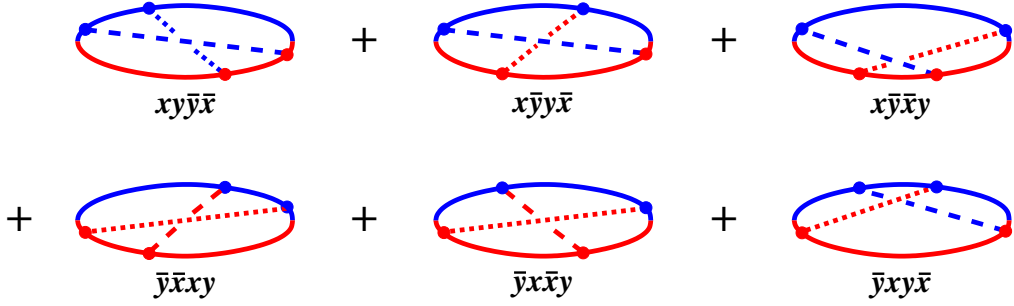}
  \caption{
     \label{fig:cancel}
     A subset of crossed diagrams for which all $1/\Delta t$ divergences
     cancel.
  }
\end {center}
\end {figure}

We save some effort in what follows by having separated (\ref{eq:usualish})
from the other terms in (\ref{eq:poleDR2A}).
Specifically, note that (\ref{eq:usualish}) is almost proportional to
the coefficient of $1/\Delta t$ in the integrand of
(\ref{eq:poled2}), which is the
$1/\Delta t$  behavior
found previously in the $d{=}2$ analysis of ref.\ \cite{2brem} and
which we'll call the ``usual'' behavior.
We say ``almost''
proportional because (\ref{eq:usualish}) has
\begin {itemize}
\item
  $(i\Omega_\ix \sgn M_\ix)^{d/2}$ and $(i\Omega_\fx \sgn M_\fx)^{d/2}$
  instead of $i\Omega_\ix \sgn M_\ix$ and $i\Omega_\fx \sgn M_\fx$;
\item
  $d$-dimensional $(\alpha,\beta,\gamma)$ instead of
  2-dimensional $(\alpha,\beta,\gamma)$.
\end {itemize}
These differences inspires our designation ``almost usual''
in (\ref{eq:usualish}).
In ref.\ \cite{2brem}, the $1/\Delta t$ pieces of
the integrand canceled between the six diagrams of fig.\
\ref{fig:cancel} and so canceled in the sum of all crossed diagrams
as well.
So, we might expect the corresponding
contribution (\ref{eq:usualish}) to similarly
cancel here.
Indeed, the replacement of $i\Omega \sgn M$ by $(i\Omega \sgn M)^{d/2}$
does not affect the cancellation, because the pieces with different
values of $i\Omega \sgn M$ canceled separately in ref.\ \cite{2brem}.%
\footnote{
  A little more argument is needed here.  If one goes through the
  cancellation of these terms in fig.\ \ref{fig:cancel},
  the cancellation relies on the fact that
  $(\tilde\Omega_\fx)_{x\leftrightarrow y}^* = \Omega_\fx$
  and thence
  $(i \tilde\Omega_\fx \sgn \tilde M_\fx)_{x\leftrightarrow y}^*
    = i \Omega_\fx \sgn M_\fx$, where $\tilde\Omega_\fx$ and
  $\tilde M_\fx$ are defined as in AI section VI.B.
  So it is important that we generalized $i \Omega \sgn M$ to
  $(i \Omega \sgn M)^{d/2}$ when separating the terms
  (\ref{eq:usualish}) from the rest and did not generalize to, e.g.,
  $i (\Omega \sgn M)^{d/2}$ or $(i \Omega)^{d/2} \sgn M$.
}
The replacement of
2-dimensional $(\alpha,\beta,\gamma)$ with $d$-dimensional
$(\alpha,\beta,\gamma)$ similarly does not affect the cancellation
because the cancellation in ref.\ \cite{2brem}
did not depend on the specific form of the functions
$(\alpha,\beta,\gamma)$.  The upshot is that the contributions
(\ref{eq:usualish}) will cancel between diagrams.  Among other things,
this means that the $\ln a$ terms which depended on our choice of
cut-off $a$ in (\ref{eq:dtintDR2}) will disappear.

So, we may ignore the ``almost usual'' contributions
of (\ref{eq:usualish}) and focus exclusively on the other
terms in (\ref{eq:poleDR2A}).%
\footnote{
  Readers uneasy about this conclusion may instead keep
  the terms (\ref{eq:usualish}) through everything that follows and
  come to our final result with a bit more algebra.
}
Those other terms are finite, and so we
may now take $(\alpha,\beta,\gamma)$ to be given by their
$d{=}2$ formulas because the discrepancy will only contribute to
the final result at $O(\epsilon)$.


\subsection {Other diagrams}

In ref.\ \cite{2brem}, it was shown that $x\bar y y\bar x$ could
be obtained from $xy\bar y\bar x$ by%
\footnote{
  Specifically, section VI of ref.\ \cite{2brem}.
}
\begin {itemize}
\item
   $\hat x_i \to x'_i$;
\item
   $(M_\ix,\Omega_\ix) \leftrightarrow (M_\fx,\Omega_\fx)$ if this
   change was overlooked when applying the $\hat x_i \to x'_i$ rule;
\item
   $\alpha(x,y) \leftrightarrow \beta(x,y)$;
\end {itemize}
with
\begin {equation}
   (x'_1,x'_2,x'_3,x'_4) = \bigl(-(1{-}y),{-}y,1{-}x,x\bigr)
   =
   \bigl(
     -(\hat x_3{+}\hat x_4),{-}\hat x_2,{-}(\hat x_1{+}\hat x_4),\hat x_4
   \bigr) .
\label {eq:xprime}
\end {equation}
So (\ref{eq:poleDR2A}) gives
\begin {align}
   &
   \left[\frac{d\Gamma}{dx\,dy}\right]^{(\Delta t < a)}_{x\bar y y\bar x}
\nonumber\\ & \qquad
   = (\mbox{almost usual})'
\nonumber\\ & \qquad\quad
   - \frac{i\CA^2 \alphas^2}{16\pi^2}
     \bigl[ \Omega_\fx \sgn M_\fx + \Omega_\ix \sgn M_\ix \bigr]
     {x'_1}^2 x'_2 {x'_3}^2 x'_4
     (x'_1+x'_4)^2 (x'_3+x'_4)^2
\nonumber\\ & \qquad\quad\qquad \times
     \left\{
       \left[
         (\beta + \alpha)
         - \frac{(\beta + \gamma) x'_2 x'_4}
                {(x'_1+x'_4)(x'_3+x'_4)}
       \right]
       \ln(x'_1 x'_2 x'_3 x'_4)
       - 2\gamma
       - \frac{2(\beta+\gamma) x'_2 x'_4}
                {(x'_1+x'_4)(x'_3+x'_4)}
    \right\}
\displaybreak[1]\nonumber\\ &\qquad
   = (\mbox{almost usual})'
\nonumber\\ & \qquad\quad
   - \frac{i\CA^2 \alphas^2}{16\pi^2}
     \bigl[ \Omega_\ix \sgn M_\ix + \Omega_\fx \sgn M_\fx \bigr]
     \hat x_1^2 \hat x_2 \hat x_3^2 \hat x_4
     (\hat x_1+\hat x_4)^2 (\hat x_3+\hat x_4)^2
\nonumber\\ &\qquad\quad\qquad \times
     \biggl\{
       \left[
         - (\alpha + \beta)
         - \frac{(\beta + \gamma) \hat x_2 \hat x_4}
                {\hat x_1 \hat x_3}
       \right]
       \ln\bigl|
          \hat x_2 \hat x_4 (\hat x_1 + \hat x_4)(\hat x_3 + \hat x_4)
       \bigr|
\nonumber\\ &\qquad\quad\qquad\qquad
       + 2\gamma
       - \frac{2(\beta+\gamma) \hat x_2 \hat x_4}
              {\hat x_1 \hat x_3}
    \biggr\} ,
\label {eq:poleDR2B}
\end {align}
where we have found it convenient to use the fact that
$x_1' x_2' x_3' x_4' > 0$ to
replace $\ln(x_1' x_2' x_3' x_4')$ by $\ln| x_1' x_2' x_3' x_4'|$ before
rewriting the $x_i'$ in terms of $\hat x_i$.

Similarly, $x\bar y \bar x y$ was shown to be obtainable
from $xy\bar y\bar x$ by
\begin {itemize}
\item
   $\hat x_i \to \tilde x_i$;
\item
   $(M_\ix,\Omega_\ix;M_\fx,\Omega_\fx)
    \to (\tilde M_\fx,\tilde\Omega_\fx;M_\ix,\Omega_\ix)$
    if this
    change was overlooked when applying the $\hat x_i \to \tilde x_i$ rule;
\item
   $\bigl(\alpha(x,y),\beta(x,y),\gamma(x,y)\bigr) \to
    \bigl(\gamma(x,y),\alpha(x,y),\beta(x,y)\bigr)$;
\end {itemize}
with
\begin {equation}
   (\tilde x_1,\tilde x_2,\tilde x_3,\tilde x_4) =
   \bigl({-}y,-(1{-}y),x,1{-}x\bigr) =
   \bigl(
     {-}\hat x_2,-(\hat x_3{+}\hat x_4),\hat x_4,{-}(\hat x_1{+}\hat x_4)
   \bigr) ,
\label {eq:xtilde}
\end {equation}
\begin {equation}
  \tilde M_\fx = \tilde x_1 \tilde x_4 (\tilde x_1{+}\tilde x_4) E
  = - y(1{-}x)(1{-}x{-}y) E ,
\end {equation}
and
\begin {equation}
   \tilde\Omega_\fx
   = \sqrt{ 
     -\frac{i \hat q_{\rm A}}{2E}
     \left( \frac{1}{\tilde x_1}
            + \frac{1}{\tilde x_4} - \frac{1}{\tilde x_1 {+} \tilde x_4}
            \right)
   }
   = \sqrt{ 
     -\frac{i \hat q_{\rm A}}{2E}
     \left( -\frac{1}{y} + \frac{1}{1{-}x} - \frac{1}{1{-}x{-}y}
            \right)
   } .
\end {equation}
So (\ref{eq:poleDR2A}) gives
\begin {align}
   &
   \left[\frac{d\Gamma}{dx\,dy}\right]^{(\Delta t < a)}_{x\bar y \bar x y}
\nonumber\\ & \qquad
   = (\mbox{almost}\widetilde{~}\mbox{usual})
\nonumber\\ & \qquad\quad
   - \frac{i\CA^2 \alphas^2}{16\pi^2}
     \bigl[ \tilde\Omega_\fx \sgn \tilde M_\fx + \Omega_\ix \sgn M_\ix \bigr]
     \tilde x_1^2 \tilde x_2 \tilde x_3^2 \tilde x_4
     (\tilde x_1+\tilde x_4)^2 (\tilde x_3+\tilde x_4)^2
\nonumber\\ & \qquad\quad\qquad \times
     \left\{
       \left[
         (\gamma + \alpha)
         - \frac{(\gamma + \beta) \tilde x_2 \tilde x_4}
                {(\tilde x_1+\tilde x_4)(\tilde x_3+\tilde x_4)}
       \right]
       \ln(\tilde x_1 \tilde x_2 \tilde x_3 \tilde x_4)
       - 2\beta
       - \frac{2(\gamma+\beta) \tilde x_2 \tilde x_4}
                {(\tilde x_1+\tilde x_4)(\tilde x_3+\tilde x_4)}
    \right\}
\displaybreak[1]\nonumber\\ &\qquad
   = (\mbox{almost}\widetilde{~}\mbox{usual})
\nonumber\\ & \qquad\quad
   - \frac{i\CA^2 \alphas^2}{16\pi^2}
     \bigl[ \Omega_\ix \sgn M_\ix + \tilde\Omega_\fx \sgn \tilde M_\fx \bigr]
     \hat x_1^2 \hat x_2 \hat x_3^2 \hat x_4
     (\hat x_1+\hat x_4)^2 (\hat x_3+\hat x_4)^2
\nonumber\\ &\qquad\quad\qquad \times
     \biggl\{
       \left[
         \frac{(\alpha {+} \gamma) \hat x_2 \hat x_4}
              {(\hat x_1{+}\hat x_4)(\hat x_3{+}\hat x_4)}
         + \frac{(\beta {+} \gamma) \hat x_2 \hat x_4}
                {\hat x_1 \hat x_3}
       \right]
       \ln\bigl|
          \hat x_2 \hat x_4 (\hat x_1 {+} \hat x_4)(\hat x_3 {+} \hat x_4)
       \bigr|
\nonumber\\ &\qquad\quad\qquad\qquad
       - \frac{2\beta \hat x_2 \hat x_4}
              {(\hat x_1{+}\hat x_4)(\hat x_3{+}\hat x_4)}
       + \frac{2(\beta+\gamma) \hat x_2 \hat x_4}
              {\hat x_1 \hat x_3}
    \biggr\}
    .
\label {eq:poleDR2C}
\end {align}
Finally, now that we are no longer making substitutions to get
one diagram from another, we can use $\hat x_1 \hat x_2 \hat x_3 \hat x_4 < 0$
to replace the $\ln(\hat x_1 \hat x_2 \hat x_3 \hat x_4)$ in
(\ref{eq:poleDR2A}) by
$\ln(e^{-i\pi} |\hat x_1 \hat x_2 \hat x_3 \hat x_4|)$ as in (\ref{eq:sheet2}).

The $\Omega_\ix$ piece arising from adding together the three
diagrams (\ref{eq:poleDR2A}),
(\ref{eq:poleDR2B}), and (\ref{eq:poleDR2C}) is
\begin {align}
   - \frac{i\CA^2 \alphas^2}{16\pi^2} &
     \bigl[ \Omega_\ix \sgn M_\ix \bigr]
     \hat x_1^2 \hat x_2 \hat x_3^2 \hat x_4
     (\hat x_1+\hat x_4)^2 (\hat x_3+\hat x_4)^2
\nonumber\\ &\times
     \biggl\{
       \left[
         (\alpha + \beta)
         - \frac{(\alpha + \gamma) \hat x_2 \hat x_4}
                {(\hat x_1+\hat x_4)(\hat x_3+\hat x_4)}
       \right]
       \ln\left(
         e^{-i\pi} \left|
           \frac{\hat x_1 \hat x_3}
                {(\hat x_1+\hat x_4)(\hat x_3+\hat x_4)}
         \right|
       \right)
\nonumber\\ &\qquad
       - \frac{2 (\alpha+\beta+\gamma) \hat x_2 \hat x_4}
              {(\hat x_1+\hat x_4)(\hat x_3+\hat x_4)}
    \biggr\}
    .
\label {eq:poleDR2ABCi}
\end {align}
In fig.\ \ref{fig:cancel}, we further added in diagrams corresponding to
swapping $x\leftrightarrow y$ while also conjugating.
If it weren't for the $e^{-i\pi}$ factor inside the logarithm, this
would have the effect, after including all $\Omega$ terms and
evaluating all $\sgn M$, of
replacing
\begin {align}
  i [ \Omega_\ix \sgn M_\ix ]
  \longrightarrow
  i & [ \Omega_\ix + \Omega_\fx ] + \{x \leftrightarrow y\}^*
  =
  i [ \Omega_\ix - \tilde\Omega_\fx ] + \{x \leftrightarrow y\}^*
\nonumber\\
  =
  i & [\Omega_{-1,1-x,x} + \Omega_{-(1-y),1-x-y,x}
        - \Omega_{-1,1-y,y}^* - \Omega_{-(1-x),1-x-y,y}^*]
\end {align}
in (\ref{eq:poleDR2ABCi}) above,
where
\begin {equation}
   \Omega_{x_1,x_2,x_3} \equiv
   \sqrt{ 
     -\frac{i \hat q_{\rm A}}{2E}
     \left( \frac{1}{x_1} 
            + \frac{1}{x_2} + \frac{1}{x_3} \right)
   } .
\end {equation}
(Recall that $\alpha$, $\beta$, and
$\gamma$ are real and symmetric in $x \leftrightarrow y$.)
But the $\ln(e^{-i\pi}) = -i \pi$ needs to be treated separately.
The result is
\begin {align}
   \left[\frac{d\Gamma}{dx\,dy}\right]^{\rm pole}_{\rm fig.~\ref{fig:cancel}}
   &=
    \frac{\CA^2 \alphas^2}{16\pi^2} \,
    x y (1{-}x)^2 (1{-}y)^2(1{-}x{-}y)^2
\nonumber\\ & \qquad \times
    \biggl\{
       -i
       [\Omega_{-1,1-x,x} + \Omega_{-(1-y),1-x-y,x}
        - \Omega_{-1,1-y,y}^* - \Omega_{-(1-x),1-x-y,y}^*]
\nonumber\\ & \qquad \qquad \times
       \biggl[
       \left(
         (\alpha + \beta)
         + \frac{(\alpha + \gamma) xy}{(1{-}x)(1{-}y)}
       \right) \ln \left( \frac{1{-}x{-}y}{(1{-}x)(1{-}y)} \right)
       + \frac{2(\alpha+\beta+\gamma) x y}{(1{-}x)(1{-}y)}
       \biggr]
\nonumber\\ &\qquad \quad
     - \pi
       [\Omega_{-1,1-x,x} + \Omega_{-(1-y),1-x-y,x}
        + \Omega_{-1,1-y,y}^* + \Omega_{-(1-x),1-x-y,y}^*]
\nonumber\\ & \qquad \qquad \times
       \left(
         (\alpha + \beta)
         + \frac{(\alpha + \gamma) xy}{(1{-}x)(1{-}y)}
       \right)
    \biggr\}
    .
\label {eq:crossedpole}
\end {align}
The last term (the $\pi/\pi^2 = 1/\pi$ term)
is the same as the (incomplete) pole term that was
found in ref.\ \cite{2brem}.%
\footnote{
  AI (7.25)
}
The other term (the $1/\pi^2$ term) is new.


\section{Sequential diagrams with dimensional regularization}
\label {sec:seq}

We now turn to the sequential diagrams of fig.\ \ref{fig:seq2}.
Ref.\ \cite{seq}
analyzed these diagrams except
for the pole pieces, for which results were quoted by reference
to this paper.  Here we present the analysis of those poles,
using techniques similar to those of the previous section.


\subsection{\boldmath$2 \Re(x\bar xy\bar y + x\bar x\bar y y)$
            minus Monte Carlo}
\label {sec:xxyy}

Consider the last two diagrams in fig.\ \ref{fig:seq2} plus
their conjugates.  As discussed in ref.\ \cite{seq}, these diagrams
factorize into separate pieces associated with $x$ emission followed
by $y$ emission, and they are {\it almost}\/ the same thing as the
corresponding idealized Monte Carlo calculation based on single splitting
rates.  The ``almost'' has to do with restrictions on the
time ordering, which cause the difference with idealized Monte Carlo
to be%
\footnote{
   ACI (2.23)
}
\begin {align}
   \left[\Delta \frac{d\Gamma}{dx\,dy}\right]_{
        \begin{subarray}{} x\bar x y\bar y + x\bar x \bar y y \\
                        + \bar xx \bar yy + \bar x xy\bar y \end{subarray}
   }
   &= - \frac{1}{1-x}
   \int_0^\infty \! d(\Delta t_x)
   \int_0^\infty \! d(\Delta t_y) \>
   \tfrac12(\Delta t_x+\Delta t_y)
\nonumber\\& \qquad \times
     \Re \left[\frac{d\Gamma}{dx\,d(\Delta t_x)}\right]_E \,
     \Re \left[\frac{d\Gamma}{d\yfrak\,d(\Delta t_y)}\right]_{(1-x)E} ,
\label {eq:DxxyyetcDR}
\end {align}
where $d\Gamma/dx\,d(\Delta t)$ is the integrand associated
with the single-splitting result (\ref{eq:dGdxQM}):
\begin {equation}
   \left[\frac{d\Gamma}{dx}\right]_E
   =
   \Re \int_0^\infty d(\Delta t) \>
   \left[\frac{d\Gamma}{dx\,d(\Delta t)}\right]_E \,,
\end {equation}
with
\begin {equation}
   \left[\frac{d\Gamma}{dx\,d(\Delta t)}\right]_E \equiv
   \frac{\alphas P(x)}{x^2(1-x)^2E^d}
   \grad_{\B^\xbx} \cdot \grad_{\B^\xx}
   \langle \B^\xbx,\Delta t | \B^\xx,0 \rangle_{E,x}
   \Bigr|_{\B^\xbx = \B^\xx = 0} .
\label {eq:dGdxdt}
\end {equation}
In this calculation, it will be important to keep in mind that
$P(x)$ is, for now, the $d$-dimensional generalization of the
DGLAP splitting function.  We have written the overall factor of
$E$ as $E^{-d}$ in order to keep $P(x)$ dimensionless, similar
to our choice of convention in section \ref{sec:first}.
The argument \cite{seq} for (\ref{eq:DxxyyetcDR}) 
was valid in any transverse dimension $d$.

We've already seen from (\ref{eq:integral2}) that
\begin {align}
   \frac{d\Gamma}{dx}
   &=
   \Re \int_0^\infty d(\Delta t) \>
   \left[\frac{d\Gamma}{dx\,d(\Delta t)}\right]_E
\nonumber\\
   &=
   \Re\left[
     - \frac{\alphas P(x)}{x^2(1-x)^2E^d}
     \frac{i d M}{2} \left( \frac{M\Omega}{2\pi} \right)^{d/2}
     \Beta(\tfrac12{+}\tfrac{d}{4},-\tfrac{d}{4})
   \right]
\label {eq:seqint1}
\end {align}
in dimensional regularization.
For (\ref{eq:DxxyyetcDR}), we also need integrals of the form
\begin {subequations}
\label{eq:otherintDR}
\begin {multline}
   \Re \int_0^\infty d(\Delta t) \>
   \Delta t
   \left[\frac{d\Gamma}{dx\,d(\Delta t)}\right]_E
\\
   =
   \Re \left[
   \frac{2d\pi\alphas P(x)}{x^2(1-x)^2E^d}
   \left( \frac{M \Omega}{2\pi i} \right)^{\frac{d}{2}+1}
   \int_0^\infty d(\Delta t) \>
   \Delta t \,
   \csc^{\frac{d}{2}+1}(\Omega\,\Delta t)
   \right] .
\end {multline}
However, for the pole contribution, we'll really be interested in the
small-$\Delta t$ contribution to this integral, which is
\begin {equation}
   \Re \int_0^a d(\Delta t) \>
   \Delta t
   \left[\frac{d\Gamma}{dx\,d(\Delta t)}\right]_E
   =
   -
   \frac{d\alphas P(x)}{2\pi E^{d-2}}
   \Re \left[
     \left( \frac{M}{2\pi i} \right)^{\frac{d}{2}-1}
     \int_0^a \frac{d(\Delta t)}{(\Delta t)^{d/2}}
   \right] ,
\label{eq:seqint2}
\end {equation}
\end {subequations}
where $a$ is a tiny cut-off on $\Delta t$ just like the one we introduced for
the crossed diagrams in (\ref{eq:dtintDR2}).

Combining (\ref{eq:seqint1}) and (\ref{eq:seqint2})
as in (\ref{eq:DxxyyetcDR}) gives
\begin {align}
   \left[\Delta \frac{d\Gamma}{dx\,dy}\right]^{(\Delta t<a)}_{
        \begin{subarray}{} x\bar x y\bar y + x\bar x \bar y y \\
                        + \bar xx \bar yy + \bar x xy\bar y \end{subarray}
   }
   &= - \frac{\alphas^2 \, P(x) \, P(\yfrak)}{4\pi^2(1-x)^{d-1} E^{2(d-2)}}
\nonumber\\ & \quad \times
   \left( \frac{d}{2} \right)^{\!2}
   \Beta(\tfrac12{+}\tfrac{d}{4},-\tfrac{d}{4})
     \Re\bigl( i \Omega_{E,x}^{d/2} + i \Omega_{(1-x)E,\yfrak}^{d/2} \bigr)
\nonumber\\ & \quad \times
   \left( \frac{M_{E,x}}{2\pi} \right)^{\frac{d}{2}-1}
   \left( \frac{M_{(1-x)E,\yfrak}}{2\pi} \right)^{\frac{d}{2}-1}
   \Re( i^{1-\frac{d}{2}} )
   \int_0^a \frac{d(\Delta t)}{(\Delta t)^{d/2}} .
\label {eq:xxyyDR}
\end {align}
(See notes in Appendix \ref{app:details}.)
Note for future reference that
\begin {equation}
   \Re( i^{1-\frac{d}{2}} )
   = 1 + O(\epsilon^2) .
\end {equation}


\subsection{\boldmath$xy\bar x\bar y$}

We now turn to the remaining diagram of fig.\ \ref{fig:seq2}.
For $d{=}2$, ref.\ \cite{seq} showed that one of the large-$\Nc$
color routings
$xy\bar x\bar y_2$
of $xy\bar x\bar y$
gives%
\footnote{
  ACI (2.36)
}
\begin {align}
  \left[\frac{d\Gamma}{dx\,dy}\right]_{xy\bar x\bar y_2} &=
   \frac{\CA^2 \alphas^2 M_\ix M_\fx^\seq}{32\pi^4 E^2} \, 
   ({-}\hat x_1 \hat x_2 \hat x_3 \hat x_4)
   \int_0^\infty d(\Delta t) \>
   \Omega_+\Omega_- \csc(\Omega_+\Delta t) \csc(\Omega_-\Delta t)
\nonumber\\ &\times
   \Bigl\{
     (\bar\beta Y_\yx^\seq Y_\xbx^\seq
        + \bar\alpha \Ybar_{\yx\xbx}^{\,\seq} Y_{\yx\xbx}^\seq) I_0^\seq
     + (\bar\alpha+\bar\beta+2\bar\gamma) Z_{\yx\xbx}^\seq I_1^\seq
\nonumber\\ &\quad
     + \bigl[
         (\bar\alpha+\bar\gamma) Y_\yx^\seq Y_\xbx^\seq
         + (\bar\beta+\bar\gamma) \Ybar_{\yx\xbx}^{\,\seq} Y_{\yx\xbx}^\seq
        \bigr] I_2^\seq
\nonumber\\ &\quad
     - (\bar\alpha+\bar\beta+\bar\gamma)
       (\Ybar_{\yx\xbx}^{\,\seq} Y_\xbx^\seq I_3^\seq + Y_\yx^\seq Y_{\yx\xbx}^\seq I_4^\seq)
   \Bigl\} ,
\label{eq:xyxyresult}
\end {align}
which is identical to the similar formula for $xy\bar y\bar x$
\cite{2brem},%
\footnote{
   AI (5.45)
}
\begin {align}
   \left[\frac{d\Gamma}{dx\,dy}\right]_{xy\bar y\bar x} = &
   \frac{\CA^2 \alphas^2 M_\ix M_\fx}{32\pi^4 E^2} \, 
   ({-}\hat x_1 \hat x_2 \hat x_3 \hat x_4)
   \int_0^{\infty} d(\Delta t) \>
   \Omega_+\Omega_- \csc(\Omega_+\Delta t) \csc(\Omega_-\Delta t)
\nonumber\\ &\times
   \Bigl\{
     (\beta Y_\bx Y_\Ax + \alpha \Ybar_{\bx\Ax} Y_{\bx\Ax}) I_0
     + (\alpha+\beta+2\gamma) Z_{\bx\Ax} I_1
\nonumber\\ &\quad
     + \bigl[
         (\alpha+\gamma) Y_\bx Y_\Ax
         + (\beta+\gamma) \Ybar_{\bx\Ax} Y_{\bx\Ax}
        \bigr] I_2
     - (\alpha+\beta+\gamma)
       (\Ybar_{\bx\Ax} Y_\Ax I_3 + Y_\bx Y_{\bx\Ax} I_4)
   \Bigl\} ,
\label {eq:IGamma}
\end {align}
except for the
addition of the superscript ``seq'' on some symbols, the bars on
$(\alpha,\beta,\gamma)$, and the purely notational change of relabeling
$\bar y$ subscripts as $\bar x$.  (See ref.\ \cite{seq} for details.)
From having seen earlier how the small-$\Delta t$ behavior of the
$xy\bar y\bar x$ diagram generalized from $d{=}2$
to (\ref{eq:IGammaDR}) for arbitrary $d$,
it is easy to see how (\ref{eq:xyxyresult})
similarly generalizes:
\begin {align}
   \left[\frac{d\Gamma}{dx\,dy}\right]_{xy\bar x\bar y_2} \simeq &
   \frac{\CA^2 \alphas^2 M_\ix M_\fx^\seq}{2^{d+3}d\pi^{2d}i^dE^d} \,
     \frac{\Gamma^2( \tfrac{d+2}{4} )}{\sin(\frac{\pi d}{4})} \,
     \bigl({-}\hat x_1 \hat x_2 \hat x_3 \hat x_4 |M_\ix| \Omega_\ix\bigr)^{d/2}
   \int \frac{d(\Delta t)}{(\Delta t)^d} \>
\nonumber\\ &\times
   \Bigl\{
     (\bar\beta Y_\yx^\seq Y_\xbx^\seq
        + \bar\alpha \Ybar_{\yx\xbx}^{\,\seq} Y_{\yx\xbx}^\seq) J_{\ix 0}^\seq
     + (\bar\alpha+\bar\beta+d\bar\gamma) Z_{\yx\xbx}^\seq J_{\ix 1}^\seq
\nonumber\\ &\quad
     + \bigl[
         (\bar\alpha+\bar\gamma) Y_\yx^\seq Y_\xbx^\seq
         + (\bar\beta+\bar\gamma) \Ybar_{\yx\xbx}^{\,\seq} Y_{\yx\xbx}^\seq
        \bigr] J_{\ix 2}^\seq
\nonumber\\ &\quad
     - (\bar\alpha+\bar\beta+\bar\gamma)
       (\Ybar_{\yx\xbx}^{\,\seq} Y_\xbx^\seq J_{\ix 3}^\seq
            + Y_\yx^\seq Y_{\yx\xbx}^\seq J_{\ix 4}^\seq)
   \Bigl\}
\nonumber\\ &
   + \{ \ix \leftrightarrow \fx^\seq \} ,
\label {eq:xyxyresultDR1}
\end {align}
where the $J^\seq$'s are defined the same as (\ref{eq:J}) but using
$(X^\seq,Y^\seq,Z^\seq)$ instead of $(X,Y,Z)$.
In the small $\Delta t$ limit,
ref.\ \cite{seq} found that%
\footnote{
  Eqs.\ (\ref{eq:XYZseqsim}) are the same as ACI (E15) except that,
  similar to footnote \ref{foot:calX} of this paper,
  our $\calX^\seq$ here does not contain the $|M_\ix|\Omega_\ix$ and
  $|M_\fx^\seq|\Omega_\fx^\seq$ terms that $X^\seq$ has in ACI.
}
\begin {subequations}
\label {eq:XYZseqsim}
\begin {align}
   \begin{pmatrix} \calX_\yx^\seq & Y_\yx^\seq \\ Y_\yx^\seq & Z_\yx^\seq \end{pmatrix}
   &=
     -\frac{i}{\Delta t}
       \begin{pmatrix} M_\ix & 0 \\ 0 & M_\fx^\seq \end{pmatrix}
     + O(\Delta t) ,
\\
   \begin{pmatrix} \calX_\xbx^\seq & Y_\xbx^\seq \\
                   Y_\xbx^\seq & Z_\xbx^\seq \end{pmatrix}
   &=
     -\frac{i}{\Delta t}
       \begin{pmatrix} M_\fx^\seq & 0 \\ 0 & M_\ix \end{pmatrix}
     + O(\Delta t) ,
\\
   \begin{pmatrix} X_{\yx\xbx}^\seq & Y_{\yx\xbx}^\seq \\
                   \Ybar_{\yx\xbx}^\seq & Z_{\yx\xbx}^\seq \end{pmatrix}
   &=
     -\frac{i}{\Delta t}
     \begin{pmatrix} 0 &  M_\ix \\
                     M_\fx^\seq &  0 \end{pmatrix}
     + O(\Delta t) .
\end {align}
\end {subequations}
Correspondingly,
(\ref{eq:xyxyresultDR1}) reduces to
\begin {align}
   \left[\frac{d\Gamma}{dx\,dy}\right]_{xy\bar x\bar y_2} \simeq &
   \frac{\CA^2 \alphas^2 M_\ix M_\fx^\seq}{2^{d+3}d\pi^{2d}i^dE^d} \,
     \frac{\Gamma^2( \tfrac{d+2}{4} )}{\sin(\frac{\pi d}{4})} \,
     \bigl({-}\hat x_1 \hat x_2 \hat x_3 \hat x_4 |M_\ix| \Omega_\ix\bigr)^{d/2}
   \int \frac{d(\Delta t)}{(\Delta t)^d} \>
\nonumber\\ &\times
     \Ybar_{\yx\xbx}^{\,\seq} Y_{\yx\xbx}^\seq
     (\bar\alpha J_{\ix 0}^\seq + \bar\beta J_{\ix 2}^\seq + \bar\gamma J_{\ix 2}^\seq)
\nonumber\\ &
   + \{ \ix \leftrightarrow \fx^\seq \} ,
\label {eq:xyxyresultDR2}
\end {align}
analogous to (\ref{eq:IGammaDR2}).
The analogs of the $J$ integrals (\ref{eq:J2}) are
\begin {subequations}
\label {eq:J2seq}
\begin {align}
   J_{\ix 0}^\seq &\simeq
   \frac{2^{\frac{d}{2}}\pi^d}{\Gamma(\frac{d}{2})}
   \left( \frac{\Delta t}{-i M_\ix} \right)^{d/2}
   \left( \frac{\Delta t}{-i} \right)^2
   \frac{2d}{M_\ix M_\fx^\seq}
   \,,
\displaybreak[0]\\
   J_{\ix 2}^\seq &\simeq
   \frac{2^{\frac{d}{2}}\pi^d}{\Gamma(\frac{d}{2})}
   \left( \frac{\Delta t}{-i M_\ix} \right)^{d/2}
   \left( \frac{\Delta t}{-i} \right)^2
   \frac{2}{M_\ix M_\fx^\seq}
   \,,
\end {align}
\end {subequations}
yielding%
\footnote{
  For $d{=}2$, this agrees with ACI (E17).
}
\begin {align}
   \left[\frac{d\Gamma}{dx\,dy}\right]_{xy\bar x\bar y_2} &\simeq
   \frac{\CA^2 \alphas^2 M_\ix M_\fx^\seq}{2^{\frac{d}{2}+2}d\pi^d E^d} \,
     \frac{\Gamma^2( \tfrac{d+2}{4} )}
          {\Gamma(\frac{d}{2}) \sin(\frac{\pi d}{4})} \,
     \bigl(i\hat x_1 \hat x_2 \hat x_3 \hat x_4\Omega_\ix \sgn M_\ix\bigr)^{d/2}
     (d\bar\alpha + \bar\beta + \bar\gamma)
   \int \frac{d(\Delta t)}{(\Delta t)^{d/2}} \>
\nonumber\\ &\quad
   + \{ \ix \leftrightarrow \fx^\seq \} ,
\label {eq:xyxyresultDR3}
\end {align}
which is the analog of (\ref{eq:poleDR0}).

The only effect of different color routings on $xy\bar x\bar y$ is
on how color is distributed during the 4-particle evolution in the
diagram, which in turn affects how the 4-particle evolution
interacts with the medium.
Because the pole terms we are interested in here
only arise from situations where $\Delta t$ is
small enough that the 4-particle propagation is effectively vacuum
propagation, the color routing has no effect.  For this
reason, adding in the other color routing $x y\bar x\bar y_1$
described in ref.\ \cite{seq} simply multiplies
(\ref{eq:xyxyresultDR3}) by two.%
\footnote{
  Alternatively, ref.\ \cite{seq} explains that the other color
  routing differs only by a permutation of the 4-particle
  $x_i$ to
  $(x_1,x_2,x_3,x_4) = (-1,1{-}x{-}y,y,x)
   = (\hat x_1,\hat x_3,\hat x_2,\hat x_4)$.
  One may check explicitly that this permutation leaves
  (\ref{eq:xyxyresultDR2}) invariant above.
}
So, summing both routings, and adding the conjugate diagram,
\begin {align}
   2 \Re \left[\frac{d\Gamma}{dx\,dy}\right]_{xy\bar x\bar y} &\simeq
   \frac{\CA^2 \alphas^2 M_\ix M_\fx^\seq}{2^{\frac{d}{2}}d\pi^d E^d} \,
     \frac{\Gamma^2( \tfrac{d+2}{4} )}
          {\Gamma(\frac{d}{2}) \sin(\frac{\pi d}{4})}
     \Re\Bigl[
        (i\hat x_1 \hat x_2 \hat x_3 \hat x_4)^{d/2}
        \bigl[
           (\Omega_\ix)^{d/2} + (\Omega_\fx^\seq)^{d/2}
        \bigr]
     \Bigr]
\nonumber\\ &\qquad \times
   (d\bar\alpha + \bar\beta + \bar\gamma)
   \int \frac{d(\Delta t)}{(\Delta t)^{d/2}} ,
\label {eq:xyxyresultDR4}
\end {align}
where we have now used the fact that $\sgn M_\ix$ and $\sgn M_\fx^\seq$ are
both +1.


\subsection{Combining sequential diagrams}

The sequential diagram results
(\ref{eq:xxyyDR}) and (\ref{eq:xyxyresultDR4}) are individually
UV divergent, and it is only when we combine them that the
divergences will cancel.  But we need to be careful because
one formula is expressed directly in terms of DGLAP splitting
functions $P(x)$ and $P(\yfrak)$, whereas the other is expressed
instead in terms of the splitting-function
combinations $(\bar\alpha,\bar\beta,\bar\gamma)$
defined in ref.\ \cite{seq}.  For $d{=}2$, the two are related
by%
\footnote{
  ACI (E5)
}
\begin {align}
   \bar\alpha + \tfrac12 \bar\beta + \tfrac12 \bar\gamma
   &=
   \frac{P(x)}{\CA x^2(1{-}x)^2} \,\,
   \frac{P\bigl(\frac{y}{1{-}x}\bigr)}
        {\CA (1{-}x) y^2 (1{-}x{-}y)^2}
   \qquad
   (d=2) .
\label {eq:abcPP2}
\end {align}
However, because the diagrams are individually UV divergent, we need
the $O(\epsilon)$ corrections to this relation.  To do that correctly
is finicky: it requires diving into the details of how the splitting
function factors are defined and normalized for $d$ dimensions.  We
find that the generalization of (\ref{eq:abcPP2}) is
\begin {equation}
   \bar\alpha + \tfrac1{d} \bar\beta + \tfrac1{d} \bar\gamma
   =
   \frac{P(x)}{\CA x^2(1{-}x)^2} \,\,
   \frac{P\bigl(\frac{y}{1{-}x}\bigr)}
        {\CA (1{-}x)^{d-1} y^2 (1{-}x{-}y)^2}
   \, .
\label {eq:abcPP2DR}
\end {equation}
However, rather than justifying this directly, we find it
easier (and less prone to error)
to simply repeat the derivation of the $x\bar x y\bar y$
diagram from the beginning directly in the language of
$d$-dimensional $(\bar\alpha,\bar\beta,\bar\gamma)$.  We carry this out in
appendix \ref{app:xxyy}.  One may then (i) identify the conversion
(\ref{eq:abcPP2DR}) by comparison with our earlier result
(\ref{eq:abcPP2}).  Alternatively, one can avoid the need for
(\ref{eq:abcPP2DR}) altogether by (ii) keeping all sequential
diagrams in terms of $(\bar\alpha,\bar\beta,\bar\gamma)$,
combining the diagrams to get a finite total, and only then taking $d\to 2$.
In the final, finite result, one only
needs known $d{=}2$ formulas for $(\bar\alpha,\bar\beta,\bar\gamma)$,
which at that point may be related to the $d{=}2$ splitting
functions using (\ref{eq:abcPP2}).

The final result is the same either way, but the intermediate
formulas are a little simpler to present by using the conversion
(\ref{eq:abcPP2DR}) to rewrite (\ref{eq:xyxyresultDR4}) as
\begin {align}
   2 \Re \left[\frac{d\Gamma}{dx\,dy}\right]^{(\Delta t<a)}_{xy\bar x\bar y}
   &\simeq \frac{\alphas^2 \, P(x) \, P(\yfrak)}{4\pi^2(1-x)^{d-1} E^{2(d-2)}}
\nonumber\\ & \quad \times
   \frac{d}{2} \,
   \Beta(\tfrac12{+}\tfrac{d}{4},-\tfrac{d}{4})
   \Re\left(
      \frac{i \Omega_{E,x}^{d/2} + i \Omega_{(1-x)E,\yfrak}^{d/2}}
           {i^{\frac{d}{2}-1}}
    \right) 
\nonumber\\ & \quad \times
   \left( \frac{M_{E,x}}{2\pi} \right)^{\frac{d}{2}-1}
   \left( \frac{M_{(1-x)E,\yfrak}}{2\pi} \right)^{\frac{d}{2}-1}
   \int_0^a \frac{d(\Delta t)}{(\Delta t)^{d/2}} \,,
\label {eq:xyxyDR}
\end {align}
where we have also used 
(\ref{eq:xhat}), (\ref{eq:GammaRewrite}),
$M_\ix = x(1{-}x)E$ and $M_\fx^\seq = y(1{-}x)(1{-}x{-}y)E$.
Combining  (\ref{eq:xxyyDR}) with (\ref{eq:xyxyDR}),
\begin {align}
   & 2 \Re \left[\Delta \frac{d\Gamma}{dx\,dy}\right]^{(\Delta t<a)}_{
         x\bar xy\bar y + x\bar x\bar y y + x y \bar x\bar y
     }
\nonumber\\
   & \qquad
   \simeq \frac{\alphas^2 \, P(x) \, P(\yfrak)}{4\pi^2(1-x)^{d-1} E^{2(d-2)}}
   \,
   \left( \frac{d}{2} \right)^{\!2}
   \Beta(\tfrac12{+}\tfrac{d}{4},-\tfrac{d}{4})
   \left( \frac{M_{E,x}}{2\pi} \right)^{\frac{d}{2}-1}
   \left( \frac{M_{(1-x)E,\yfrak}}{2\pi} \right)^{\frac{d}{2}-1}
   \int_0^a \frac{d(\Delta t)}{(\Delta t)^{d/2}}
\nonumber\\ & \qquad\qquad \times
   \left[
      \frac{2}{d}
      \Re\left(
        \frac{i \Omega_{E,x}^{d/2} + i \Omega_{(1-x)E,\yfrak}^{d/2}}
             {i^{\frac{d}{2}-1}}
      \right) 
      - \Re(i \Omega_{E,x}^{d/2} + i \Omega_{(1-x)E,\yfrak}^{d/2})
        \Re\left( \frac{1}{i^{\frac{d}{2}-1}} \right)
   \right]
   .
\end {align}
Using (\ref{eq:dtintDR}) and taking $\epsilon \to 0$,
\begin {align}
   & 2 \Re \left[\Delta \frac{d\Gamma}{dx\,dy}\right]^{\rm pole}_{
         x\bar xy\bar y + x\bar x\bar y y + x y \bar x\bar y
     }
\nonumber\\ & \quad
   = \frac{\alphas^2 \, P(x) \, P(\yfrak)}{\pi^2(1-x)}
   \biggl(
      - \tfrac12
           \Re(i \Omega_{E,x} + i \Omega_{(1-x)E,\yfrak})
      + \tfrac{\pi}{4}
           \Re(\Omega_{E,x} + \Omega_{(1-x)E,\yfrak})
   \biggr)
   .
\label {eq:seqDR}
\end {align}
As promised, the final answer is finite, and so one may now use ordinary
$d{=}2$ results for the splitting functions $P$.


\section{Dimensional regularization passes diagnostic}
\label {sec:DRtest}

We should check our methods by checking that the diagnostic test
(\ref{eq:test}) works when we apply dimensional regularization to
computing individual time-ordered diagrams in the QED independent
emission model.  That calculation is carried out in appendix
\ref{app:DRtest}.  Taking the $y \ll x$ limit for simplicity,
the results for the total contribution of crossed vs.\ sequential diagrams
to the diagnostic (\ref{eq:test}) are
\begin {subequations}
\label {eq:DRtest}
\begin {align}
  2\Re \left[\frac{d\Gamma}{dx\,dy}\right]_{\rm crossed}^{\rm non-pole}
  &=
   \left[ 0.6173 - \frac{\ln(x/y)}{\pi^2} \right]
   \frac{\alphaEM^2}{x^{1/2} y} \sqrt{\frac{\hat q}{E}} \,,
\\
  2\Re \left[\frac{d\Gamma}{dx\,dy}\right]_{\rm crossed}^{\rm pole} ~~
  &=
   \frac{4\alphaEM^2}{\pi^2 x y}
   \left[ - \Re(i\Omega_x) - \frac{\pi}{2} \Re(\Omega_x) \right]
  =
  -0.5210 \, \frac{\alphaEM^2}{x^{1/2} y} \sqrt{\frac{\hat q}{E}} \,,
\\
  2\Re \left[\Delta\frac{d\Gamma}{dx\,dy}\right]_{\rm seq}^{\rm non-pole}
  &=
   \left[ -0.2120 + \frac{\ln(x/y)}{\pi^2} \right]
   \frac{\alphaEM^2}{x^{1/2} y} \sqrt{\frac{\hat q}{E}} \,,
\\
  2\Re \left[\Delta\frac{d\Gamma}{dx\,dy}\right]_{\rm seq}^{\rm pole} ~~~~~
  &=
   \frac{4\alphaEM^2}{\pi^2 x y}
   \left[ - \Re(i\Omega_x) + \frac{\pi}{2} \Re(\Omega_x) \right]
  =
  0.1157 \, \frac{\alphaEM^2}{x^{1/2} y} \sqrt{\frac{\hat q}{E}} \,,
\end {align}
\end {subequations}
where we have split each case into the pole contribution (which requires
UV regularization) and the non-pole contribution.
The contributions of (\ref{eq:DRtest}) indeed sum to zero, and so
dimensional regularization passes this test.
It had to because dimensional regularization is a fully consistent
regularization scheme and the test (\ref{eq:test}) was ultimately
a tautology, but it is reassuring to verify that we can correctly
carry out the technical procedures of calculating the pole terms.

There is also a reassurance test, of sorts, for QCD results.
Ref.\ \cite{seq}, which makes use of the QCD pole terms derived
in this paper, shows that%
\footnote{
  See ACI appendix B and ACI footnote 20.
}
(i) the pole terms computed with
dimensional regularization have the right behavior to effect a
Gunion-Bertsch-like cancellation of logarithmic enhancements to
$\Delta \, d\Gamma/dx\,dy$, which (ii) would not occur were one
to use the naive $i\epsilon$ prescription to determine those poles
(i.e.\ neglect the $1/\pi^2$ pole terms).


\section{Summary of QCD results}
\label {sec:summary}

Our final results for QCD are that the crossed diagrams, accounting for
all permutations, have
pole contribution
\begin {equation}
   \left[\frac{d\Gamma}{dx\,dy}\right]^{\rm pole}_{\rm crossed}
   =
   A^{\rm pole}(x,y)
   + A^{\rm pole}(z,y)
   + A^{\rm pole}(x,z) ,
\end {equation}
where $z \equiv 1{-}x{-}y$ and
where $A_{\rm pole}(x,y)$ is twice the real part of (\ref{eq:crossedpole}):
\begin {align}
   A_{\rm pole}(x,y)
   &=
    \frac{\CA^2 \alphas^2}{16\pi^2} \,
    x y (1{-}x)^2 (1{-}y)^2(1{-}x{-}y)^2
\nonumber\\ & \qquad \times
    2\Re\biggl\{
       -i
       [\Omega_{-1,1-x,x} + \Omega_{-(1-y),1-x-y,x}
        - \Omega_{-1,1-y,y}^* - \Omega_{-(1-x),1-x-y,y}^*]
\nonumber\\ & \qquad \qquad \times
       \biggl[
       \left(
         (\alpha + \beta)
         + \frac{(\alpha + \gamma) xy}{(1{-}x)(1{-}y)}
       \right) \ln \left( \frac{1{-}x{-}y}{(1{-}x)(1{-}y)} \right)
       + \frac{2(\alpha+\beta+\gamma) x y}{(1{-}x)(1{-}y)}
       \biggr]
\nonumber\\ &\qquad \quad
     - \pi
       [\Omega_{-1,1-x,x} + \Omega_{-(1-y),1-x-y,x}
        + \Omega_{-1,1-y,y}^* + \Omega_{-(1-x),1-x-y,y}^*]
\nonumber\\ & \qquad \qquad \times
       \left(
         (\alpha + \beta)
         + \frac{(\alpha + \gamma) xy}{(1{-}x)(1{-}y)}
       \right)
     \biggr\} .
\end {align}
Using the same notation as ref.\ \cite{seq}, the sequential diagrams
give
\begin {align}
   \left[ \Delta \frac{d\Gamma}{dx\>dy} \right]_{\rm sequential}^{\rm pole}
   = \quad
   & {\cal A}^{\rm pole}_\seq(x,y)
     + {\cal A}^{\rm pole}_\seq(z,y)
     + {\cal A}^{\rm pole}_\seq(x,z)
\nonumber\\
   + ~ &
   {\cal A}^{\rm pole}_\seq(y,x)
     + {\cal A}^{\rm pole}_\seq(y,z)
     + {\cal A}^{\rm pole}_\seq(z,x)
   ,
\end {align}
where ${\cal A}^{\rm pole}_\seq(x,y)$ is half of the
$y\leftrightarrow z$ symmetric result (\ref{eq:seqDR}):
\begin {equation}
   {\cal A}_\seq^{\rm pole}(x,y)
   = \frac{\alphas^2 \, P(x) \, P(\yfrak)}{2\pi^2(1-x)}
   \biggl(
      - \tfrac12
           \Re(i \Omega_{E,x} + i \Omega_{(1-x)E,\yfrak})
      + \tfrac{\pi}{4}
           \Re(\Omega_{E,x} + \Omega_{(1-x)E,\yfrak})
   \biggr) .
\end {equation}
 

\begin{acknowledgments}

The work of PA and HC was supported, in part, by the U.S. Department
of Energy under Grant No.~DE-SC0007984.

\end{acknowledgments}

\appendix

\section {More details on some formulas}
\label {app:details}

\paragraph*{Eq.\ (\ref{eq:integral1}):}
The change of integration variable $\tau \equiv i\Omega\,\Delta t$
in (\ref{eq:integral0}) actually gives
\begin {equation}
   =
   - i d M \left( \frac{M\Omega}{2\pi} \right)^{d/2}
   \int_0^{\infty\,e^{i\pi/4}} \frac{d\tau}{\sh^{1+\frac{d}{2}}\tau}
\label {eq:integral1a}
\end {equation}
since $\Omega \propto \sqrt{-i} = e^{-i\pi/4}$
as in (\ref{eq:OmegaSingle}).%
\footnote{
  Why not the other square root, $-e^{-i\pi/4}$?  This would give
  $\Omega$ a positive imaginary part instead of a negative one,
  which would lead to exponential growth in time of the propagator
  instead of exponential decay.  Exponential decay is the
  physically relevant case: decoherence due to random collisions
  with the medium causes interference contributions to decay as
  $\Delta t \to \infty$.
}
The behavior of the integrand at infinity is such that
we can extend the contour to additionally
circle from $\infty e^{i\pi/4}$ to $+\infty$ at infinity without
changing the integral.  There are no singularities of the integrand
that obstruct then deforming the entire contour to rewrite
(\ref{eq:integral1a}) as an integral (\ref{eq:integral1})
along the positive real axis.

\paragraph*{Eq.\ (\ref{eq:integral2}):}
One way
to do the integral in (\ref{eq:integral1}) by hand is to change
integration variable to $u = \cth\tau$ to get
$\int_1^\infty (u^2-1)^{-\frac12+\frac{d}{4}} \,du$
and then proceed from there.  There are a number
of equivalent ways to write the result, which can be manipulated into
each other using $\Gamma$ function identities.

\paragraph*{Eq.\ (\ref{eq:dimint}):}
This integral can also be obtained%
\footnote{
  This is an example of an integral that 
  Mathematica version 10.4.1 \cite{mathematica}
  cannot manage directly without first changing variables.
}
by first changing to integration variable $u = \cth\tau$.
Note that convergence of the integral requires $\Re z > 0$.
This is satisfied in the application
to single splitting in section \ref{sec:single} since there $M > 0$
and $\Omega \propto e^{-i\pi/4}$, so that
$\Re z = \Re(\tfrac12 M \Omega B^2) > 0$.

\paragraph*{Eq.\ (\ref{eq:IIxyyxDR}):} With regard to the factors of
$E$, AI (4.29--30) wrote
$\langle \p_j,\p_k | \delta H | \p_i \rangle
   = {g \bcalT_{i \to jk} \cdot \P_{jk}}$
with
$\bcalT_{i \to jk} =
  T^{\rm color}_{i \to jk} \bcalP_{i \to jk} / 2 E^{3/2}$
and $\bcalP_{i\to jk}$ dimensionless (which is what made the
later definitions of $\alpha$, $\beta$, and $\gamma$ dimensionless).
The fact that the power $E^{-3/2}$ of $E$ in the last formula
is necessitated by
dimensional analysis was explained in notes on AI (4.29--31) in
AI appendix A.  That same argument, applied to $d$ dimensions, gives
$\langle \p_j,\p_k | \delta H | \p_i \rangle \propto
  g T^{\rm color}_{i \to jk} \bcalP_{i \to jk}\cdot\P_{jk} / 2 E^{(d+1)/2}$.
The four factors of $E^{-(d+1)/3}$ then combine with a factor of
$E^2$ in AI (4.10), whose origin is not dimension dependent
[see the notes on AI (4.16) in AI appendix A], to give the
$E^{-2d}$ of (\ref{eq:IIxyyxDR}) in this paper.  Alternatively, one
could just determine the power by dimensionally analyzing
(\ref{eq:IIxyyxDR}) or later formulas.

With regard to the factors of $|x_1 + x_4|^{-1}$ and $|x_3 + x_4|^{-1}$,
these came from the vertex formula AI (B32), which originated from
the normalizations of states in AI Appendix B and AI (4.22--25).
Repeating the derivations of AI Appendix B in $d$ dimensions, one
finds that AI (B11) changes to
\begin {equation}
   \langle \{\B'_{ij}\} | \{\B_{ij}\} \rangle =
   |x_2|^d
   \prod_{i=3}^N
   \delta^{(d)}(\b_{i1}-\b'_{i1})
   =
   |x_2|^d
   \prod_{i=3}^N
   |x_i+x_1|^{-d} \delta^{(d)}(\B_{i1}-\B'_{i1}) ,
\label {eq:BsBs}
\end {equation}
with consequence that AI (4.22) is replaced by
\begin {subequations}
\begin {align}
  \bigl\langle\{\C_{ij}\}\big|\{\C_{ij}'\}\bigr\rangle
  &= |\hat x_3+\hat x_4|^{-d} \, \delta^{(d)}(\C_{34}{-}\C'_{34}) \,
                     \delta^{(d)}(\C_{12}{-}\C'_{12})
  \qquad (N{=}4),
\label {eq:Csnorm}
\\
  \bigl\langle\{\B_{ij}\}\big|\{\B_{ij}'\}\bigr\rangle
  &= \delta^{(d)}(\B_{12}{-}\B'_{12})
  \qquad (N{=}3),
\label {eq:Bsnorm}
\\
  \langle|\rangle &= |x_1|^d
  \qquad (N{=}2) .
\label {eq:N2norm}
\end {align}
\end {subequations}
This means we need to replace AI (4.23) by
\begin {equation}
  |\C_{34},\C_{12}\rangle
  \equiv |\hat x_3+\hat x_4|^{d/2} \, \bigl| \{\C_{ij}\} \bigr\rangle
\label{eq:C3412def}
\end {equation}
in order to get the desired normalization of AI (4.24), and make
a similar replacement for AI (B31).  This change propagates to replacing
the vertex formula AI (4.15) by
\begin {align}
   \langle\C_{41},\C_{23}|\delta H|\B\rangle
   &=
     \langle\C_{23}|\delta H|\B\rangle \,
     |\hat x_4+\hat x_1|^{-d/2} \, \delta^{(d)}(\C_{41}-\B)
\nonumber\\
   &=
     -i g \bcalT_{i \to 23} \cdot \grad \delta^{(d)}(\C_{23}) \,
     |\hat x_4+\hat x_1|^{-d/2} \, \delta^{(d)}(\C_{41}-\B) ,
\label {eq:dH43}
\end {align}
which is the source of the factor $|\hat x_1 + \hat x_4|^{-d/2}$ and
the corresponding permutation $|\hat x_3 + \hat x_4|^{-d/2}$ in
our (\ref{eq:IIxyyxDR}).

\paragraph*{Eq.\ (\ref{eq:t1inti}):}
If $M < 0$ and $\Omega \propto \sqrt{+i}$, then $\Re(M\Omega) < 0$,
which means the integral in (\ref{eq:tauint}) does not converge
at large $\tau$.  To get a convergent integral,
we should change variables from $\Delta t$ in
(\ref{eq:tauint0}) to $\tau \equiv -i\Omega\Delta t$ instead of
$\tau \equiv i\Omega\Delta t$.  This yields
\begin {equation}
   - i M\B \left( - \frac{M\Omega}{2\pi} \right)^{d/2}
   \int_0^{\infty\,e^{-i\pi/4}} \frac{d\tau}{\sh^{1+\frac{d}{2}}\tau} \,
   e^{+\frac12 M\Omega B^2\cth\tau}
\end {equation}
instead of (\ref{eq:tauint}).  Similar to this appendix's
notes for (\ref{eq:integral1}) above, we may deform the upper limit
of the $\tau$ contour from $\infty\, e^{-i\pi/4}$ to $+\infty$.
Recognizing that $M = -|M|$ in the $M<0$ case discussed here, we
see that both this case and the $M>0$ case of (\ref{eq:tauint})
can be written in the convergent form
\begin {equation}
   - i M\B \left( \frac{|M|\Omega}{2\pi} \right)^{d/2}
   \int_0^{\infty} \frac{d\tau}{\sh^{1+\frac{d}{2}}\tau} \,
   e^{-\frac12 |M|\Omega B^2\cth\tau} ,
\end {equation}
with result (\ref{eq:t1inti}).
This result agrees with AI (5.9a) in the special case $d=2$.

\paragraph*{Eq.\ (\ref{eq:CpropSmall}):}
The variables $(\C_{34},\C_{12})$ and $(\C_{41},\C_{23})$ are related
by AI (5.31),
\begin {equation}
   \begin{pmatrix} \C'_{41} \\ \C'_{23} \end{pmatrix}
   =
   \frac{1}{(x_1+x_4)}
   \begin{pmatrix}
       -x_3 & -x_2 \\
        \phantom{-}x_4 &  \phantom{-}x_1
   \end {pmatrix}
   \begin{pmatrix} \C'_{34} \\ \C'_{12} \end{pmatrix} ,
\label{eq:Cchange}
\end {equation}
which follows from the definition (\ref{eq:CijDef})
and the fact that
momentum conservation implies $x_1+x_2+x_3+x_4=0$ (in our sign
conventions).  In addition to making this change for
$\C'$ on the right-hand side of (\ref{eq:Cprop00}), it is necessary
to account for the change in normalization of the ket
$|\C'_{34},\C'_{12}\rangle$ vs.\ $|\C'_{41},\C'_{23}\rangle$.
The former is normalized so that
$\langle\C_{34},\C_{12}|\C'_{34},\C'_{12}\rangle
  = \delta^{(d)}(\C_{34}{-}\C'_{34}) \, \delta^{(d)}(\C_{12}{-}\C'_{12})$
and the latter so that
$\langle\C_{41},\C_{23}|\C'_{41},\C'_{23}\rangle
  = \delta^{(d)}(\C_{41}{-}\C'_{41}) \, \delta^{(d)}(\C_{23}{-}\C'_{23})$.
The relationship between these $\delta$ functions is determined
by the Jacobean of the transformation (\ref{eq:Cchange}).
Specifically, the normalization change accounts for an overall factor
of
$|(x_1 x_3 - x_2 x_4)/(x_1+x_4)^2|^{-d/2}
   = |(x_3 + x_4)/(x_1+x_4)|^{-d/2}
   = (\det{\mathfrak M}/\det{\mathfrak M}')^{-d/4}$
in going from (\ref{eq:Cprop00}) to (\ref{eq:CpropSmall}).

\paragraph*{Eq.\ (\ref{eq:xxyyDR}):}
For $d{=}2$, this agrees with ACI (2.28).  Note that in dimensional
regularization (unlike for $i\epsilon$ prescriptions), the
$\Delta t$ integral in (\ref{eq:seqint2}) is real valued.
When using (\ref{eq:seqint1}) and (\ref{eq:seqint2}) in
(\ref{eq:DxxyyetcDR}), remember that $E$ is replaced
by $(1{-}x)E$ in the expression
$[d\Gamma/d\yfrak\,d(\Delta t_y)]_{(1-x)E}$.

\paragraph*{Eq.\ (\ref{eq:DRint1}):}
The last integral in (\ref{eq:DRint1a}) is the same one as in
(\ref{eq:integral1}).  One way to do the first integral is to
make the same change of variables described in the notes for
(\ref{eq:integral2}) above to get
\begin {align}
  \int_1^\infty du \> (u^2-1)^{-\tfrac12+\tfrac{d}{4}} \cth^{-1}u
  &=
  \frac12
  \int_1^\infty du \> (u^2-1)^{-\tfrac12+\tfrac{d}{4}}
  \ln\Bigl( \frac{u+1}{u-1} \Bigr)
\nonumber\\
  &=
  \frac12
  \left( \frac{\partial}{\partial a} - \frac{\partial}{\partial b} \right)
  \int_1^\infty du\> (u+1)^a(u-1)^b
  \Bigr|_{a=b=-\tfrac12+\tfrac{d}{4}}
\nonumber\\
  &=
  \left( \frac{\partial}{\partial a} - \frac{\partial}{\partial b} \right)
  2^{a+b} \Beta(-a{-}b{-}1,b{+}1)
  \Bigr|_{a=b=-\tfrac12+\tfrac{d}{4}}
\end {align}
and then simplify from there using
$\psi(1{-}z)-\psi(z) = \pi\cot(\pi z)$, where
$\psi(z) \equiv \Gamma'(z)/\Gamma(z)$ is the digamma
function.


\section {Quick review of naive \boldmath$i\epsilon$ prescription}
\label {app:epsnaive}

\begin {figure}[t]
\begin {center}
  \includegraphics[scale=0.5]{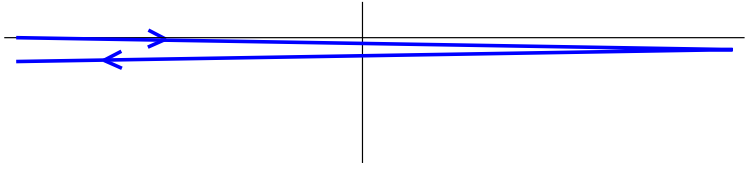}
  \caption{
     \label{fig:SK}
     A Schwinger-Keldysh contour shown in the complex time plane.
     The contour runs from $t=-\infty$ to $t=+\infty$ for times in the
     amplitude and then turns around and runs back again from
     $t=+\infty$ to $t=-\infty$ for the times in the conjugate amplitude.
     The imaginary axis has been vastly exaggerated to show an
     {\it infinitesimal}\/ downward slope along the contour above.
  }
\end {center}
\end {figure}

The underlying $i\epsilon$ prescriptions used in
ref.\ \cite{2brem} can be summarized by the Schwinger-Keldysh
contour shown in fig.\ \ref{fig:SK}.  The infinitesimal downward
slope from left to right along the top half of the contour
implements the usual time-ordering prescription associated with
propagators in amplitudes.  The infinitesimal downward slope
in the opposite direction along the bottom half of the contour
implements the complex conjugate of that:
anti-time-ordering in conjugate amplitudes.
Specifically, the $i\epsilon$ prescriptions
are
\begin {center}
\begin {tabular}{ll}
  $
     \frac{1}{t_2-t_1} ~\rightarrow~
     \frac{1}{t_2-t_1 - i\epsilon\sgn(t_2{-}t_1)} \,,
   $ & \qquad
   $t_1$ and $t_2$ both in amplitude; \\[5pt]
  $
     \frac{1}{t_2-t_1} ~\rightarrow~
     \frac{1}{t_2-t_1 + i\epsilon\sgn(t_2{-}t_1)} \,,
   $ & \qquad
   $t_1$ and $t_2$ both in conjugate amplitude; \\[5pt]
  $
     \frac{1}{t_2-t_1} ~\rightarrow~
     \frac{1}{t_2-t_1 - i\epsilon} \,,
   $ & \qquad
   $t_1$ in amplitude and $t_2$ in conjugate amplitude; \\[5pt]
  $
     \frac{1}{t_2-t_1} ~\rightarrow~
     \frac{1}{t_2-t_1 + i\epsilon} \,,
   $ & \qquad
   $t_1$ in conjugate amplitude and $t_2$ in amplitude.
\end {tabular}
\end{center}
See AI section VII.A.2 and AI appendix D.3 \cite{2brem}.
The $i\epsilon$ prescription above corresponds
to time ordering in the first case, anti-time ordering in the second,
and Wightman correlators (no time ordering) in the remaining cases.

However, there is a difficulty.  Double bremsstrahlung interference
diagrams such as fig.\ \ref{fig:pole} involve four different times,
and, in the method of ref.\ \cite{2brem}, the first and last of those
times are already integrated over.  The time $\Delta t$ left after
these integrations represents the separation between the middle
two times in the diagrams, but by then the effects of the
$i\epsilon$'s that should have been associated with the first and
last times were lost.  AI appendix D.3 \cite{2brem} sorted out this
issue in detail in order to figure out what the net $\pm i\epsilon$
should be in denominators.  If one naively applies the results
of that appendix to $1/\Delta t$ factors, one obtains the results
listed in the last column of table \ref{tab:epstest} of this paper.
As explained in section \ref{sec:epsbad} of this paper,
that argument was flawed because there can also be importantly different
$i\epsilon$ factors in numerators of $\Delta t/(\Delta t)^2$.
Additionally,
the situation is confusingly complicated by the fact that
both amplitude and conjugate-amplitude particles are interacting
with the medium between high-energy splitting times, linking
(after medium-averaging) the evolution of one to the evolution of
the other.  As described in appendix \ref{app:epsgeneral} of this
paper, it is unclear
how to fully incorporate the $i\epsilon$ prescription into this
evolution in the case of double bremsstrahlung.

\section {Test of \boldmath$i\epsilon$ prescription for
          independent emission model}
\label {app:epstest}

In this appendix, we will take a look at how the QED independent
emission test of section \ref{sec:diagnostic} works out when we
regulate each single splitting amplitude with an $i\epsilon$
prescription as in (\ref{eq:dGdxeps}), which we will write here
as
\begin {equation}
   \frac{d\Gamma}{dx}
   =
   2\Re\left[\frac{d\Gamma}{dx}\right]_{x\bar x}
   =
   \Re \int_0^\infty d(\Delta t_x) \>
   \frac{d\Gamma}{dx\,d(\Delta t_x)}
\end {equation}
with
\begin {equation}
   \frac{d\Gamma}{dx\,d(\Delta t_x)} \equiv
   - \frac{\alphaEM P(x)}{\pi} \,
   \Omega_x^2 \csc^2\bigl(\Omega_x(\Delta t_x-i\epsilon)\bigr) .
\label {eq:dGdxdteps}
\end {equation}
Remembering that the QED independent emission approximation
is only relevant for small $x$ and $y$, we will use the
corresponding limit of the splitting function and just write
\begin {equation}
   \frac{d\Gamma}{dx\,d(\Delta t_x)}
   =
   - \frac{2\alphaEM}{\pi x} \,
   \Omega_x^2 \csc^2\bigl(\Omega_x(\Delta t_x-i\epsilon)\bigr) .
\label {eq:dGdx9}
\end {equation}
In this appendix, all calculations will be in the context of the
independent emission model, and so we will just use equal signs
(rather than $\simeq$) in
our formulas with that understanding.  For QED (and not QCD!),
the relevant complex frequency $\Omega$ in the
small $x$ limit is
\begin {equation}
   \Omega_x \equiv \sqrt{-\frac{i x \hat q}{2E}}
   \qquad
   (\mbox{QED}).
\label {eq:OmxQED}
\end {equation}

In general, the double emission
rate is given in the independent emission model by
\begin {equation}
  \frac{d\Gamma}{dx\,dy} =
  \int dt_\xx \, dt_\yx \, dt_\ybx \>
  \Re \left[ \frac{d\Gamma}{dx\,d(\Delta t_x)} \right]
  \Re \left[ \frac{d\Gamma}{dy\,d(\Delta t_y)} \right] ,
\end {equation}
where $\Delta t_x = |t_\xbx - t_\xx|$, $\Delta t_y \equiv |t_\ybx-t_\yx|$,
and the time integral is (using overall time translation invariance) over any
three of the four times $t_\xx$, $t_\xbx$, $t_\yx$, $t_\ybx$.
To isolate different time-ordered contributions such as
$xy\bar y\bar x$, $x\bar y y\bar x$, etc.\ in the diagnostic
(\ref{eq:test}), we just need to
impose the appropriate restrictions on the time integrals and
take the appropriate pieces of the real parts.


\subsection{\boldmath$xy\bar y\bar x + x\bar y y\bar x$}

As an example, we begin by computing the $xy\bar y\bar x$ contribution
to the diagnostic (\ref{eq:test}).  This corresponds to
\begin {equation}
  \left[\frac{d\Gamma}{dx\,dy}\right]_{xy\bar y\bar x}
  =
  \int_{t_\xx < t_\yx < t_\ybx < t_\xbx} dt_\xx \, dt_\yx \, dt_\ybx \>
  \frac12 \, \frac{d\Gamma}{dx\,d(\Delta t_x)} \,
  \frac12 \, \frac{d\Gamma}{dy\,d(\Delta t_y)}
  \,.
\end {equation}
The time ordering requires $\Delta t_y < \Delta t_x$.
Using the fact that the integrand depends only on $\Delta t_x$ and
$\Delta t_y$, the time integral corresponding to where the interval
$(t_\yx,t_\ybx)$ lies within the interval $(t_\xx,t_\xbx)$ is
trivial and gives a factor of $\Delta t_x-\Delta t_y$, so that
\begin {equation}
  \left[\frac{d\Gamma}{dx\,dy}\right]_{xy\bar y\bar x}
  =
   \int_0^\infty d(\Delta t_y) \int_{\Delta t_y}^\infty d(\Delta t_x) \>
   (\Delta t_x-\Delta t_y) \,
   \frac12 \, \frac{d\Gamma}{dx\,d(\Delta t_x)} \,
   \frac12 \, \frac{d\Gamma}{dy\,d(\Delta t_y)}
  \,.
\end {equation}
The corresponding equation for
$x y \bar y \bar x + x \bar y y \bar x
 + \bar x  y \bar y x + \bar x \bar y y x$ is
\begin {equation}
  2\Re \left[\frac{d\Gamma}{dx\,dy}\right]_{xy\bar y\bar x + x\bar y y\bar x}
  =
   \int_0^\infty d(\Delta t_y) \int_{\Delta t_y}^\infty d(\Delta t_x) \>
   (\Delta t_x-\Delta t_y)
   \Re \frac{d\Gamma}{dx\,d(\Delta t_x)}
   \Re \frac{d\Gamma}{dy\,d(\Delta t_y)} \,.
\label {eq:xyyxINDEP0}
\end {equation}
Using (\ref{eq:dGdx9}),
the $\Delta t_x$ integral is
\begin {multline}
   \int_{\Delta t_y}^\infty d(\Delta t_x) \>
   (\Delta t_x-\Delta t_y) \,
   \frac{d\Gamma}{dx\,d(\Delta t_x)}
   =
   \int_{\Delta t_y^-}^\infty d(\Delta t_x^-) \>
   (\Delta t_x^--\Delta t_y^-) \,
   \frac{d\Gamma}{dx\,d(\Delta t_x)}
\\
   =
   -\frac{2 \alphaEM}{\pi x} \left[
     i \Omega_x \Delta t_y^-
     - \ln\bigl( 2 i \sin(\Omega_x \Delta t_y^-) \bigr)
   \right]
   =
   \frac{2 \alphaEM}{\pi x} \,  \ln\bigl( 1 - e^{-2 i \Omega_x \Delta t_y^-} \bigr)
   ,
\label {eq:int1}
\end {multline}
where
\begin {equation}
   \Delta t^\pm \equiv \Delta t_\pm \equiv \Delta t \pm i\epsilon
\end {equation}
as in (\ref{eq:Deltatpm}).
Using this in (\ref{eq:xyyxINDEP0}) yields
\begin {equation}
  2\Re \left[\frac{d\Gamma}{dx\,dy}\right]_{xy\bar y\bar x + x\bar y y\bar x}
  =
   -\frac{4\alphaEM^2}{\pi^2 x y}
   \int_0^\infty d(\Delta t) \>
   \Re\left[
     \Omega_y^2 \csc^2(\Omega_y\Delta t_-)
   \right]
   \Re \ln\bigl( 1 - e^{-2 i \Omega_x \Delta t_-} \bigr) ,
\label {eq:xyyxINDEP}
\end {equation}
where we have now relabeled $\Delta t_y$ as simply $\Delta t$.


\subsection{\boldmath$x\bar y\bar x y$}

In the case of $x\bar y\bar x y$, we want
$t_\xx < t_\ybx < t_\xbx < t_\yx$.
In keeping with the conventions of our full analysis of crossed
diagrams, we will
want to write this contribution as an integral over the
time separation $\Delta t = t_\xbx-t_\ybx$ of the two intermediate
times.  The corresponding constrained time integral is
\begin {equation}
  \left[\frac{d\Gamma}{dx\,dy}\right]_{x\bar y\bar x y}
  =
  \int_0^\infty d(\Delta t)
   \int_{\Delta t}^\infty d(\Delta t_y) \int_{\Delta t}^\infty d(\Delta t_x) \>
   \frac12 \, \frac{d\Gamma}{dx\,d(\Delta t_x)} \,
   \frac12 \left( \frac{d\Gamma}{dy\,d(\Delta t_y)} \right)^* ,
\label {eq:xyxyINDEP0}
\end {equation}
where the conjugation of the last factor is because the $y$ emissions
appear in the order $\bar y y$ in $x\bar y\bar x y$.
The basic integral needed is
\begin {equation}
   \int_{\Delta t}^\infty d(\Delta t_x) \>
   \Omega_x^2 \csc^2(\Omega_x\Delta t_x^-)
   =
   \Omega_x \bigl(-i + \cot(\Omega_x \Delta t_-) \bigr)
   =
     \frac{2i \Omega_x e^{-2 i \Omega_x \Delta t_-}}{1 - e^{-2 i \Omega_x \Delta t_-}}
   \,,
\label {eq:int3}
\end {equation}
giving
\begin {equation}
  2\Re \left[\frac{d\Gamma}{dx\,dy}\right]_{x\bar y\bar x y}
  =
   \frac{2\alphaEM^2}{\pi^2 x y}
   \Re
   \int_0^\infty d(\Delta t) \>
   \frac{2i \Omega_x e^{-2 i \Omega_x \Delta t_-}}{1 - e^{-2 i \Omega_x \Delta t_-}}
   \left(
     \frac{2i \Omega_y e^{-2 i \Omega_y \Delta t_-}}{1 - e^{-2 i \Omega_y \Delta t_-}}
   \right)^*
   .
\label {eq:xyxyINDEP}
\end {equation}


\subsection{Total crossed diagrams}

Adding together (\ref{eq:xyyxINDEP}), (\ref{eq:xyxyINDEP}) and their
$x\leftrightarrow y$ permutations gives the total contribution
from QED crossed diagrams in the independent emission model:
\begin {align}
  2\Re \left[\frac{d\Gamma}{dx\,dy}\right]_{\rm crossed}
  =
   \frac{4\alphaEM^2}{\pi^2 x y} &
   \int_0^\infty d(\Delta t) \> \Biggl\{
   -
   \Re\left[
     \Omega_y^2 \csc^2(\Omega_y\Delta t_-)
   \right]
   \Re \ln\bigl( 1 - e^{-2 i \Omega_x \Delta t_-} \bigr)
\nonumber\\
   &-
   \Re\left[
     \Omega_x^2 \csc^2(\Omega_x\Delta t_-)
   \right]
   \Re \ln\bigl( 1 - e^{-2 i \Omega_y \Delta t_-} \bigr)
\nonumber\\
   &+
   \Re \left[
   \frac{2i \Omega_x e^{-2 i \Omega_x \Delta t_-}}{1 - e^{-2 i \Omega_x \Delta t_-}}
   \left(
     \frac{2i \Omega_y e^{-2 i \Omega_y \Delta t_-}}{1 - e^{-2 i \Omega_y \Delta t_-}}
   \right)^{\!*}\,
   \right]
   \Biggr\} .
\label {eq:crossedINDEP0}
\end {align}

In order to make contact with the discussion in the text, take
the simplifying limit $y \ll x$, in which case the QED frequencies
(\ref{eq:OmxQED}) are ordered as $|\Omega_y| \ll |\Omega_x|$.
The integrand in (\ref{eq:crossedINDEP0}) is negligible unless
$|\Omega_x|\,\Delta t\lesssim 1$, which therefore means we can assume
$|\Omega_y|\,\Delta t\ll 1$:
\begin {align}
  2\Re \left[\frac{d\Gamma}{dx\,dy}\right]_{\rm crossed}
  \simeq
   \frac{4\alphaEM^2}{\pi^2 x y} &
   \int_0^\infty d(\Delta t) \> \Biggl\{
   -
   \Re\left[
     (\Delta t_-)^{-2}
   \right]
   \Re \ln\bigl( 1 - e^{-2 i \Omega_x \Delta t_-} \bigr)
\nonumber\\
   &-
   \Re\left[
     \Omega_x^2 \csc^2(\Omega_x\Delta t_-)
   \right]
   \Re \ln( 2 i \Omega_y \Delta t_- )
\nonumber\\
   &+
   \Re \left[
   \frac{2i \Omega_x e^{-2 i \Omega_x \Delta t_-}}{1 - e^{-2 i \Omega_x \Delta t_-}}
   \left( \frac{1}{\Delta t_-} \right)^{\!*}\,
   \right]
   \Biggr\} .
\label {eq:crossedINDEP}
\end {align}
Now separate out the UV divergent pieces by writing
\begin {align}
  2\Re \left[\frac{d\Gamma}{dx\,dy}\right]_{\rm crossed}
  \simeq
   \frac{4\alphaEM^2}{\pi^2 x y} &
   \int_0^\infty d(\Delta t)
   \bigl[ f(\Delta t) + D(\Delta t;\epsilon) + P(\Delta t;\epsilon) \bigr] ,
\label {eq:crosssplit}
\end {align}
where $D+P$ represent the divergent pieces from the
$\Delta t\to 0$ behavior of the integrand and
$f$ is everything else (so that the integral of $f$ will be finite).
Specifically, we split the divergent pieces into%
\begin {subequations}
\label {eq:DPcross}
\begin {equation}
  D(\Delta t;\epsilon) =
   -
   \Re\left[
     \frac{1}{(\Delta t_-)^2}
   \right]
   \Re\bigl[
      \ln( 2 i \Omega_x \Delta t_- ) + \ln( 2 i \Omega_y \Delta t_- )
   \bigr]
   +
   \Re\left[
     \frac{1}{\Delta t_- \, \Delta t_+}
   \right] ,
\end {equation}
which represents vacuum contributions ($\hat q = 0$) as well as
double pole terms of the form $(\Delta t)^{-2}\ln\Omega$, and
\begin {equation}
  P(\Delta t;\epsilon) =
   \Re\left[
     \frac{1}{(\Delta t_-)^2}
   \right]
   \Re[
     i\Omega_x\Delta t_-
   ]
   -
   \Re\left[
     \frac{i\Omega_x}{\Delta t_+}
   \right]
,
\label {eq:Pcross}
\end {equation}
\end {subequations}
which represents the (regulated) simple pole terms of the
form $\Omega/\Delta t$.
In order for (\ref{eq:crosssplit}) to reproduce (\ref{eq:crossedINDEP}),
we then have the remaining
\begin {align}
  f(\Delta t)
  = 
   {}&
   \frac{1}{(\Delta t)^2}
   \Re\left[
     \ln\left( \frac{2i\Omega_x\Delta t}{1 - e^{-2 i \Omega_x \Delta t}} \right)
   \right]
\nonumber\\
   &-
   \Re\Bigl[
     \Omega_x^2 \csc^2(\Omega_x\Delta t) - (\Delta t)^{-2}
   \Bigr]
   \Re \ln( 2 i \Omega_y \, \Delta t )
\nonumber\\
   &+
   \frac{1}{\Delta t}
   \Re \left[
     \frac{2i \Omega_x e^{-2 i \Omega_x \Delta t}}{1 - e^{-2 i \Omega_x \Delta t}}
   - \frac{1}{\Delta t}
   \right] .
\end {align}
$D(\Delta t;\epsilon)$ turns out to integrate to zero, and so we
may drop $D$ altogether.
The regulated small-$\Delta t$ behavior of
(\ref{eq:Pcross}) is equal to the sum of the crossed diagram
entries (the entries above the line) in the second column of
Table\ \ref{tab:epstest}.
Reviewing the above derivation of (\ref{eq:Pcross}) and isolating
the separate contribution from each diagram will give the
individual entries for crossed diagrams in the table.


\subsection{Sequential diagrams}
\label {app:seqINDEPeps}

The sequential diagrams proceed similarly.  $xy\bar x\bar y$ is
the same as the earlier $x\bar y\bar x y$ except that the $y$
factor is not conjugated.  The analogs to (\ref{eq:xyxyINDEP0})
and (\ref{eq:xyxyINDEP}) are
\begin {equation}
  \left[\frac{d\Gamma}{dx\,dy}\right]_{xy\bar x\bar y}
  =
  \int_0^\infty d(\Delta t)
   \int_{\Delta t}^\infty d(\Delta t_y) \int_{\Delta t}^\infty d(\Delta t_x) \>
   \frac12 \, \frac{d\Gamma}{dx\,d(\Delta t_x)} \,
   \frac12 \, \frac{d\Gamma}{dy\,d(\Delta t_y)}
\end {equation}
and
\begin {equation}
   2\Re \left[\frac{d\Gamma}{dx\,dy}\right]_{xy\bar x\bar y}
   =
   \frac{2\alphaEM^2}{\pi^2 x y}
   \Re
   \int_0^\infty d(\Delta t) \>
   \frac{2i \Omega_x e^{-2 i \Omega_x \Delta t_-}}{1 - e^{-2 i \Omega_x \Delta t_-}}
   \,
   \frac{2i \Omega_y e^{-2 i \Omega_y \Delta t_-}}{1 - e^{-2 i \Omega_y \Delta t_-}}
   \,.
\label {eq:seq1INDEP}
\end {equation}
For the other sequential diagrams shown in fig.\ \ref{fig:seq2},
the starting point is just (\ref{eq:DxxyyetcDR}) but taking the
small $x$ and $y$ limit appropriate to the independent emission
model:
\begin {align}
   \left[\Delta \frac{d\Gamma}{dx\,dy}\right]_{x\bar x y\bar y}
   &= -
   \int_0^\infty \! d(\Delta t_x)
   \int_0^\infty \! d(\Delta t_y) \>
   \tfrac12(\Delta t_x+\Delta t_y)
     \frac12 \, \frac{d\Gamma}{dx\,d(\Delta t_x)} \,
     \frac12 \, \frac{d\Gamma}{dy\,d(\Delta t_y)} .
\label {eq:factorized}
\end {align}
Adding in related diagrams and also
performing those integrals which correspond to simple factors
of the single splitting rate,
\begin {multline}
   2\Re \left[\Delta \frac{d\Gamma}{dx\,dy}
        \right]_{x\bar x y\bar y + x\bar x\bar y y}
   =
   \frac{2\alphaEM^2}{\pi^2 x y}
   \biggl[
     \Re(i\Omega_x)
     \Re \int_0^\infty d(\Delta t_y) \>
       \Omega_y^2 \, \Delta t_y \csc^2(\Omega_y\Delta t_y^-)
\\
     +
     \Re(i\Omega_y)
     \Re \int_0^\infty d(\Delta t_x) \>
       \Omega_x^2 \, \Delta t_x \csc^2(\Omega_x\Delta t_x^-)
   \biggr] .
\label {eq:seq2INDEP}
\end {multline}
Totaling (\ref{eq:seq1INDEP}) and (\ref{eq:seq2INDEP}) and
adding in $x\leftrightarrow y$ permutations,
\begin {align}
  2\Re \left[\Delta \frac{d\Gamma}{dx\,dy}\right]_{\rm seq}
  =
   \frac{4\alphaEM^2}{\pi^2 x y} &
   \int_0^\infty d(\Delta t) \> \Biggl\{
   \Delta t \, \Re(i\Omega_x) \,
     \Re[\Omega_y^2 \csc^2(\Omega_y\Delta t_-)]
\nonumber\\
   &+
   \Delta t \, \Re(i\Omega_y) \,
     \Re[\Omega_x^2 \csc^2(\Omega_x\Delta t_-)]
\nonumber\\
   &+
   \Re \left[
   \frac{2i \Omega_x e^{-2 i \Omega_x \Delta t_-}}{1 - e^{-2 i \Omega_x \Delta t_-}} \,
   \frac{2i \Omega_y e^{-2 i \Omega_y \Delta t_-}}{1 - e^{-2 i \Omega_y \Delta t_-}}
   \right]
   \Biggr\} .
\end {align}
Again focusing on the $y \ll x$ limit,
\begin {align}
  2\Re \left[\Delta \frac{d\Gamma}{dx\,dy}\right]_{\rm seq}
  \simeq
   \frac{4\alphaEM^2}{\pi^2 x y} &
   \int_0^\infty d(\Delta t) \> \Biggl\{
   \Delta t \, \Re(i\Omega_x) \,
     \Re[\Omega_y^2 \csc^2(\Omega_y\Delta t_-)]
\nonumber\\
   &+
   \Re \left[
   \frac{2i \Omega_x e^{-2 i \Omega_x \Delta t_-}}{1 - e^{-2 i \Omega_x \Delta t_-}}
   \,\frac{1}{\Delta t_-}
   \right]
   \Biggr\} .
\label {eq:seqINDEP}
\end {align}
Separating UV divergences as before,
\begin {align}
  2\Re \left[\Delta \frac{d\Gamma}{dx\,dy}\right]_{\rm seq}
  \simeq
   \frac{4\alphaEM^2}{\pi^2 x y} &
   \int_0^\infty d(\Delta t)
   \bigl[ f_\seq(\Delta t)
           + D_\seq(\Delta t;\epsilon) + P_\seq(\Delta t;\epsilon) \bigr] ,
\label {eq:seqsplit}
\end {align}
with
\begin {subequations}
\label {eq:DPseq}
\begin {equation}
  D_\seq(\Delta t;\epsilon) =
   \Re\left[
     \frac{1}{(\Delta t_-)^2}
   \right] ,
\label {eq:Dseq}
\end {equation}
\begin {equation}
  P_\seq(\Delta t;\epsilon) =
   \Delta t \Re(i\Omega_x)
   \Re\left[
     \frac{1}{(\Delta t_-)^2}
   \right]
   -
   \Re\left[
     \frac{i\Omega_x}{\Delta t_-}
   \right]
   ,
\label {eq:Pseq}
\end {equation}
\end {subequations}
and
\begin {equation}
  f_\seq(\Delta t)
  =
   \Delta t \, \Re(i\Omega_x) \,
     \Re \bigl[\Omega_y^2 \csc^2(\Omega_y\Delta t)
     \bigr]
   +
   \frac{1}{\Delta t}
   \Re \left[
     \frac{2i \Omega_x e^{-2 i \Omega_x\Delta t}}{1 - e^{-2 i \Omega_x\Delta t}}
   - \frac{1}{\Delta t}
   \right] .
\end {equation}
$D_\seq(\Delta t;\epsilon)$ integrates to zero, and
the remaining divergences (\ref{eq:Pseq}) correspond to the
sequential diagram
entries (the entries below the line) in the second column of
Table \ref{tab:epstest}.


\subsection{Checking the diagnostic}

For the sake of completeness, we should verify that the independent
emission model results pass the diagnostic (\ref{eq:test}), as they
must.
Using (\ref{eq:OmxQED}), we find (for $y \ll x$),
that the various integrals above give
\begin {align}
  2\Re \left[\frac{d\Gamma}{dx\,dy}\right]_{\rm crossed}^{\rm non-pole}
  \equiv &
   \frac{4\alphaEM^2}{\pi^2 x y}
   \int_0^\infty d(\Delta t)
   \> f(\Delta t; \epsilon)
  =
   \left[0.6173 - \frac{\ln(x/y)}{\pi^2} \right]
   \frac{\alphaEM^2}{x^{1/2} y} \sqrt{\frac{\hat q}{E}} \,,
\label {eq:crossNoPole}
\\
  2\Re \left[\frac{d\Gamma}{dx\,dy}\right]_{\rm crossed}^{\rm pole} ~~
  \equiv &
   \frac{4\alphaEM^2}{\pi^2 x y}
   \int_0^\infty d(\Delta t)
   \> P(\Delta t; \epsilon)
\nonumber\\ & \quad
  =
   \frac{4\alphaEM^2}{\pi^2 x y}
   \left[ - \Re(i\Omega_x) - \frac{\pi}{2} \Re(\Omega_x) \right]
  =
   -0.5210 \, \frac{\alphaEM^2}{x^{1/2} y} \sqrt{\frac{\hat q}{E}} \,,
\label {eq:crossPole}
\\
  2\Re \left[\Delta \frac{d\Gamma}{dx\,dy}\right]_{\rm seq}^{\rm non-pole}
  \equiv &
   \frac{4\alphaEM^2}{\pi^2 x y}
   \int_0^\infty d(\Delta t)
   \> f_\seq(\Delta t; \epsilon)
  =
   \left[-0.2120 + \frac{\ln(x/y)}{\pi^2} \right]
   \frac{\alphaEM^2}{x^{1/2} y} \sqrt{\frac{\hat q}{E}} \,,
\label {eq:seqNoPole}
\\
  2\Re \left[\Delta \frac{d\Gamma}{dx\,dy}\right]_{\rm seq}^{\rm pole} ~~~~~
  \equiv &
   \frac{4\alphaEM^2}{\pi^2 x y}
   \int_0^\infty d(\Delta t)
   \> P_\seq(\Delta t; \epsilon)
\nonumber\\ & \quad
  =
   \frac{4\alphaEM^2}{\pi^2 x y}
   \left[ - \Re(i\Omega_x) + \frac{\pi}{2} \Re(\Omega_x) \right]
  =
  0.1157 \, \frac{\alphaEM^2}{x^{1/2} y} \sqrt{\frac{\hat q}{E}} \,.
\label {eq:seqPole}
\end {align}
As promised, these sum to zero.


\section {Test of dimensional regularization for
          independent emission model}
\label {app:DRtest}

Testing the diagnostic (\ref{eq:test}) for dimensional regularization
will be very similar to appendix \ref{app:epstest} except that now
the regularized single splitting rate is given in terms of
(\ref{eq:integral0}) and (\ref{eq:dGdxdt}), so that
\begin {equation}
   \frac{d\Gamma}{dx\,d(\Delta t_x)}
   =
   - \frac{\alphaEM P(x)}{x^2(1-x)^2E^d}
   \left( \frac{M\Omega_x\csc(\Omega_x \Delta t_x)}{2\pi i} \right)^{d/2}
   i d M\Omega_x\csc(\Omega_x \Delta t_x)
\label {eq:xxxx}
\end {equation}
instead of (\ref{eq:dGdxdteps}).
Here, $P(x)$ is the $d$-dimensional DGLAP splitting function.
Remember that the test corresponds to the case of small $x$.
The dimension dependence of time-independent factors will not be
interesting, and so we rewrite
(\ref{eq:xxxx}) as
\begin {equation}
   \frac{d\Gamma}{dx\,d(\Delta t)}
   =
   N_x \, \frac{2\alphaEM}{\pi x} \,
   \bigl[ -i \Omega_x \csc(\Omega_x\Delta t_x) \bigr]^{1+\frac{d}{2}} .
\label {eq:dGdx9DR}
\end {equation}
where (for small $x$, for which $M = x(1{-}x)E \simeq xE$)
\begin {equation}
   N_x \equiv
   \frac{x d P(x)}{4}
   \left( \frac{x}{2\pi E} \right)^{-1+\frac{d}{2}}
\label {eq:Nx}
\end {equation}
is normalized so that
\begin {equation}
  N_x(d{=}2) = 1 \qquad \mbox{(in small-$x$ limit)}.
\end {equation}
Note also that $N_x$ is dimensionful when $d\not=2$.


\subsection{\boldmath$xy\bar y\bar x + x\bar y y\bar x$}

For $xy\bar y\bar x$ and related diagrams, we need the analog
of the integral (\ref{eq:int1}):
\begin {multline}
   \int_{\Delta t_y}^\infty d(\Delta t_x) \>
   (\Delta t_x-\Delta t_y) \,
   \frac{d\Gamma}{dx\,d(\Delta t_x)}
   =
\\
   N_x \, \frac{2\alphaEM}{\pi x} \,
   \int_{\Delta t_y}^\infty d(\Delta t_x) \>
   (\Delta t_x-\Delta t_y) \,
   \bigl[ -i \Omega_x \csc(\Omega_x\Delta t_x) \bigr]^{1+\frac{d}{2}}
   .
\label {eq:DRint0}
\end {multline}

We can make the integrals we need to do a little easier, however.
The UV regularization is only required for the small-$\Delta t$
behavior of our later results, where, for the crossed diagrams,
$\Delta t$ is the separation between
the two intermediate times in the diagram.
What we need to do is compute the analog
for dimensional regularization of the regulated small-$\Delta t$
divergences $D(\Delta t) + P(\Delta t)$, which were given by
(\ref{eq:DPcross}) and (\ref{eq:DPseq}) for the $i\epsilon$
prescription.  The other contributions,
(\ref{eq:crossNoPole}) and (\ref{eq:seqNoPole}), will be exactly
the same in the two regularization methods.

For the diagrams $xy\bar y x + x\bar y y\bar x$, $\Delta t$ is
specifically $\Delta t_y$.
So, in order to
isolate the divergences, we will only need to know (\ref{eq:DRint0})
when $\Delta t_\yx$ is small.  We can take advantage of this by
rewriting (\ref{eq:DRint0}) as
\begin {equation}
   \int_0^\infty d(\Delta t_x) \>
     (\Delta t_x-\Delta t) \,
     \frac{d\Gamma}{dx\,d(\Delta t_x)}
   -
   \int_0^{\Delta t} d(\Delta t_x) \>
     (\Delta t_x-\Delta t) \,
     \frac{d\Gamma}{dx\,d(\Delta t_x)} .
\end {equation}
We may then make a small-$\Delta t_x$ approximation to the last
integrand above (since $\Delta t_x < \Delta t$ there),
\begin {align}
   \int_0^{\Delta t} d(\Delta t_x) \>
     (\Delta t_x-\Delta t) \,
     \frac{d\Gamma}{dx\,d(\Delta t_x)}
   &\simeq
   N_x \, \frac{2\alphaEM}{\pi x} \,
   \int_{0}^{\Delta t} d(\Delta t_x) \>
   (\Delta t_x-\Delta t) \,
   (i\Delta t_x)^{-1-\frac{d}{2}}
\nonumber\\
   &=
   - N_x \, \frac{2\alphaEM}{\pi x} \,
   \frac{ (i\Delta t)^{1-\frac{d}{2}} }{ \frac{d}{2}(1-\frac{d}{2}) } .
\end {align}
The other integral we need is
\begin {align}
   &
   \int_0^\infty d(\Delta t_x) \>
   (\Delta t_x-\Delta t) \,
   \frac{d\Gamma}{dx\,d(\Delta t_x)}
\nonumber\\ & \qquad
   =
   N_x \, \frac{2\alphaEM}{\pi x}
   \int_0^\infty d(\Delta t_x) \>
   (\Delta t_x-\Delta t) \,
   \bigl[ -i \Omega_x \csc(\Omega_x\Delta t_x) \bigr]^{1+\frac{d}{2}}
\nonumber\\ & \qquad
   =
   N_x \, \frac{2\alphaEM}{\pi x} \, \Omega_x^{-1+\frac{d}{2}} \left[
     - \int_0^\infty \frac{\tau \, d\tau}{\sh^{1+\frac{d}{2}}\tau}
     + i \Omega_x \Delta t
       \int_0^\infty \frac{d\tau}{\sh^{1+\frac{d}{2}}\tau}
   \right] ,
\label {eq:DRint1a}
\end {align}
where $\tau \equiv i\Omega_x\Delta t_x$.
This evaluates to
\begin {equation}
   \int_0^\infty d(\Delta t_x) \>
   (\Delta t_x-\Delta t) \,
   \frac{d\Gamma}{dx\,d(\Delta t_x)}
   =
   N_x \, \frac{2\alphaEM}{\pi x} \, \Omega_x^{-1+\frac{d}{2}}
     \left[
        \tfrac{\pi}{4} \tan(\tfrac{\pi d}{4})
        + \tfrac{i}{2} \Omega_x\Delta t
     \right]
     \Beta(\tfrac12{+}\tfrac{d}{4},-\tfrac{d}{4})
\label {eq:DRint1}
\end {equation}
(see appendix \ref{app:details}).
Combining the above formulas and (\ref{eq:xyyxINDEP0}),
and restricting the final
$\Delta t$ integration to small $\Delta t < a$,
\begin {align}
  &
  2\Re \left[\frac{d\Gamma}{dx\,dy}
       \right]_{xy\bar y\bar x + x\bar y y\bar x}^{(\Delta t<a)}
\nonumber\\ &\qquad
  \simeq
  N_x N_y \, \frac{4\alphaEM^2}{\pi^2 x y} \,
  \int_0^a d(\Delta t) \>
  \Re \left[ \left( \frac{-i}{\Delta t} \right)^{1+\frac{d}{2}} \right]
\nonumber\\ &\qquad\qquad \times
  \Re
  \left[
     \Omega_x^{-1+\frac{d}{2}}
     \left[
        \tfrac{\pi}{4} \tan(\tfrac{\pi d}{4})
        + \tfrac{i}{2} \Omega_x\Delta t
     \right]
     \Beta(\tfrac12{+}\tfrac{d}{4},-\tfrac{d}{4})
     +
     \frac{ (i\Delta t)^{1-\frac{d}{2}} }{ \frac{d}{2}(1-\frac{d}{2}) }
  \right]
\nonumber\\ &\qquad
  =
  N_x N_y \, \frac{4\alphaEM^2}{\pi^2 x y} \,
  \Re\bigl[(-i)^{1+\frac{d}{2}}\bigr]
\nonumber\\ &\qquad\qquad \times
  \Re
  \left[
     \Omega_x^{-1+\frac{d}{2}}
     \left(
        - \frac{\pi \tan(\tfrac{\pi d}{4}) a^{-d/2}}{2d}
        + \frac{ i \Omega_x a^{1-\frac{d}{2}} }{2-d} 
     \right)
     \Beta(\tfrac12{+}\tfrac{d}{4},-\tfrac{d}{4})
     +
     \frac{ i^{1-\frac{d}{2}} a^{1-d} }
          { \frac{d}{2} (1-\frac{d}{2}) (1-d) }
  \right]
  .
\end {align}
Writing $d = 2-\epsilon$ and expanding in $\epsilon$
(except for the normalization factor $N_x N_y$),%
\footnote{
  The fact that the argument of the last logarithm
  in (\ref{eq:xyyxINDEP1DR}) is not dimensionless
  is an artifact of
  (i) choosing a normalization $N_x$ (\ref{eq:Nx}) whose dimension
  depends on $d$ and
  (ii) not having expanded $N_x$ and $N_y$ in $\epsilon$.
}
\begin {equation}
  2\Re \left[\frac{d\Gamma}{dx\,dy}
       \right]_{xy\bar y\bar x + x\bar y y\bar x}^{(\Delta t<a)}
  \simeq
  N_x N_y \, \frac{4\alphaEM^2}{\pi^2 x y} \,
  \Re \left[
     \frac{\ln(2\Omega_x a)+1}{a}
     + i\Omega_x \left(
         \frac{2}{\epsilon}
         + \ln\Bigl( \frac{a}{2\Omega_x} \Bigr)
         + 1
       \right)
  \right] .
\label {eq:xyyxINDEP1DR}
\end {equation}


\subsection{\boldmath$x\bar y\bar x y$}

For $x\bar y\bar x y$, we need the analog
of the integral (\ref{eq:int3}):
\begin {equation}
   \int_{\Delta t}^\infty d(\Delta t_x) \>
   \bigl[-i \Omega_x \csc(\Omega_x\Delta t_x) \bigr]^{1+\tfrac{d}{2}} .
\end {equation}
Using the same trick as above of rewriting the integral as
$\int_0^\infty - \int_0^{\Delta t}$ and focusing on the case of small
$\Delta t$, we find
\begin {equation}
   \int_{\Delta t}^\infty d(\Delta t_x) \>
     \bigl[-i\Omega_x \csc(\Omega_x\Delta t_x) \bigr]^{1+\tfrac{d}{2}}
   \simeq
   -i
   \left[
     \tfrac12 \Omega_x^{d/2} \Beta(\tfrac12{+}\tfrac{d}{4},-\tfrac{d}{4})
     + \tfrac{2}{d} (i\Delta t)^{-d/2}
   \right] .
\end {equation}
Then (\ref{eq:xyxyINDEP0}) gives
\begin {align}
  &
  2\Re \left[\frac{d\Gamma}{dx\,dy}\right]_{x\bar y\bar x y}^{(\Delta t<a)}
\nonumber\\ & \qquad
  \simeq
  N_x N_y \, \frac{2\alphaEM^2}{\pi^2 x y} \Re
  \int_0^a d(\Delta t) \>
   \left[
     \tfrac12 \Omega_x^{d/2} \Beta(\tfrac12{+}\tfrac{d}{4},-\tfrac{d}{4})
     + \tfrac{2}{d} (i\Delta t)^{-d/2}
   \right]
   \bigl[x \leftrightarrow y\bigr]^*
\nonumber\\ & \qquad
  \simeq
  N_x N_y \, \frac{2\alphaEM^2}{\pi^2 x y} \Re
  \biggl[
    \frac{4 a^{1-d}}{d^2(1-d)}
    +
    \frac{a^{1-\frac{d}{2}}}{d(1-\frac{d}{2})}
      \Beta(\tfrac12{+}\tfrac{d}{4},-\tfrac{d}{4}) \,
      \bigl[ (i\Omega_x)^{d/2}
               + {(i\Omega_y)^*\,}^{d/2} \bigr]
  \biggr] 
\nonumber\\ & \qquad
  \simeq
  N_x N_y \, \frac{2\alphaEM^2}{\pi^2 x y} \Re
  \biggl[
    - \frac{1}{a}
    - i\Omega_x \ln\Bigl( \frac{a}{2\Omega_x} \Bigr)
    - i\Omega_y \ln\Bigl( \frac{a}{2\Omega_y} \Bigr)
    - i(\Omega_x+\Omega_y)
      \left(
        \frac{2}{\epsilon}
        + 2
        - \frac{i\pi}{2}
      \right)
  \biggr]
  .
\label {eq:xyxyONDEP1DR}
\end {align}


\subsection{Total crossed diagrams}

The sum of (\ref{eq:xyyxINDEP1DR}) and (\ref{eq:xyxyONDEP1DR}) and
their $x\leftrightarrow y$ permutations gives
\begin {equation}
  2\Re \left[\frac{d\Gamma}{dx\,dy}\right]_{\rm crossed}^{(\Delta t < a)}
  =
  \frac{4\alphaEM^2}{\pi^2 x y} \Re
  \biggl[
    \frac{\ln(2\Omega_x a) + \ln(2\Omega_y a) + 1}{a}
    - i(\Omega_x+\Omega_y)
      \left(
        1
        - \frac{i\pi}{2}
      \right)
  \biggr]
\label {eq:DRcrossINDEP0}
\end {equation}
for the sum of all QED crossed diagrams in the independent emission model.
Because the
answer is finite as $d\to 2$, we have now been able to set
$d{=}2$ for $N_x N_y$ as well.

The $a \to 0$ divergences of (\ref{eq:DRcrossINDEP0})
are canceled by divergent contributions
from the region of integration $\Delta t > a$, which was not included
above.  Specifically, our goal here has been to evaluate the
analog, in dimensional regularization, of the
$\int_0^\infty d(\Delta t) \, (D+P)$ of appendix \ref{app:epstest}
(and to then reuse that appendix's result for the remaining
$\int_0^\infty d(\Delta t) \, f$).  What (\ref{eq:DRcrossINDEP0})
gives us is $\int_0^a d(\Delta t) \, (D+P)$.  So, to compute
the total contribution from $D+P$ in dimensional regularization,
we need to add in $\int_a^\infty d(\Delta t) \, (D+P)$.  Since
this integral involves $\Delta t > a$ rather than $\Delta t\to 0$,
it does not require further UV regularization and there is no
reason not to use the $d{=}2$ expressions for $D$ and $P$.
We can take the latter from (\ref{eq:DPcross}), ignoring the
$i\epsilon$ prescriptions since $\Delta t > a$.  But ignoring the
$i\epsilon$ prescriptions in (\ref{eq:Pcross}) gives $P = 0$,
leaving
\begin {align}
   \frac{4\alphaEM^2}{\pi^2 xy}
   \int_a^\infty d(\Delta t)\>D(\Delta t,0)
   &=
   \frac{4\alphaEM^2}{\pi^2 xy}
   \int_a^\infty \frac{d(\Delta t)}{(\Delta t)^2}
   \Re\bigl[
      1
      - \ln( 2 i \Omega_x \Delta t) - \ln( 2 i \Omega_y \Delta t)
   \bigr]
\nonumber\\
   &=
   \frac{4\alphaEM^2}{\pi^2 xy} \, \Re
    \biggl[
      - \frac{\ln(2\Omega_x a) + \ln(2\Omega_y a) + 1}{a}
    \biggr] .
\end {align}
Adding this to (\ref{eq:DRcrossINDEP0}) gives, finally,
\begin {equation}
  2\Re \left[\frac{d\Gamma}{dx\,dy}\right]_{\rm crossed}^{\rm pole}
  =
  \frac{4\alphaEM^2}{\pi^2 x y} \Re
  \bigl[
    (- i-\tfrac{\pi}{2})(\Omega_x+\Omega_y)
  \bigr] .
\label {eq:crossDRpole}
\end {equation}


\subsection{Sequential diagrams}

As observed in appendix \ref{app:epstest},
$xy\bar x\bar y$ is
the same as the earlier $x\bar y\bar x y$ except that the $y$
factor is not conjugated.  The corresponding analog of
(\ref{eq:xyxyONDEP1DR}) is
\begin {align}
  &
  2\Re \left[\frac{d\Gamma}{dx\,dy}\right]_{x\bar y\bar x y}^{(\Delta t < a)}
\nonumber\\ & \qquad
  \simeq
  - N_x N_y \, \frac{2\alphaEM^2}{\pi^2 x y} \Re
  \int_0^a d(\Delta t) \>
   \left[
     \tfrac12 \Omega_x^{d/2} \Beta(\tfrac12{+}\tfrac{d}{4},-\tfrac{d}{4})
     + \tfrac{2}{d} (i\Delta t)^{-d/2}
   \right]
   \bigl[x \leftrightarrow y\bigr]
\nonumber\\ & \qquad
  \simeq
  - N_x N_y \, \frac{2\alphaEM^2}{\pi^2 x y} \Re
  \biggl[
    \frac{4 i^{-d} a^{1-d}}{d^2(1-d)}
    +
    \frac{a^{1-\frac{d}{2}}}{d(1-\frac{d}{2})}
      \Beta(\tfrac12{+}\tfrac{d}{4},-\tfrac{d}{4}) \,
      \bigl[ (-i\Omega_x)^{d/2}
               + (-i\Omega_y)^{d/2} \bigr]
  \biggr] 
\nonumber\\ & \qquad
  \simeq
  N_x N_y \, \frac{2\alphaEM^2}{\pi^2 x y} \Re
  \biggl[
    - \frac{1}{a}
    - i\Omega_x \ln\Bigl( \frac{a}{2\Omega_x} \Bigr)
    - i\Omega_y \ln\Bigl( \frac{a}{2\Omega_y} \Bigr)
    - i(\Omega_x+\Omega_y)
      \left(
        \frac{2}{\epsilon}
        + 2
        + \frac{i\pi}{2}
      \right)
  \biggr]
  .
\label {eq:seq1DR}
\end {align}
For $x\bar xy\bar y$, we take (\ref{eq:factorized}) but use
(\ref{eq:xxxx}) for $d\Gamma/dx\,d(\Delta t_x)$.  The analog
of (\ref{eq:seq2INDEP}) is then
\begin {align}
   &
   2\Re \left[\frac{d\Gamma}{dx\,dy}\right]_{x\bar x y\bar y+x\bar x\bar y y}
   =
\nonumber\\ & \qquad
   N_x N_y \frac{\alphaEM^2}{\pi^2 x y}
     \Beta(\tfrac12{+}\tfrac{d}{4},-\tfrac{d}{4}) \,
     \Re(i\Omega_x^{d/2})
     \Re \int_0^\infty d(\Delta t_y) \> \Delta t_y
     \bigl[-i\Omega_y \csc(\Omega_y\Delta t_y) \bigr]^{1+\tfrac{d}{2}}
\nonumber\\ & \qquad
  + (x\leftrightarrow y) .
\end {align}
The corresponding contribution from small $\Delta t$ is
\begin {align}
   2\Re \left[\frac{d\Gamma}{dx\,dy}
          \right]_{x\bar x y\bar y+x\bar x\bar y y}^{(\Delta t < a)}
   &\simeq
   N_x N_y \frac{\alphaEM^2}{\pi^2 x y}
     \Beta(\tfrac12{+}\tfrac{d}{4},-\tfrac{d}{4}) \,
     \Re(i\Omega_x^{d/2})
     \Re \int_0^a d(\Delta t_y) \> \Delta t_y \,
     (i\Delta t_y)^{-1-\tfrac{d}{2}}
\nonumber\\ & \qquad
  + (x\leftrightarrow y)
\nonumber\\
   &\simeq
   N_x N_y \frac{\alphaEM^2}{\pi^2 x y}
     \Beta(\tfrac12{+}\tfrac{d}{4},-\tfrac{d}{4}) \,
     \Re(i\Omega_x^{d/2} + i\Omega_y^{d/2})
     \frac{ \Re(i^{-1-\frac{d}{2}}) \, a^{1-\frac{d}{2}} }{ 1 - \frac{d}{2} }
\nonumber\\
   &\simeq
  N_x N_y \, \frac{2\alphaEM^2}{\pi^2 x y} \Re
  \biggl[
      i\Omega_x \ln\Bigl( \frac{a}{2\Omega_x} \Bigr)
    + i\Omega_y \ln\Bigl( \frac{a}{2\Omega_y} \Bigr)
\nonumber\\ & \qquad\hspace{10em}
    + i(\Omega_x+\Omega_y)
      \left(
        \frac{2}{\epsilon}
        + 1
      \right)
  \biggr] .
\label {eq:DRseq2INDEP}
\end {align}
Adding this to (\ref{eq:seq1DR}) and to $x\leftrightarrow y$
permutations of the diagrams gives
\begin {equation}
  2\Re \left[\frac{d\Gamma}{dx\,dy}\right]_\seq^{\Delta t < a}
  =
  \frac{4\alphaEM^2}{\pi^2 x y} \Re
  \biggl[
    - \frac{1}{a}
    - i(\Omega_x+\Omega_y)
      \left(
        1
        + \frac{i\pi}{2}
      \right)
  \biggr]
\label {eq:DRseqINDEP0}
\end {equation}
for the sum of all QED sequential diagrams.
Finally, similar to the case of crossed diagrams, we need to also add
\begin {equation}
   \frac{4\alphaEM^2}{\pi^2 xy}
   \int_a^\infty d(\Delta t)\>D_\seq(\Delta t,0)
   =
   \frac{4\alphaEM^2}{\pi^2 xy}
   \int_a^\infty \frac{d(\Delta t)}{(\Delta t)^2}
   =
   \frac{4\alphaEM^2}{\pi^2 xy} \, \frac{1}{a}
\end {equation}
[where $D_\seq$ is taken from (\ref{eq:Dseq})] to get
\begin {equation}
  2\Re \left[\frac{d\Gamma}{dx\,dy}\right]_{\rm seq}^{\rm pole}
  =
  \frac{4\alphaEM^2}{\pi^2 x y} \Re
  \bigl[
    (- i + \tfrac{\pi}{2})(\Omega_x+\Omega_y)
  \bigr] .
\label {eq:seqDRpole}
\end {equation}


\subsection{Summary in the limit \boldmath$y \ll x$}

In the limit $y \ll x$, which was taken to simplify the discussion
in both the main text and in appendix \ref{app:epstest},
the pole pieces (\ref{eq:crossDRpole}) and (\ref{eq:seqDRpole})
are identical to the ones computed with the $i\epsilon$
prescription given by (\ref{eq:crossPole}) and (\ref{eq:seqPole}).
Since the non-pole pieces are unaffected by regularization,
those will be the same too, as quoted in the main text in
(\ref{eq:DRtest}).


\section {Difficulties generalizing the
          naive \boldmath$i\epsilon$ prescription}
\label {app:epsgeneral}

In this appendix, we briefly outline the difficulties we encountered
attempting to generalize the $i\epsilon$ prescription method to
work outside of the independent emission model of
Table \ref{tab:epstest}.


\subsection {First attempt}

Consider the short-time behavior of the 4-particle propagator
shown in (\ref{eq:Cprop00}), setting $d{=}2$ here:
\begin {multline}
  \langle\C_{34},\C_{12},\Delta t|\C'_{34},\C'_{12},0\rangle
  \simeq
\\
  (2\pi i \, \Delta t)^{-2} (\det{\mathfrak M})
  \exp\Biggl[
     -\frac{1}{2}
     \begin{pmatrix} \C_{34}{-}\C'_{34} \\ \C_{12}{-}\C'_{12} \end{pmatrix}^\top
     \frac{\mathfrak M}{i\,\Delta t}
     \begin{pmatrix} \C_{34}{-}\C'_{34} \\ \C_{12}{-}\C'_{12} \end{pmatrix}
  \Biggr] ,
\label {eq:Cprop01}
\end {multline}
where ${\mathfrak M}$ is given by (\ref{eq:frakM}) as
\begin {equation}
   {\mathfrak M}
   =
   \begin{pmatrix}
      x_3 x_4 & \\ & -x_1 x_2
   \end {pmatrix} (x_3+x_4) E .
\end {equation}
A technical difficulty with this
expression is that ${\mathfrak M}/(i\,\Delta t)$ is purely
imaginary and so (\ref{eq:Cprop01}) becomes infinitely oscillatory
as $\Delta t \to 0$.  However, this can be fixed by choosing
appropriate $i\epsilon$ prescriptions.
For example, for $xy\bar y\bar x$, the appropriate $x_i$ are
given by (\ref{eq:xhat}), which give
\begin {equation}
   \frac{\mathfrak M}{i\,\Delta t}
   =
   \begin{pmatrix}
      \frac{x(1{-}x{-}y)}{i \, \Delta t} & \\ & \frac{y}{i \,\Delta t}
   \end {pmatrix} (1{-}y) E .
\end {equation}
If we replace both of the $\Delta t$ by $\Delta t_- = \Delta - i\epsilon$
this becomes
\begin {equation}
   \frac{\mathfrak M}{i\,\Delta t}
   =
   \begin{pmatrix}
      \frac{x(1{-}x{-}y)}{i \, \Delta t_-} & \\ & \frac{y}{i \,\Delta t_-}
   \end {pmatrix} (1{-}y) E ,
\end {equation}
which for $\Delta t = 0$ is real and positive definite,
\begin {equation}
   \frac{\mathfrak M}{i\,\Delta t}
   \to
   \begin{pmatrix}
      \frac{x(1{-}x{-}y)}{\epsilon} & \\ & \frac{y}{\epsilon}
   \end {pmatrix} (1{-}y) E ,
\end {equation}
making (\ref{eq:Cprop01}) sensible at $\Delta t = 0$.

As another example, consider $x\bar y y\bar x$, for which the
appropriate $x_i$ are
given by (\ref{eq:xprime}).
The choice of $i\epsilon$ prescription which makes (\ref{eq:Cprop01})
sensible at $\Delta t = 0$ is then instead
\begin {equation}
   \frac{\mathfrak M}{i\,\Delta t}
   =
   \begin{pmatrix}
      \frac{x(1{-}x)}{i \, \Delta t_-} & \\ & -\frac{y(1{-}y)}{i \,\Delta t_+}
   \end {pmatrix} E
   \to
   \begin{pmatrix}
      \frac{x(1{-}x)}{\epsilon} & \\ & \frac{y(1{-}y)}{\epsilon}
   \end {pmatrix} E .
\end {equation}
Note the mixture of $\Delta t_-$ and $\Delta t_+$ in this expression.

By such considerations, one could seemingly determine the necessary
$i\epsilon$ prescriptions for all $\Delta t$'s in 4-particle
propagators.  Once the prescription for
$\langle\C_{34},\C_{12},\Delta t|\C'_{34},\C'_{12},0\rangle$ above
is fixed,
we could use the change of variables (\ref{eq:Cchange})
to get the version
$\langle\C_{34},\C_{12},\Delta t|\C'_{41},\C'_{23},0\rangle$
needed for expressions like (\ref{eq:IIxyyx2DR}).

Carrying out this procedure for a full calculation of QED double
bremsstrahlung (not the simple independent emission model),
we have found that for $y \ll x \ll 1$ we correctly obtain the
small-$\Delta t$ behavior shown in the second column of
table \ref{tab:epstest}.  So, what's not to like?

The problem is that, {\it except}\/ for the limiting case of QED with
$y \ll x \ll 1$,
we find different results if we instead determine
our prescriptions by doing a similar analysis starting from
\begin {multline}
  \langle\C_{41},\C_{23},\Delta t|\C'_{41},\C'_{23},0\rangle
  \simeq
\\
  (2\pi i \, \Delta t)^{-2} (\det{\mathfrak M}')
  \exp\Biggl[
     -\frac{1}{2}
     \begin{pmatrix} \C_{41}{-}\C'_{41} \\ \C_{23}{-}\C'_{23} \end{pmatrix}^\top
     \frac{{\mathfrak M}'}{i\,\Delta t}
     \begin{pmatrix} \C_{41}{-}\C'_{41} \\ \C_{23}{-}\C'_{23} \end{pmatrix}
  \Biggr] ,
\end {multline}
where ${\mathfrak M}'$ is given by (\ref{eq:frakMprime}),
instead of starting from (\ref{eq:Cprop01}).
We'll discuss why results can be different in a moment.


\subsection {Second attempt}

Alternatively, we considered giving up on trying to
determine the $i\epsilon$ prescriptions in a basis like
$(\C_{34},\C_{12})$ or $(\C_{41},\C_{23})$ above and instead
using a more natural basis: the normal mode basis
for the 4-particle propagation.
From AI (5.30) \cite{2brem}, this propagator is
\begin {multline}
   \langle \A_+,\A_-,\Delta t | \A_+',\A_-',0 \rangle
   =
\\
   \prod_\pm \left[
      \frac{\Omega_\pm\csc(\Omega_\pm \Delta t)}{2\pi i} \,
      \exp\Bigl(
        i \bigl[
          \tfrac12 (\A_\pm^2+\A_\pm'^2) \Omega_\pm\cot(\Omega_\pm \Delta t)
          - \A_\pm\cdot\A_\pm' \Omega_\pm \csc(\Omega_\pm \Delta t)
        \bigr]
      \Bigr)
   \right]
\label {eq:Aprop}
\end {multline}
for $xy\bar y\bar x$, where $\Omega_\pm$ are complex normal mode
frequencies associated with the 4-particle propagation.
In the small $\Delta t$ limit, this becomes
\begin {multline}
  \langle \A_+,\A_-,\Delta t | \A_+',\A_-',0 \rangle
  \simeq
\\
  (2\pi i \, \Delta t)^{-2}
  \exp\Biggl[
     -\frac{1}{2}
     \begin{pmatrix} \A_+{-}\A_+' \\ \A_-{-}\A_-' \end{pmatrix}^\top
     \begin {pmatrix}
       \frac{1}{i\,\Delta t} & \\ & \frac{1}{i\,\Delta t}
     \end {pmatrix} 
     \begin{pmatrix} \A_+{-}\A_+' \\ \A_-{-}\A_-' \end{pmatrix}
  \Biggr] .
\end {multline}
The prescription which makes this sensible for $\Delta t = 0$ is
$\Delta t \to \Delta t_-$, as before.  This is equivalent to the
previous method after changing the basis.

For $x\bar y y\bar x$, the corresponding normal mode propagator is
instead (see appendix E of ref.\ \cite{2brem})
\begin {multline}
   \langle \A_+,\A_-,\Delta t | \A_+',\A_-',0 \rangle
   =
\\
   \prod_\pm \left[
      \frac{\Omega_\pm\csc(\Omega_\pm \Delta t)}{\pm 2\pi i} \,
      \exp\Bigl(
        \pm
        i \bigl[
          \tfrac12 (\A_\pm^2+\A_\pm'^2) \Omega_\pm\cot(\Omega_\pm \Delta t)
          - \A_\pm\cdot\A_\pm' \Omega_\pm \csc(\Omega_\pm \Delta t)
        \bigr]
      \Bigr)
   \right]
\label {eq:Apropmodified}
\end {multline}
with limit
\begin {multline}
  \langle \A_+,\A_-,\Delta t | \A_+',\A_-',0 \rangle
  \simeq
\\
  (2\pi \Delta t)^{-2}
  \exp\Biggl[
     -\frac{1}{2}
     \begin{pmatrix} \A_+{-}\A_+' \\ \A_-{-}\A_-' \end{pmatrix}^\top
     \begin {pmatrix}
       \frac{1}{i\,\Delta t} & \\ & - \frac{1}{i\,\Delta t}
     \end {pmatrix} 
     \begin{pmatrix} \A_+{-}\A_+' \\ \A_-{-}\A_-' \end{pmatrix}
  \Biggr] .
\label {eq:Apropeps}
\end {multline}
As in the previous method, we need to replace one $\Delta t$ by
$\Delta t_-$ and the other by $\Delta t_+$ to make
(\ref{eq:Apropeps}) sensible when $\Delta t = 0$.
This prescription is {\it not}\/
equivalent to what we did before, once we change basis back
to $(\C_{34},\C_{12})$ and/or $(\C_{41},\C_{23})$.
It does, though, at least have the virtue of treating the two
ends of the 4-particle propagation symmetrically.

The new prescription also gives
the same results in the $y\ll x\ll 1$ limit of QED.
Unfortunately, outside of this special case, it does not agree with
the result derived in this paper using dimensional regularization
(nor with the first proposal in this appendix).
Which method should we trust?

There is a dirty secret shared by all of the $i\epsilon$ methods just
proposed: they only attempted to regulate the 4-particle
propagation.  In the $d{=}2$ derivation of AI \cite{2brem},
there was a step at AI (5.7) where an infinitely oscillatory term
associated with time-integrated 3-particle evolution was discarded.
The argument given there
was that a finite infinitely-oscillatory function would
give zero when later integrated against a smooth function.
However, the pole pieces of our calculation are coming from
infinitesimal $\Delta t \sim \epsilon$ when we regulate the 4-particle
propagation with an $i\epsilon$ prescription.  So, when we use
the $i\epsilon$ prescription, it is important that we treat arbitrarily
short-time features of our expressions correctly.
That's inconsistent with requiring the infinitely oscillatory piece of
AI (5.7) to be integrated only against smooth functions.
So we need to keep that piece and regulate it as well.
Could we regulate it with yet another $i\epsilon$ prescription?
We attempted this but were unable to find a method that gave a
convincing, unique answer independent of details of exactly how
we chose the magnitudes of the various $\epsilon$ factors.

In contrast, the advantage of dimensional regularization is that it
simultaneously treats all UV problems in a consistent manner.


\section {4-particle propagator in medium}
\label {app:4particle}

In this appendix, we discuss the $d$-dimensional generalization
of the 4-particle propagator
{\it without}\/ first taking the $\Delta t \to 0$ limit
relevant to the main text.

In ref.\ \cite{2brem}, the relationship between normal modes
$\A_\pm$ and the transverse momentum variables $(\C_{34},\C_{12})$
or $(\C_{41},\C_{23})$ was written%
\footnote{
  AI (5.27) and AI (5.32)
}
\begin {equation}
   \begin{pmatrix} \C_{34} \\ \C_{12} \end{pmatrix}
   = a_\ybx \begin{pmatrix} \A_+ \\ \A_- \end{pmatrix}
\label {eq:changef}
\end {equation}
and
\begin {equation}
   \begin{pmatrix} \C_{41} \\ \C_{23} \end{pmatrix}
   = a_\yx \begin{pmatrix} \A_+ \\ \A_- \end{pmatrix} ,
\label {eq:changei}
\end {equation}
where $a_\ybx$ and $a_\yx$ are $2\times2$ matrices whose explicit
formulas will not be needed here.
The important point is that, because of rotation invariance in
the transverse plane, (\ref{eq:changef}) and (\ref{eq:changei}) are
satisfied component by component for the various transverse $\C$ and
$\A$ vectors.  In consequence, these equations apply without change to the
more general situation of $d$ transverse dimensions.

The $d{=}2$ product of normal mode propagators is (\ref{eq:Aprop}),
which generalizes to
\begin {multline}
   \langle \A_+,\A_-,\Delta t | \A_+',\A_-',0 \rangle
   =
\\
   \prod_\pm \left[
      \left( \frac{\Omega_\pm\csc(\Omega_\pm \Delta t)}{2\pi i} \right)^{d/2}
      \exp\Bigl(
        i \bigl[ \tfrac12 (\A_\pm^2+\A_\pm'^2) \Omega_\pm\cot(\Omega_\pm \Delta t)
        - \A_\pm\cdot\A_\pm' \Omega_\pm \csc(\Omega_\pm \Delta t) \bigr]
      \Bigr)
   \right] .
\label {eq:ApropDR}
\end {multline}
(The explicit formulas for the 4-particle normal mode frequencies
$\Omega_\pm$ will also not be relevant here.)  Using (\ref{eq:changef})
and (\ref{eq:changei}) to change variables,
\begin {equation}
   \langle\C_{34}^\Ax,\C_{12}^\Ax,\Delta t|\C_{41}^\bx,\C_{23}^\bx,0\rangle
   =
   \frac{
     \langle \A_+^\Ax,\A_-^\Ax,\Delta t | \A_+^\bx,\A_-^\bx,0 \rangle
   }{
     | \det a_\Ax |^{d/2} | \det a_\bx |^{d/2}
   } .
\label{eq:CvsAprop}
\end {equation}
Ref.\ \cite{2brem} finds a generic result for the determinants which
only depends on the normalization convention for the normal modes:%
\footnote{
  AI (5.36--37)
}
\begin {align}
   |\det a_\ybx|^{-1}
   &= |x_1 x_2 x_3 x_4|^{1/2} |x_3{+}x_4| E ,
\nonumber\\
   \qquad
   |\det a_\yx|^{-1}
   &= |x_1 x_2 x_3 x_4|^{1/2} |x_1{+}x_4| E ,
\end {align}
and so%
\footnote{
  The $d{=}2$ cases of (\ref{eq:CvsAprop}), (\ref{eq:Cprop0DR}),
  (\ref{eq:CpropDR}) and (\ref{eq:iamacow})
  reproduce AI (5.34), AI (5.38), AI (5.39) and AI (5.43) respectively.
}
\begin {equation}
   \langle\C_{34}^\Ax,\C_{12}^\Ax,\Delta t|\C_{41}^\bx,\C_{23}^\bx,0\rangle
   =
     |x_1 x_2 x_3 x_4|^{d/2} |x_1{+}x_4|^{d/2} |x_3{+}x_4|^{d/2} E^d
     \langle \A_+^\Ax,\A_-^\Ax,\Delta t | \A_+^\bx,\A_-^\bx,0 \rangle .
\label {eq:Cprop0DR}
\end {equation}
Put altogether,
\begin {align}
   \langle\C_{34}^\Ax,\C_{12}^\Ax,\Delta t&|\C_{41}^\bx,\C_{23}^\bx,0\rangle
   =
\nonumber\\ &
   (2\pi i)^{-d}
     ({-}x_1 x_2 x_3 x_4)^{d/2}
     |x_1{+}x_4|^{d/2} |x_3{+}x_4|^{d/2} E^d
   \bigl[
      \Omega_+\Omega_- \csc(\Omega_+\Delta t) \csc(\Omega_-\Delta t)
   \bigr]^{d/2}
\nonumber\\ &\times
   \exp\Biggl[
     \frac{i}2
     \begin{pmatrix} \C^\bx_{41} \\ \C^\bx_{23} \end{pmatrix}^\top
       a_\bx^{-1\top} \uOmega \cot(\uOmega\,\Delta t) \, a_\bx^{-1}
       \begin{pmatrix} \C^\bx_{41} \\ \C^\bx_{23} \end{pmatrix}
\nonumber\\ &\qquad\quad
     +
     \frac{i}2
     \begin{pmatrix} \C^\Ax_{34} \\ \C^\Ax_{12} \end{pmatrix}^\top
       a_\Ax^{-1\top} \uOmega \cot(\uOmega\,\Delta t) \, a_\Ax^{-1}
       \begin{pmatrix} \C^\Ax_{34} \\ \C^\Ax_{12} \end{pmatrix}
\nonumber\\ &\qquad\quad
     - i
     \begin{pmatrix} \C^\bx_{41} \\ \C^\bx_{23} \end{pmatrix}^\top
       a_\bx^{-1\top} \uOmega \csc(\uOmega\,\Delta t) \, a_\Ax^{-1}
       \begin{pmatrix} \C^\Ax_{34} \\ \C^\Ax_{12} \end{pmatrix}
   \Biggr]
\label {eq:CpropDR}
\end {align}
with
$\uOmega \equiv
 \bigl(
 \begin{smallmatrix} \Omega_+ & \\ & \Omega_- \end{smallmatrix}
 \bigr)$.
Using this in (\ref{eq:IIxyyx2DR}) gives
\begin {align}
   \left[\frac{d\Gamma}{dx\,dy}\right]_{xy\bar y\bar x}
   = &
   - \frac{\CA^2 \alphas^2 M_\ix M_\fx}{2^{d+3}\pi^{2d+1}i^dE^d} \,
     \Gamma^2( \tfrac12{+}\tfrac{d}{4} ) \,
     ({-}\hat x_1 \hat x_2 \hat x_3 \hat x_4)^{d/2}
     ( \alpha \delta^{\bar n n} \delta^{\bar m m}
     {+} \beta \delta^{\bar n \bar m} \delta^{nm}
     {+} \gamma \delta^{\bar n m} \delta^{n \bar m} )
\nonumber\\ & \times
   \int_0^{\infty} d(\Delta t) \>
   \bigl[
     \Omega_+\Omega_- \csc(\Omega_+\Delta t) \csc(\Omega_-\Delta t)
   \bigr]^{d/2}
\nonumber\\ & \times
   \int_{\B^\Ax,\B^\bx}
   B^{\ybx}_{\bar n} \left(\frac{|M_\fx|\Omega_\fx}{(B^\ybx)^2}\right)^{d/4}
     K_{d/4}\bigl(\tfrac12 |M_\fx| \Omega_\fx (B^\ybx)^2\bigr)
\nonumber\\ & \times
   B^{\yx}_{m} \left(\frac{|M_\ix|\Omega_\ix}{(B^\yx)^2}\right)^{d/4}
     K_{d/4}\bigl(\tfrac12 |M_\ix| \Omega_\ix (B^\yx)^2\bigr)
\nonumber\\ & \times
   \bigl[
     (Y_\bx \B^\bx - \Ybar_{\bx\Ax} \B^\Ax)_n
     (Y_\Ax \B^\Ax - Y_{\bx\Ax} \B^\bx)_{\bar m}
     + Z_{\bx\Ax} \delta_{n\bar m}
   \bigr]
\nonumber\\ & \times
   \exp\Bigl[
     - \tfrac12
     \calX_\bx (B^\bx)^2
     -
     \tfrac12
     \calX_\Ax (B^\Ax)^2
     +
     X_{\bx\Ax} \B^\bx \cdot \B^\Ax
   \Bigr]
\label {eq:iamacow}
\end {align}
with
\begin {subequations}
\label {eq:XYZdefDR}
\begin {align}
   \begin{pmatrix} \calX_\bx & Y_\bx \\ Y_\bx & Z_\bx \end{pmatrix}
   &\equiv
     - i a_\bx^{-1\top} \uOmega \cot(\uOmega\,\Delta t)\, a_\bx^{-1} ,
\\
   \begin{pmatrix} \calX_\Ax & Y_\Ax \\ Y_\Ax & Z_\Ax \end{pmatrix}
   &\equiv
     - i a_\Ax^{-1\top} \uOmega \cot(\uOmega\,\Delta t)\, a_\Ax^{-1} ,
\\
   \begin{pmatrix} X_{\bx\Ax} & Y_{\bx\Ax} \\ \Ybar_{\bx\Ax} & Z_{\bx\Ax} \end{pmatrix}
   &\equiv
   - i a_\bx^{-1\top} \uOmega \csc(\uOmega\,\Delta t) \, a_\Ax^{-1} .
\end {align}
\end {subequations}
Expanding in $\Delta t$ gives (\ref{eq:calXYZsmalldt}) and
(\ref{eq:DRgeneral}).


\section{\boldmath$J$ integrals}
\label {app:Jints}

Formally, the various integrals $J_{\ix n}$ defined in
(\ref{eq:J}) may be obtained by taking
derivatives of the single master integral
\begin {equation}
   {\cal J} \equiv
   \int
   \frac{d^dB_1 \> d^dB_2}{B_2^d}
   \exp\Bigl[
     - \tfrac12 a B_1^2
     + b \B_1 \cdot \B_2
     - \tfrac12 c B_2^2
   \Bigr]
\label {eq:calJdef0}
\end {equation}
with respect to the parameters $(a,b,c)$.  Unfortunately, the above
integral is UV divergent (even in $d$ dimensions)
from the region of integration $B_2 \to 0$.
However, the divergent part must vanish when we take the derivatives
necessary to get the integrals (\ref{eq:J}) because the latter integrals
are all UV convergent.  We will see that this works out.  For now, let us
just temporarily regulate the UV divergence by
replacing (\ref{eq:calJdef0}) by
\begin {equation}
   {\cal J}_\delta \equiv
   \int
   \frac{d^dB_1 \> d^dB_2}{B_2^{d-2\delta}}
   \exp\Bigl[
     - \tfrac12 a B_1^2
     + b \B_1 \cdot \B_2
     - \tfrac12 c B_2^2
   \Bigr] ,
\end {equation}
with the understanding that $\delta$ is infinitesimal.

Use the Schwinger trick to rewrite $B_2^{-(d-2\delta)}$ as an exponential:
\begin {align}
   {\cal J}_\delta
   &=
   \frac{1}{\Gamma(\frac{d}{2}-\delta)}
   \int_0^\infty s^{\frac{d}{2}-\delta-1} \>ds
   \int d^dB_1 \> d^dB_2
   \exp\Bigl[
     - \tfrac12 a B_1^2
     + b \B_1 \cdot \B_2
     - \tfrac12 (c+2s) B_2^2
   \Bigr]
\nonumber\\
   &=
   \frac{(2\pi)^d}{\Gamma(\frac{d}{2}-\delta)}
   \int_0^\infty
   \frac{ s^{\frac{d}{2}-\delta-1} \>ds }{ (ac-b^2 + 2 a s)^{d/2} }
\nonumber\\
   &=
   \frac{(2\pi)^d}{\Gamma(\frac{d}{2}-\delta)\,(2a)^{d/2}}
   \left( \frac{ac-b^2}{2a} \right)^{-\delta}
   \int_0^\infty
   \frac{ u^{\frac{d}{2}-\delta-1} \>du }{ (1 + u)^{d/2} } ,
\label {eq:calJdelta}
\end {align}
where $u \equiv 2as/(ac-b^2)$.  The $u$ integral can be done exactly,
but we don't even need too: The $u \gg 1$ behavior of the integrand
is $u^{-\delta-1}$,
and so
\begin {equation}
   \int_0^\infty
     \frac{ u^{\frac{d}{2}-\delta-1} \>du }{ (1 + u)^{d/2} }
   =
   \int_1^\infty u^{-\delta-1} \>du
   + O(1)
   = \frac{1}{\delta} + O(1) .
\end {equation}
That's enough to figure out that the small-$\delta$ expansion of
(\ref{eq:calJdelta}) is
\begin {equation}
   {\cal J}_\delta =
   (\mbox{something independent of $b$ and $c$})
   - \frac{2^{d/2} \pi^d}{\Gamma(\frac{d}{2})\, a^{d/2}} \ln(ac-b^2)
   + O(\delta) .
\end {equation}
The results for the
integrals $J_{\ix n}$ of (\ref{eq:J}) may now be obtained from
\begin {align}
   J_{\ix 0} &= 4 \partial_a \partial_c {\cal J}_\delta ,
\nonumber\\
   J_{\ix 1} &= \partial_b {\cal J}_\delta ,
\nonumber\\
   J_{\ix 2} &= \partial_b J_{\ix 1} ,
\nonumber\\
   J_{\ix 3} &= -2 \partial_c J_{\ix 1} ,
\nonumber\\
   J_{\ix 4} &= -2 \partial_a J_{\ix 1} ,
\end {align}
with $\delta \to 0$ and $(a,b,c) = (\calX_\yx, X_{\yx\ybx}, \calX_\ybx)$.

The small $\Delta t$ limit of $J_{\ix 0}$, $J_{\ix 1}$, and $J_{\ix 2}$
are given in (\ref{eq:J2}).  In the main text,
we did not require similar expressions for $J_{\ix 3}$ and $J_{\ix 4}$,
but they are
\begin {subequations}
\label {eq:J34small}
\begin{align}
   J_{\ix 3} &\simeq
   \frac{2^{\frac{d}{2}}\pi^d}{\Gamma(\frac{d}{2})}
   \left( \frac{\Delta t}{-i M_\ix} \right)^{d/2}
   \left( \frac{\Delta t}{-i E} \right)^2
   \left[
     -\frac{4 (x_1+x_4)}{x_2^2 x_3 x_4^4}
   \right]
   ,
\\
   J_{\ix 4} &\simeq
   \frac{2^{\frac{d}{2}}\pi^d}{\Gamma(\frac{d}{2})}
   \left( \frac{\Delta t}{-i M_\ix} \right)^{d/2}
   \left( \frac{\Delta t}{-i E} \right)^2 \left[
     - \frac{4\bigl(x_1x_3 - (1+\frac{d}{2}) x_2x_4\bigr)}
            {x_1 x_2^2 x_4^4 (x_1+x_4)}
   \right]
   .
\end {align}
\end {subequations}


\section{Branch cuts}
\label {app:sheet}

In this appendix, we identify the origin of all complex phases
in results for the various crossed diagrams and verify that
they are reproduced by the factor (\ref{eq:sheet1}) in
(\ref{eq:poleDR1}).
For readers who would rather not think carefully about branch cuts
in dimensional regularization,
one could instead just take some reassurance from the fact that these
details only affect the $1/\pi$ terms, and our results in the main
text reproduce the $1/\pi$ pole terms found earlier in ref.\ \cite{2brem}
by other means.


\subsection{\boldmath$x_1 x_2 x_3 x_4 < 0$}

In diagrams such as $xy\bar y\bar x$ where both emissions happen first
in the amplitude (as opposed to the conjugate amplitude), we have
$x_1 x_2 x_3 x_4 < 0$.
From (\ref{eq:Cprop00}) and (\ref{eq:CpropSmall})
for the corresponding 4-particle propagator
[or from (\ref{eq:ApropDR}) and (\ref{eq:CpropDR})],
we get an overall factor of $i^{-d}$,
which later appears in the first factor in
(\ref{eq:IGammaDR2}).  This $i^{-d}$ should be understood as
$(e^{i\pi/2})^{-d} = e^{-id\pi/2}$, because that's the interpretation
that normalizes (\ref{eq:Cprop00}) to be a representation of
$\delta^{(d)}(\C_{34} - \C'_{34}) \, \delta^{(d)}(\C_{12} - \C'_{12})$
as $\Delta t \to 0$. 

On a related note, the factor $(-x_1 x_2 x_3 x_4)^{d/2}$ in
(\ref{eq:DRgeneral}) originated from the normalization factor
\begin {equation}
  (\det{\mathfrak M})^{d/4} (\det{\mathfrak M}')^{d/4}
  =
  \bigl[-x_1 x_2 x_3 x_4(x_3{+}x_4)^2\bigr]^{d/4}
  \bigl[-x_1 x_2 x_3 x_4(x_1{+}x_4)^2\bigr]^{d/4}
\label {eq:Mfactors}
\end {equation}
in the 4-particle propagator (\ref{eq:CpropSmall}).
Since $-x_1 x_2 x_3 x_4$ is positive real for $xy\bar y\bar x$,
there is no phase here.

Another factor with branch cut is the $(|M|\Omega)^{d/2}$
factor in (\ref{eq:Kfactor}).  The branch can be determined here
simply by rotating the 3-particle frequency $\Omega$ from positive
real values (where the correct result in unambiguous)
to $\Omega \propto e^{\pm i\pi/4}$.  In the case of $xy\bar y\bar x$,
$\Omega \propto e^{-i\pi/4}$ and so the phase of
$(|M|\Omega)^{d/2}$ is $(e^{-i\pi/4})^{d/2} = e^{-id\pi/8}$.

Finally, we have the
$\calX_\yx^{-d/2} \propto (-iM/\Delta t)^{-d/2}$
of (\ref{eq:J}) and (\ref{eq:J2}).
Remembering that the 3-particle $M$'s for
$x y \bar y\bar x$ are positive, we have
$\calX_\yx \propto -i$, which should be interpreted as
$\calX_\yx \propto e^{-i\pi/2}$.

Combining all of the above phases, $d\Gamma/dx\,dy$ in
(\ref{eq:poleDR0}) should have the phase of
\begin {multline}
  i^{-d}
    \times (-\hat x_1 \hat x_2 \hat x_3 \hat x_4)^{d/2}
    \times (|M|\Omega)^{d/2}
    \times \calX^{-d/2}
\\
  \propto
  (e^{i\pi/2})^{-d}
    \times 1
    \times (e^{-i\pi/4})^{d/2}
    \times (e^{-i\pi/2})^{-d/2}
  =
  e^{-i 3\pi d/8} .
\label {eq:sheetxyyx}
\end {multline}
This is indeed the same as the conventional sheet for
the factor $(i x_1 x_2 x_3 x_4 \Omega \sgn M)^{d/2}$ written in
that equation, which in the case at hand is
\begin {equation}
   (i x_1 x_2 x_3 x_4 \Omega \sgn M)^{d/2}
   \propto (-i \Omega)^{d/2}
   \propto (e^{-i 3\pi/4})^{d/2} .
\label {eq:sheetxyyxB}
\end {equation}
It is also the same as the right-hand side of
(\ref{eq:sheet1}) with (\ref{eq:sheet2}), which in this case is
\begin {equation}
  (i \Omega \sgn M)^{d/2} (x_1 x_2 x_3 x_4)^{d/2}
  \propto (i \Omega)^{d/2} (e^{-i\pi})^{d/2}
  \propto (e^{i\pi/4})^{d/2} (e^{-i\pi})^{d/2} .
\label {eq:sheetxyyxC}
\end {equation}


\subsection{\boldmath$x_1 x_2 x_3 x_4 > 0$ with \boldmath$M > 0$}

This case occurs for $x\bar y y\bar x$ and for the $\Omega_\ix \sgn M_\ix$
piece of
$x\bar y\bar x y$.  Here the 4-particle frequencies have conjugate
phases: $\Omega_+ \propto \sqrt{-i} = e^{-i\pi/4}$ and
$\Omega_- \propto \sqrt{+i} = e^{+i\pi/4}$.
As discussed in appendix E of ref.\ \cite{2brem},
and in this paper when introducing (\ref{eq:Apropmodified}),
that means that the $\A_-$ normal mode propagator
will be a conjugate propagator and so have a factor of
$[1/(-2\pi i)]^{-d/2}$ in (\ref{eq:ApropDR}) instead of $[1/(2\pi i)]^{d/2}$.
Its phase will therefore cancel with that of the $\A_+$ propagator,
and so there is no overall $i^{-d}$ associated with the 4-particle
propagator in this case.

Since $x_1 x_2 x_3 x_4$ is now positive,
$\det{\mathfrak M}$ and $\det{\mathfrak M}'$ are now negative in
(\ref{eq:Mfactors}).  However, this is actually irrelevant.
When we make the change of variables from the normal modes
variables to $(\C_{34},\C_{12})$ and $(\C_{41},\C_{23})$, the
Jacobean actually involves the absolute value of the transformation,
and so
$|\det{\mathfrak M}|^{d/4} |\det{\mathfrak M}'|^{d/4}$ instead
of (\ref{eq:Mfactors}).  The upshot is that the total overall
phase of the 4-particle propagator in this case is actually
that of
\begin {equation}
   i^{-d/2} (i^{-d/2})^* \, |\det{\mathfrak M}|^{d/4} |\det{\mathfrak M}'|^{d/4}
   \propto 1 .
\end {equation}

It did not matter for the case of $xy\bar y\bar x$,
but why did we not look ahead
and write absolute value signs in the factors
$(\det{\mathfrak M})^{d/4} (\det{\mathfrak M}')^{d/4}$
of (\ref{eq:CpropSmall}) in
the main text?  We knew with hindsight that leaving them off
would lead to a series of steps in the main text generating a final
expression that actually works in all cases.  The role of this
appendix is to verify that claim.

The rest of the discussion is similar to the $xy \bar y\bar x$ case
above, but now (\ref{eq:sheetxyyx}) is replaced by
\begin {equation}
  1
    \times |x_1 x_2 x_3 x_4|^{d/2}
    \times (|M|\Omega)^{d/2}
    \times \calX^{-d/2}
  \propto
  1
    \times 1
    \times (e^{-i\pi/4})^{d/2}
    \times (e^{-i\pi/2})^{-d/2}
  =
  e^{i \pi d/8} .
\label {eq:sheetxyyx2}
\end {equation}
(Here, $\Omega$ refers to a 3-particle $\Omega$, not to $\Omega_+$ or
$\Omega_-$.)
This matches the analogs of (\ref{eq:sheetxyyxB}) and (\ref{eq:sheetxyyxC}),
which are
\begin {equation}
   (i x_1 x_2 x_3 x_4 \Omega \sgn M)^{d/2}
   \propto (i \Omega)^{d/2}
   \propto (e^{i\pi/4})^{d/2}
\label {eq:sheetxyyx2B}
\end {equation}
and
\begin {equation}
  (i \Omega \sgn M)^{d/2} (x_1 x_2 x_3 x_4)^{d/2}
  \propto (i \Omega)^{d/2} 1^{d/2}
  \propto (e^{i\pi/4})^{d/2} .
\label {eq:sheetxyyx2C}
\end {equation}


\subsection{\boldmath$x_1 x_2 x_3 x_4 > 0$ with \boldmath$M < 0$}

This case applies to the $\tilde\Omega_\fx \sgn\tilde M_\fx$
piece of $x\bar y\bar x y$.
Note that $M<0$ implies that the corresponding 3-particle
$\tilde\Omega_\fx \propto e^{i\pi/4}$.

Finally, consider the factor
$\calX^{-d/2} \propto (-iM/\Delta t)^{-d/2}$.
Since the relevant $M < 0$, we have
$\calX \propto e^{i\pi/2}$.
The analogs of (\ref{eq:sheetxyyx2}--\ref{eq:sheetxyyx2C}) are
then
\begin {equation}
  1
    \times |x_1 x_2 x_3 x_4|^{d/2}
    \times (|M|\Omega)^{d/2}
    \times \calX^{-d/2}
  \propto
  1
    \times 1
    \times (e^{i\pi/4})^{d/2}
    \times (e^{i\pi/2})^{-d/2}
  =
  e^{-i \pi d/8} ,
\label {eq:sheetxyxy}
\end {equation}
\begin {equation}
   (i x_1 x_2 x_3 x_4 \Omega \sgn M)^{d/2}
   \propto (-i \Omega)^{d/2}
   \propto (e^{-i\pi/4})^{d/2}
\label {eq:sheetxyxyB}
\end {equation}
and
\begin {equation}
  (i \Omega \sgn M)^{d/2} (x_1 x_2 x_3 x_4)^{d/2}
  \propto (-i \Omega)^{d/2} 1^{d/2}
  \propto (e^{-i\pi/4})^{d/2} .
\label {eq:sheetxyxyC}
\end {equation}

The upshot is that everything works: the overall phase is given
by the factors of (\ref{eq:sheet}) in all cases relevant to our
calculation.


\section{\boldmath$x\bar x y\bar y$ in terms of
         \boldmath$(\bar\alpha,\bar\beta,\bar\gamma)$}
\label{app:xxyy}

In this appendix, we rederive the result for $x\bar x y\bar y$
of section \ref{sec:xxyy} except that we keep the entire discussion
in terms of $d$-dimensional $(\bar\alpha,\bar\beta,\bar\gamma)$
instead of $d$-dimensional $P(x)$ and $P(\yfrak)$.
Initially, we will follow the method of the $d{=}2$ derivation given
in Appendix C of ref.\ \cite{seq}, which starts with the general expression
\begin {multline}
   \left[\frac{dI}{dx\,dy}\right]_{x\bar x y\bar y}
   =
   \left( \frac{E}{2\pi} \right)^2
   \int_{t_\xx < t_\xbx < t_\yx < t_\ybx}
   \sum_{\rm pol.}
   \langle|i\,\overline{\delta H}|\B^\ybx\rangle
   \langle\B^\ybx,t_\ybx|\B^\yx,t_\yx\rangle
   \langle\B^\yx|{-}i\,\delta H|\rangle
\\ \times
   \langle|\rangle^{-1}
   \langle|i\,\overline{\delta H}|\B^\xbx\rangle
   \langle\B^\xbx,t_\xbx|\B^\xx,t_\xx\rangle
   \langle\B^\xx|{-}i\,\delta H|\rangle .
\label {eq:xxyy0}
\end {multline}
In $d$ dimensions, this gives
\begin {align}
   \left[\frac{dI}{dx\,dy}\right]_{x\bar xy\bar y}
   &=
   \frac{C_R^2 \alphas^2 }{4 E^{2d}} \,
   (1-x)^{-d}
   \int_{t_\xx < t_\xbx < t_\yx < t_\ybx}
   \sum_{h_\xx,h_\yx,h_\zx}
\nonumber\\ &\times
   \Bigl[
   \sum_{\bar h}
   {\cal P}^{\bar n}_{\bar h \to h_\zx,h_\yx}\bigl(1{-}x \to 1{-}x{-}y,y\bigr) \,
   {\cal P}^{\bar m}_{h_\ix \to \bar h, h_\xx}\bigl(1 \to 1{-}x,x\bigr)
   \Bigr]^*
\nonumber\\ &\times
   \Bigl[
   \sum_h
   {\cal P}^n_{h \to h_\zx,h_\yx}\bigl(1{-}x \to 1{-}x{-}y,y\bigr) \,
   {\cal P}^m_{h_\ix \to h,h_\xx}\bigl(1 \to 1{-}x,x\bigr)
   \Bigr]
\nonumber\\ &\times
   \nabla^{\bar n}_{\B^\ybx}
   \nabla^n_{\B^\yx}
   \langle\B^\ybx,t_\ybx|\B^\yx,t_\yx\rangle
   \Bigr|_{\B^\ybx=0=\B^\yx}
   \nabla^{\bar m}_{\B^\xbx}
   \nabla^m_{\B^\xx}
   \langle\B^\xbx,t_\xbx|\B^\xx,t_\xx\rangle
   \Bigr|_{\B^\xbx=0=\B^\xx}
\label {eq:xxyy1DR}
\end {align}
and thence, using the definition of $(\bar\alpha,\bar\beta,\bar\gamma)$
from ACI (E2--E3),
\begin {align}
   \left[\frac{dI}{dx\,dy}\right]_{x\bar xy\bar y}
   &=
   \frac{C_R^2 \alphas^2 }{4 E^{2d}} \,
   \frac{
     ( \bar\alpha \delta^{\bar n n} \delta^{\bar m m}
     {+} \bar\beta \delta^{\bar n \bar m} \delta^{nm}
     {+} \bar\gamma \delta^{\bar n m} \delta^{n \bar m} )
   }{(1-x)^d}
   \int_{t_\xx < t_\xbx < t_\yx < t_\ybx}
\nonumber\\ &\times
   \nabla^{\bar n}_{\B^\ybx}
   \nabla^n_{\B^\yx}
   \langle\B^\ybx,t_\ybx|\B^\yx,t_\yx\rangle
   \Bigr|_{\B^\ybx=0=\B^\yx}
   \nabla^{\bar m}_{\B^\xbx}
   \nabla^m_{\B^\xx}
   \langle\B^\xbx,t_\xbx|\B^\xx,t_\xx\rangle
   \Bigr|_{\B^\xbx=0=\B^\xx} .
\end {align}
The only difference here is the normalization of $\langle|\rangle^{-1}$
according to (\ref{eq:N2norm}), the use of $d$-dimensional
$(\bar\alpha,\bar\beta,\bar\gamma)$, and the overall
factor of $E^{-2d}$ to keep $(\bar\alpha,\bar\beta,\bar\gamma)$
dimensionless.
Next, use rotation invariance to write
\begin {equation}
   \nabla^{\bar n}_{\B^\ybx}
   \nabla^n_{\B^\yx}
   \langle\B^\ybx,t_\ybx|\B^\yx,t_\yx\rangle
   \Bigr|_{\B^\ybx=0=\B^\yx}
   =
   \frac{\delta^{n\bar n}}{d} \,
   \grad_{\B^\ybx} \cdot \grad_{\B^\yx}
   \langle\B^\ybx,t_\ybx|\B^\yx,t_\yx\rangle
   \Bigr|_{\B^\ybx=0=\B^\yx}
\end {equation}
(and similarly for the other factor)
to give
\begin {align}
   \left[\frac{dI}{dx\,dy}\right]_{x\bar xy\bar y}
   &=
   \frac{C_R^2 \alphas^2 }{4 E^{2d}} \,
   \frac{
     ( \bar\alpha {+} \tfrac{1}{d} \bar\beta {+} \tfrac{1}{d} \bar\gamma )
   }{(1-x)^d}
   \int_{t_\xx < t_\xbx < t_\yx < t_\ybx}
\nonumber\\ &\times
   \grad_{\B^\ybx} \cdot \grad_{\B^\yx}
   \langle\B^\ybx,t_\ybx|\B^\yx,t_\yx\rangle
   \Bigr|_{\B^\ybx=0=\B^\yx}
   \grad_{\B^\xbx} \cdot \grad_{\B^\xx}
   \langle\B^\xbx,t_\xbx|\B^\xx,t_\xx\rangle
   \Bigr|_{\B^\xbx=0=\B^\xx} .
\end {align}
Using the formula for
$\grad_{\B} \cdot \grad_{\B'}
     \langle \B,\Delta t | \B',0 \rangle
     \Bigr|_{\B = \B' = 0}$
from (\ref{eq:integral0}),
\begin {align}
   \left[\frac{dI}{dx\,dy}\right]_{x\bar xy\bar y}
   &=
   \frac{d^2\pi^2 C_R^2 \alphas^2 }{E^{2d}} \,
   \left( \frac{M_\ix \Omega_\ix}{2\pi i} \right)^{\frac{d}{2}+1}
   \left( \frac{M_\fx^\seq \Omega_\fx^\seq}{2\pi i} \right)^{\frac{d}{2}+1}
   \frac{
     ( \bar\alpha {+} \tfrac{1}{d} \bar\beta {+} \tfrac{1}{d} \bar\gamma )
   }{(1-x)^d}
   \int_{t_\xx < t_\xbx < t_\yx < t_\ybx}
\nonumber\\ &\times
   \csc^{\frac{d}{2}+1}(\Omega_\ix\,\Delta t_\xx)
   \csc^{\frac{d}{2}+1}(\Omega_\fx^\seq\,\Delta t_\yx) .
\end {align}
Taking $2\Re$ of each factor in order to add in the related diagrams
$x\bar x\bar y y + \bar x x y \bar y + \bar x x\bar y y$,
and then subtracting the corresponding Monte Carlo contribution
[as in ACI (2.19--21)],
\begin {align}
   \left[\Delta \frac{d\Gamma}{dx\,dy}\right]_{
        \begin{subarray}{} x\bar x y\bar y + x\bar x \bar y y \\
                        + \bar xx \bar yy + \bar x xy\bar y \end{subarray}
   }
   &=
   - \frac{4d^2\pi^2 C_R^2 \alphas^2 }{E^{2d}}
   \frac{
     ( \bar\alpha {+} \tfrac{1}{d} \bar\beta {+} \tfrac{1}{d} \bar\gamma )
   }{(1-x)^d}
   \int_0^\infty d(\Delta t_\xx) \int_0^\infty d(\Delta t_\yx) \>
     \tfrac12 (\Delta t_\xx + \Delta t_\yx)
\nonumber\\ &\times
   \Re\left[
     \left( \frac{M_\ix \Omega_\ix}{2\pi i} \right)^{\frac{d}{2}+1}
     \csc^{\frac{d}{2}+1}(\Omega_\ix\,\Delta t_\xx)
   \right]
\nonumber\\ &\times
   \Re\left[
     \left( \frac{M_\fx^\seq \Omega_\fx^\seq}{2\pi i} \right)^{\frac{d}{2}+1}
     \csc^{\frac{d}{2}+1}(\Omega_\fx^\seq\,\Delta t_\yx)
   \right].
\end {align}
Using%
\footnote{
  ACI (2.25), ACI (2.27), and ACI footnote 21.
}
\begin {align}
   M_\ix &= x(1{-}x)E = M_{E,x} ,
\\[5pt]
   M_\fx^\seq &= (1{-}x)y(1{-}x{-}y)E
\nonumber\\ &\qquad
  = (1{-}x)^2 \times \yfrak(1{-}\yfrak)(1{-}x)E
  = (1{-}x)^2 M_{(1-x)E,\yfrak} ,
\\[5pt]
  \Omega_\ix &= \Omega_{E,x},
\\[5pt]
  \Omega_\fx &= \Omega_{(1-x)E,\yfrak},
\end {align}
gives
\begin {align}
   \left[\Delta \frac{d\Gamma}{dx\,dy}\right]_{
        \begin{subarray}{} x\bar x y\bar y + x\bar x \bar y y \\
                        + \bar xx \bar yy + \bar x xy\bar y \end{subarray}
   }
   &=
   - (1{-}x)^2
   ( \bar\alpha {+} \tfrac{1}{d} \bar\beta {+} \tfrac{1}{d} \bar\gamma )
   \int_0^\infty d(\Delta t_\xx) \int_0^\infty d(\Delta t_\yx) \>
     \tfrac12 (\Delta t_\xx + \Delta t_\yx)
\nonumber\\ &\times
   \Re\left[
     \frac{2d\pi C_R \alphas}{E^{d}}
     \left( \frac{M_{E,x} \Omega_{E,x}}{2\pi i} \right)^{\frac{d}{2}+1}
     \csc^{\frac{d}{2}+1}(\Omega_{E,x}\,\Delta t_\xx)
   \right]
\nonumber\\ &\times
   \Re\left[
     \frac{2d\pi C_R \alphas}{E^{d}}
     \left( \frac{M_{(1-x)E,\yfrak} \Omega_{(1-x)E,\yfrak}}{2\pi i}
       \right)^{\frac{d}{2}+1}
     \csc^{\frac{d}{2}+1}(\Omega_{(1-x)E,\yfrak}\,\Delta t_\yx)
   \right].
\end {align}
As promised in the main text, that's the same as
(\ref{eq:DxxyyetcDR}) with (\ref{eq:otherintDR}) but here
with the replacement (\ref{eq:abcPP2DR}).


\end {document}